\documentclass[%
 reprint,
superscriptaddress,
 amsmath,amssymb,
 aps,
prx,
]{revtex4-2}

\usepackage{graphicx}%
\usepackage{multirow}%
\usepackage{amsmath,amssymb,amsfonts}%
\usepackage{amsthm}%
\usepackage{mathrsfs}%
\usepackage[title]{appendix}%
\usepackage{xcolor}%
\usepackage[normalem]{ulem}
\usepackage{braket}
\usepackage[T1]{fontenc}
\usepackage{float}
\usepackage{dcolumn}
\usepackage{bm}

\newcommand{\removed}[1]{\sout{Removed.}}

\newcommand{\br}{\mathbf{r}}
\newcommand{\bR}{\mathbf{R}}

\newcommand{\bP}{\mathbf{P}}
\newcommand{\bK}{\mathbf{K}}
\newcommand{\bq}{\mathbf{q}}
\newcommand{\bk}{\mathbf{k}}

\newcommand{\bG}{\mathbf{G}}
\newcommand{\bn}{\mathbf{n}}
\newcommand{\bu}{\mathbf{u}}
\newcommand{\bg}{\mathbf{g}}
\newcommand{\Gammapp}{\tilde{\Gamma}}

\newcommand{\txa}{\text{a}}
\newcommand{\txb}{\text{b}}
\newcommand{\txc}{\text{c}}
\newcommand{\txd}{\text{d}}
\newcommand{\txe}{\text{e}}
\newcommand{\txn}{\text{n}}
\newcommand{\txp}{\text{p}}
\newcommand{\txr}{\text{r}}
\newcommand{\txs}{\text{s}}
\newcommand{\txt}{\text{t}}
\newcommand{\txD}{\text{D}}
\newcommand{\txH}{\text{H}}
\newcommand{\txL}{\text{L}}
\newcommand{\txF}{\text{F}}
\newcommand{\txR}{\text{R}}
\newcommand{\txen}{\text{en}}
\newcommand{\txnn}{\text{nn}}
\newcommand{\txph}{\text{ph}}
\newcommand{\txxc}{\text{xc}}
\newcommand{\txuu}{\text{uu}}
\newcommand{\txud}{\text{ud}}
\newcommand{\txion}{\text{ion}}
\newcommand{\txeff}{\text{eff}}
\newcommand{\txCT}{\text{CT}}

\newcommand{\txRPA}{\text{RPA}}

\newcommand{\txUEG}{\text{UEG}}
\newcommand{\txFS}{\text{FS}}
\newcommand{\txKS}{\text{KS}}
\newcommand{\txeph}{\text{e-ph}}

\newcommand{\pcf}{\Lambda} 
\newcommand{\verthree}{\Gamma_{3}} 
\newcommand{\psp}{\mu} %
\newcommand{\nambu}[1]{\hat{#1}}

\begin{document}

\title{Superconductivity in Electron Liquids:  Precision Many-Body Treatment of Coulomb Interaction}

\author{Xiansheng Cai}
\affiliation{Institute of Theoretical Physics, Chinese Academy of Sciences, Beijing 100190, China}
\affiliation{Department of Physics, University of Massachusetts, Amherst, MA 01003, USA}
\author{Tao Wang}
\affiliation{Department of Physics, University of Massachusetts, Amherst, MA 01003, USA}
\affiliation{Institute of Physics, Chinese Academy of Sciences, Beijing 100190, China}
\author{Shuai Zhang}
\affiliation{Institute of Theoretical Physics, Chinese Academy of Sciences, Beijing 100190, China}
\author{Tiantian Zhang}
\email{ttzhang@itp.ac.cn}
\affiliation{Institute of Theoretical Physics, Chinese Academy of Sciences, Beijing 100190, China}
\author{Andrew Millis}
\affiliation{Center for Computational Quantum Physics, Flatiron Institute}
\affiliation{Department of Physics, Columbia University, New York}
\author{Boris V. Svistunov}
\affiliation{Department of Physics, University of Massachusetts, Amherst, MA 01003, USA}
\author{Nikolay V. Prokof'ev}
\affiliation{Department of Physics, University of Massachusetts, Amherst, MA 01003, USA}
\author{Kun Chen}
\email{chenkun@itp.ac.cn}
\affiliation{Institute of Theoretical Physics, Chinese Academy of Sciences, Beijing 100190, China}
\affiliation{Center for Computational Quantum Physics, Flatiron Institute}

\date{\today}
\begin{abstract}
More than a  century after its discovery, the theory of superconductivity in conventional metals has remained incomplete. While the crucial importance of the electron-phonon coupling is understood,  a theoretically controlled first-principles treatment of the Coulomb interaction has yet to be formulated. The downfolding approximation widely employed in existing \emph{ab initio} calculations of conventional superconductors is based on a phenomenological replacement of the Coulomb interaction by a repulsive pseudopotential, $\mu^*$, while ambiguities in approximating the electron-phonon coupling in the presence of dynamical Coulomb interactions have remained unresolved. We address these limitations through an effective field theory approach based on integrating out high-energy electronic degrees of freedom using variational Diagrammatic Monte Carlo. Applying the theory to the uniform electron gas establishes a quantitative microscopic procedure to implement the downfolding approximation, define the pseudopotential, and express the effect of the dynamical Coulomb interaction on the electron-phonon coupling through the electron vertex function. We find that the bare pseudopotential is significantly larger than the conventional, phenomenologically defined values. These results provide improved estimates of the Coulomb pseudopotential in simple metals and enable tests of the accuracy of the density functional perturbation theory in describing the effective electron-phonon coupling. We present an \emph{ab initio} workflow for computing the superconducting $T_c$ from the precursory Cooper flow of the anomalous vertex, that allows us to infer the superconducting transition temperature from  normal state calculations, enabling reliable estimates even of very low $T_c$ values (including superconductivity in the proximity of quantum phase transition points) beyond the reach of conventional methods. We validate our approach by computing $T_c$ for simple metals without empirical tuning of parameters, resolve long-standing discrepancies between the theory and experiment, and predict a pressure-induced quantum phase transition from a superconducting to a non-superconducting state in Al as the pressure is increased above a critical value $\sim$ 60 GPa. We propose that ambient-pressure Mg and Na are proximal to a similar critical point. Our work establishes a controlled \emph{ab initio} framework for electron-phonon superconductivity beyond the weak electron correlation limit, paving the way for reliable $T_c$ calculations and design of novel superconducting materials.
\end{abstract}

\maketitle


\section{Introduction}

Superconductivity (SC), a macroscopic quantum phenomenon with far-reaching fundamental physics and technological implications, is a focal point of condensed matter research. 
For the first half century after its discovery, lack of a theoretical basis for understanding the physics of interacting electrons inhibited the development of a theory of superconductivity. 
Following the work of Fröhlich~\cite{frohlich1950} and the experimental confirmation by the Rutgers and NBS groups~\cite{reynolds1950,maxwell1950},
the importance of the electron phonon interaction was recognized, but how it competed with Coulomb repulsion was not clear. Bardeen, Cooper and Schrieffer (BCS) then provided our basic understanding of superconductivity in terms of a phonon-mediated attraction leading to a pairing instability~\cite{cooper_bound_1956-1,bardeen_theory_1957,bardeen_microscopic_1957}. 
The original BCS paper took the radical step of ignoring Coulomb effects altogether but among many other important concepts introduced the idea that if a weak pairing interaction existed at some energy scale, its importance increased as the scale was lowered.  This meant that superconductivity should be understood as a low energy instability of the electron gas and that the competition was between a phonon-mediated interaction and an effective electron-electron repulsion defined at the scale of the phonon frequency.  This insight  motivated studies of the effective Coulomb interaction in the pairing (Cooper) channel at low energy, obtained by ``downfolding'' the fundamental Coulomb interaction. The first downfolding argument was provided by Bogoliubov, Tolmachev, and Shirkov \cite{bogoliubov59,tolmachev1961logarithmic} who showed that the same Cooper logarithm that leads to SC for attractive interactions will renormalize a frequency independent local Coulomb repulsion to the weak coupling limit at low energies.  The resulting dimensionless parameter, $\mu^* > 0$ describing the effect of Coulomb interactions in the pairing channel at the scale of the phonon frequency, was later named the ``pseudopotential'' by Morel and Anderson \cite{morel_anderson}).

The microscopic description of phonon-mediated attraction is encoded in the Migdal-Eliashberg  theory, 
which goes beyond the original BCS theory by rigorously treating the dynamic electron-phonon (e-ph) interaction and provides
the foundational framework for understanding phonon-mediated superconductors~\cite{migdal1958,eliashberg1960interactions,eliashberg1961temperature}.
While the original formulation~\cite{migdal1958,eliashberg1960interactions} focused on the e-ph interaction, subsequent developments~\cite{tolmachev1961logarithmic,morel_anderson} incorporated Coulomb repulsion via the static pseudopotential $\mu^*$ and following modern convention we refer to this combined theory as ``Migdal-Eliashberg'' or ME theory.
ME theory as conventionally implemented relies on a ``downfolding'' approximation \cite{eliashberg1960interactions, eliashberg1961temperature, tolmachev1961logarithmic, mcmillan1968, allen1975, EPW1,EPW2,AEPW} that takes advantage of the large energy scale separation between a typical Fermi energy, $E_\txF$, and a typical phonon frequency $\omega_\txD$ to project the equations into the low-energy subspace defined by an effective Fermi-surface theory of quasiparticles, replacing bare e-ph interactions with screened effective e-ph coupling, $\lambda$ \cite{eliashberg1960interactions,eliashberg1961temperature}.
It is important to note that although the adiabatic limit (when $\omega_\txD/E_\txF \ll 1$) justifies the formulation in terms of a downfolded theory, as it is currently implemented many aspects of the ME theory remain semi-phenomenological.
Coulomb-induced quasiparticle renormalizations are neglected, as are possible renormalizations of the electron-phonon coupling by Coulomb fluctuations {as well as non-local effects coming from a scale dependence of screening, particularly pronounced in two dimensional materials \cite{Simonato23}.}
The Coulomb pseudopotential $\mu^*$ is chosen in an ad-hoc or phenomenological manner ~\cite{RAINER1986371}.
This creates critical challenges both for first-principles predictions of transition temperatures and for quantifying unconventional, i.e. not involving e-ph coupling, pairing mechanisms.

Perhaps the most controversial issue is the magnitude and sign of $\mu^*$.  The first theoretical estimates were based on the unphysical approximation that Coulomb interactions are screened at all frequencies. When this approximation was subsequently  removed by employing the ``random phase'' approximation (RPA) to compute the more physical dynamically screened interaction, Takada \cite{takada1978plasmon, takada1993} and Rietschel and Sham \cite{rietschel} found that the dynamical effects of the Coulomb potential may lead to \textit{attractive} $\mu^* < 0$ and s-wave superconductivity even in the absence of phonons, if the electron-electron interaction is strong enough. Recent studies again within the RPA, now extended to all paring channels \cite{23tao, xiansheng, pengcheng} confirm the validity of this theoretical conclusion. 

The RPA-based finding of purely electronically mediated superconductivity is inconsistent with a large body of empirical evidence, and is also internally inconsistent. For a recent discussion see Ref.~\cite{dassarma2025}.  In the electron gas, the strength of the Coulomb interaction is parametrized by the dimensionless Wigner-Seitz radius $r_\txs=\left(\frac{9\pi}{4}\right)^{1/3} \frac{m e^2}{\hbar^2 k_{\text F}}$, in effect the ratio of interaction to kinetic energies at the scale of the interelectron spacing. The RPA is a good approximation for $r_\txs \lesssim 1$ while the Coulombic pairing was found at $r_\txs\gtrsim 2$ where beyond-RPA effects, including vertex corrections and renormalization of the single-particle propagators \cite{RPA_mu2, 23tao}, may be important.

Due to these theoretical uncertainties, $\mu^*$ is often treated as an adjustable parameter, with values empirically chosen between 0.1 and 0.2 to reconcile theory with experiments. This range leads to large (sometime orders of magnitude) uncertainty when predicting $T_c$ in materials where SC depends on a delicate balance between phonon-mediated attraction and $\mu^*$.  For instance, ME theory predictions for $T_c$ in aluminum, a workhorse of superconducting electronics for transition-edge sensors~\cite{irwin_transition-edge_2005} and qubits~\cite{qubit_poisoning}, deviate from experiments by a factor of two~\cite{pellegrini2024ab}, while 
for elemental lithium the theory overestimates $T_c$ by three orders of magnitude~\cite{ashcroft1,hubbard_mu, tuoriniemi_superconductivity_2007}. These discrepancies severely limit the possibility of reliable predictive
design for quantum devices requiring precise control of energy gaps and thermal noise. Even more strikingly, some transition-metal based compounds (e.g., V, Nb$_3$Sn \cite{RevModPhys.62.1027}), alkali-doped picene \cite{PhysRevB.84.020508}, and high-pressure hydrides \cite{kostrzewa_anomalously_2018} are best described by 
$\mu^*$ values from $0.2$ to $0.5$, well outside of the limits following from the static \cite{morel_anderson} and dynamic RPA \cite{margine2016electron,RPA_mu,RPA_mu2}. While more advanced theories such as the T-matrix approximation \cite{scalapino1969electron, gladstone2018superconductivity} have been proposed to address the discrepancy, their development rather underscores the need for a systematic precise microscopic theory of $\mu^*$ as opposed to the case-selective use of uncontrolled approximations. 

Equally important is accurate treatment of the effective coupling, $\lambda$, which quantifies the phonon-mediated attraction between quasiparticles at the Fermi surface within the downfolding procedure. 
State-of-the-art \emph{ab initio} methods, such as the density functional perturbation theory (DFPT) \cite{giustino_2017,EPW1, EPW2, AEPW, pellegrini2024ab}, calculate $\lambda$ from the ground-state energy response to lattice distortions, an approximation validated for weakly correlated superconductors \cite{RAINER1986371}. However, it remains unknown 
how accurate this proxy is for correlated systems, where strong renormalization effects are expected to alter effective e-ph interactions and thus SC~\cite{RPA_mu,RPA_mu2,zfactor_2022,23tao}. Removing this uncertainty is crucial for extending the predictive power
of such a method as density functional theory plus dynamical mean-field theory (DFT+DMFT)~\cite{dftdmft1,dftdmft2,dftdmft3,dftdmft4,dftdmft5,dftdmftreview,dmft_sc1,dmft_sc2} to strongly correlated superconductors. 

The downfolded ME framework also faces conceptual challenges on top of practical difficulties with precise evaluation of $\mu^*$ and $\lambda$. The conventional approach becomes inadequate in the strong e-ph coupling regime, e.g. in high-pressure hydrides with large quantum nuclear effects \cite{kostrzewa_anomalously_2018}, 
near structural transitions, or due to the formation of bipolarons~\cite{bipolaron1,bipolaron2,bipolaron3}. 

These limitations have motivated efforts to transcend the downfolding procedure altogether. Richardson and Ashcroft suggested solving the system of ME equations ``as is'' with full momentum-frequency dependence of both the e-ph and screened (at the RPA level)  e-e interactions~\cite{ashcroft1, ashcroft2}.  Their approach successfully predicted $T_c$ for lithium~\cite{tuoriniemi_superconductivity_2007}, but its reliance on the RPA casts doubt on its applicability for electron densities relevant to most metals. This and the large computational cost have 
prevented this method from becoming widely adopted {(but see \cite{veld23} for recent applications in two dimensions)}.
Superconducting DFT~\cite{SCDFT1,SCDFT2,profeta_superconductivity_2006} offers an alternative by generalizing DFT to include the superconducting order parameter, with screening effects approximated by the RPA-based ansatz. Recently, it was adapted within the ME framework~\cite{scdft-eliashberg}, offering a potentially fruitful avenue for future research.  However, such issues as contributions coming from spin fluctuations~\cite{kawamura_benchmark_2020} and vertex corrections to the effective coupling remain unresolved, {and the method as implemented to date relies on the RPA.  To our knowledge, a consistent and quantitatively accurate microscopic treatment of Coulomb effects in the theory of SC has not yet been achieved.}

In this paper we present a theory of the Cooper (pairing) instability in an interacting electron gas coupled to phonons. The theory is founded on the separation of scales ideas that justify the Migdal-Eliashberg and Fermi Liquid theories but takes advantages of modern developments in the effective field theory of interacting fermion systems and of recent progress in diagrammatic Monte Carlo that enables calculations of relevant quantities, so that we are able to obtain a downfolded ME formulation beyond the limit of weak correlations. Our effective low-energy theory involves Fermi-surface quasiparticles coupled to phonons with precise expressions for Coulomb pseudopotential $\mu^*$ and electron-phonon coupling $\lambda$  in terms of the electron vertex functions computed via modern many-body techniques. Our work reconciles the contradiction between phenomenological and RPA treatments of the problem  by validating the use of the local, instantaneous, and universal $\mu^*$ for conventional materials and resolves issues of the renormalization of  electron-phonon coupling and the proper placement of the quasiparticle weight factor.

Importantly, we also provide limits of validity of the theory, which fails at extremely high electron densities when the  plasmon frequency, $\omega_\txp$,  softens, in 2D when the plasmon mode is no longer gapped, in materials with soft collective excitations emerging from strong correlations, and systems with low conduction electron density,  revealing regimes where the dynamic nature of screening is most relevant.
While 3D electron densities corresponding to $\omega_\txp \ll E_\txF$ are beyond current experimental reach terrestrially, they can be found in ultra-dense astrophysical objects~\cite{trubnikov1968white, ginzburg_superconductivity_1968}. 

As a critical application, we compute Coulomb pseudopotentials for the uniform electron gas (UEG) using recently developed variational diagrammatic Monte Carlo (DiagMC) method \cite{diagmc1,diagmc2,diagmc2010,detdiagmc,diagmclatest,diagmckozik,vdiagmc1,vdiagmc2}. These results provide precise parameterization of $\mu^*$ as a function of electron density to be used in ME treatment of real materials. Our estimates of the ``bare'' pseudopotential, obtained by  first computing the quasiparticle scattering amplitude at the Fermi surface  and then renormalizing it via the Cooper/Tolmachev logarithm, are significantly larger than estimates based on static screening or RPA in the moderate-density regime, bridging a crucial gap between 
theory and experiment. We also compute the quasiparticle e-ph vertex for the UEG, finding that DFPT yields remarkably accurate electron-phonon coupling values for simple metals.

By combining these advances with calculations of the precursory Cooper flow (PCF) of the anomalous vertex function 
in the normal state~\cite{pengcheng}, we propose an \emph{ab initio} workflow, capable of predicting low-temperature SC and quantum phase transition points well beyond the reach of conventional methods. 
With this approach, we revisited the problem of superconducting $T_\txc$ in various simple metals under ambient and high-pressure conditions. Our method yields orders-of-magnitude improvements in predicting $T_\txc$ for sub-Kelvin superconductors compared to previous estimates based on a phenomenological $\mu^*$.  We find that Mg and Na are close  to a normal-SC quantum phase transition, offering an opportunity to study quantum critical scaling below $10$K. We also predict that Al will undergo a transition to the normal state under pressure exceeding $60$GPa.  In the UEG context, this work establishes a complete and controlled first-principles framework for understanding the interplay between the electron correlations and e-ph interactions in SC. If rigorously followed in simulations of real materials, it will radically improve our ability to design next-generation superconducting materials.

The rest of the paper is organized as follows. In Sec.~\ref{sec:model}, we establish the effective field theory for the coupled electron-phonon system and introduce the Bethe-Salpeter equation (BSE) formalism used to detect superconducting instabilities. Sec.~\ref{sec:downfolding} provides the microscopic derivation of the downfolding approximation, rigorously integrating out high-energy degrees of freedom to establish the exact relations between the quasiparticle vertex functions and the Coulomb pseudopotential $\mu^*$ and the effective coupling $\lambda$. Building on this framework, Sec.~\ref{sec:pseudopotential} focuses on the quantitative determination of $\mu^*$ using high-order Variational Diagrammatic Monte Carlo (VDiagMC) simulations of the Uniform Electron Gas. In Sec.~\ref{sec:electronphonon}, we address the effective coupling $\lambda$ by benchmarking standard Density Functional Perturbation Theory (DFPT) against our many-body theory within the UEG model, thereby establishing the precise correspondence between the DFPT-derived interaction and the effective coupling in our theory. Finally, in Sec.~\ref{sec:realmaterial}, we combine these \textit{ab initio} inputs within the precursory Cooper flow framework to compute $T_{c}$ for a series of elemental metals, resolving discrepancies in lithium and predicting quantum critical behavior in magnesium and sodium. 
Appendices contain technical details of the calculations.


\section{The Model and Basic Relations}
\label{sec:model}
\subsection{Electron-Phonon Problem}  

Our goal is to address electron-phonon SC in crystals using effective field theory (EFT)~\cite{polchinski_effective_1999} derived on the assumption that the electron mass $m$ is much less than an ionic mass $M$.
The mass difference has three essential consequences: the typical phonon frequency $ \omega_\txD $ is suppressed relative to $ E_\txF $ by a factor of $ (m/M)^{1/2} $, ensuring that electrons adiabatically adjust to the ionic motion;  the momentum transferred by an electron to an ion in collision is very small, justifying a linearization of the electron-ion coupling; and the separation of spatial and temporal scales of electronic and phononic physics justifies a controlled EFT treatment--in particular, once properties of the pure Coulomb system (no phonons) are established at energies greater than $\omega_\txD$, further corrections to the e-ph vertex are suppressed by $ (m/M)^{1/2} $  allowing for simple perturbative analysis of the low-energy theory.   

For transparency, we present the formalism for systems with a near-spherical Fermi surface confined to a single Brillouin zone (e.g., alkali metals) such that all low-energy processes can be parameterized by the crystal momentum $\bk$. This formalism also applies to simple multivalent metals (e.g., Mg, Al): while their Fermi surfaces span multiple Brillouin zones, an approximate rotation symmetry in the extended zone scheme is still present due to weak ionic potential and Umklapp scattering. For these systems, the total momentum~$\bK = \bk + \bG_m$ (where~$\bG_m$ is a reciprocal lattice vector) is approximately conserved and can be used in place of the crystal momentum $\bk$.
This correspondence simplifies the presentation while retaining physical clarity. For materials with strong Umklapp scattering and lattice potentials, we provide a generalized~$(\bk, \bG)$ formalism in Appendix~\ref{sec:app_epi}, ensuring broad applicability without obscuring the core physics.   

By keeping terms up to quadratic order in ionic displacements (see Appendix \ref{sec:app_epi}), we obtain an effective 
action accurate up to $O(\sqrt{m/M})$ corrections:
\begin{equation}
\label{eq:full_action}
    S=S_{txe}[\bar{\psi}, \psi] + S_{\txph}[u] + S_{\txeph}[\bar{\psi}, \psi, u] +S_{\txCT}[u]+O\left(\sqrt{\frac{m}{M}}\right),
\end{equation}
where $\psi$, $\bar{\psi}$ are Grassmann fields for electrons, and $u$ is the ionic displacement field rescaled by $\sqrt{M}$ (this standard procedure applies to any number of ions in the unit cell). Here, $S_\txe$ stands for full many-electron action without any approximations. We distinguish our theory from the conventional DFT approach, which replaces $S_\txe$ with a non-interacting reference Hamiltonian governed by an approximate exchange-correlation potential~\cite{kohn1964,kohn1965}. The DFT approach does not capture the dynamical effects of electron-electron interactions.

For simplicity we take the e-ph coupling $S_{\txeph}$ to have the density-displacement form 
\begin{equation}
    S_{\txeph} =\sum_{\kappa}\int_{\bq\nu} g^{(0)}_{\kappa}(\bq)
    n_{\bq\nu} u_{\kappa\bq\nu},
    \label{eq:phaction}
\end{equation}
summed over phonon branches $\kappa$, momenta $\bq$ in the Brillouin zone,
and Matsubara frequencies $\nu$. This approximation captures the coupling of the electronic density to longitudinal phonon modes and is adequate for the simple case of nearly free electron materials with electronic eigenstates that are close to plane waves; the generalization to more complicated situations is straightforward but notationally cumbersome and will not be considered here.

$S_{\txph}$ describes phonons with  physical dispersion $\omega_{\kappa\bq}$,
\begin{equation}
\label{eq:phonon_action}
   S_{\txph} = \frac{1}{2}\sum_{\kappa}\int_{\bq\nu} D^{-1}_{\kappa\bq\nu} \left|u_{\kappa\bq\nu}\right|^2,
\end{equation}
where $D_{\kappa\bq\nu}=-1/(\nu^2+\omega_{\kappa\bq}^2)$ is the phonon propagator. The physical dispersion $\omega_{\kappa\bq}$ is characterized by the Debye frequency, $\omega_\txD$. In the case we consider, the Debye frequency is assumed to be much smaller than the Fermi energy, $E_\txF$, i.e. $\omega_\txD\ll E_\txF$. This separation of energy scales will be the basis for the approximations made later.

Defining $S_{\txph}$ in terms of the physical (experimentally measured) dispersion $\omega_{\kappa\bq}$ is convenient, because within the Migdal approximation $\omega_{\kappa\bq}$ is determined by the ion masses and interionic forces defined from the change in total energy with respect to static displacements of ions from their equilibrium positions; the changes in energy may  in many cases be accurately computed using known \emph{ab-initio} methods. However, this choice means that in the theoretical analysis one must take care of double-counting: since the
renormalized phonon spectrum already includes screening effects from the static limit of the electron polarization. When treating the e-ph coupling $S_{\txph}$ perturbatively the static limit of the electronic polarization contribution to the phonon spectrum should be excluded. Formally, this is done by introducing a counterterm $S_{\txCT}$, which subtracts the corresponding screening contributions. By construction, it is given by the zero frequency limit of the electron charge susceptibility $\chi_{\bq}^\txe$, which quantifies how electrons respond to density fluctuations:
\begin{equation}
\label{eq:ct_action}
S_{\text{CT}} = -\frac{1}{2}\sum_{\kappa}\int_{\bq\nu} \left(g^{(0)}_{\kappa\bq}\right)^2 \chi_{\bq}^\text{e} \left|u_{\kappa\bq\nu}\right|^2.
\end{equation}
This formulation ensures that the phonon propagator retains its physical properties and no double-counting takes place when performing perturbative expansion in terms of the e-ph interaction.


\subsection{Bethe-Salpeter Equation (BSE)}

\begin{figure}
    \centering
    \includegraphics[width=1.0\linewidth]{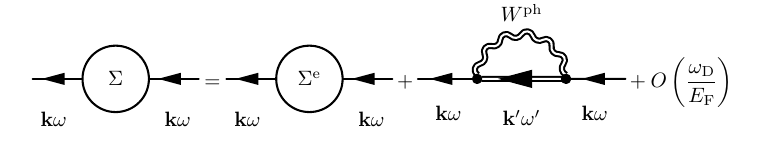}
    \caption{
        Normal component of the electron self-energy approximated by the self-consistent Fock diagram with the phonon-mediated e-e interaction $W^{ph}$. According to Migdal's theorem, higher-order vertex corrections based on $W^{ph}$ are suppressed by $O(\omega_\txD/E_\txF)$.
}
    \label{fig:sigma_ph}
\end{figure}

\begin{figure}
    \centering
    \includegraphics[width=1.0\linewidth]{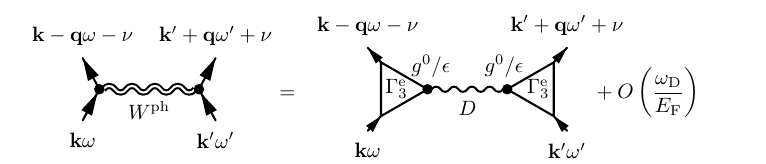}
    \caption{
       Diagrammatic representation of the phonon-mediated e-e interaction, $W^{\rm ph}$, composed of the phonon propagator, $D$, bare coupling 
       $g^{(0)}$, vertex function $\Gamma_3^e$, and the dielectric function
       $\epsilon_{\bq\nu}$. The last two are combined to form the 
       screened electron-phonon coupling.
    } 
    \label{fig:wph_def}
\end{figure}

\begin{figure}
    \centering
    \includegraphics[width=1.0\linewidth]{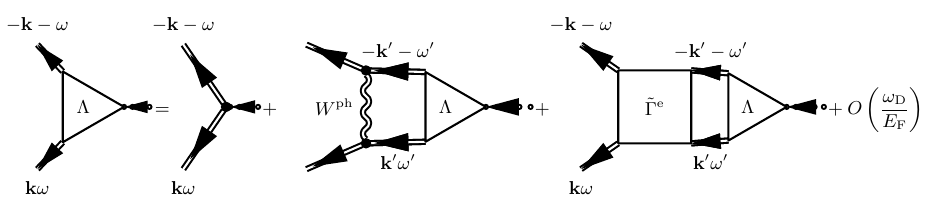}
    \caption{Self-consistent Bethe-Salpeter equation for the anomalous vertex $\Lambda(\bk,-\bk;\bq=0)$ in momentum space, where $\bk$ and $-\bk$ are the momenta of the outgoing electrons. The total momentum $\bq$ is set to zero, as this corresponds to the leading Cooper instability in our case.
    The kernel consists of the particle-particle irreducible 4-point vertex $\tilde{\Gamma}^{e}$, which is a purely electronic contribution, and the phonon-mediated interaction $W^{ph}$ described in Fig.~\ref{fig:wph_def}; higher-order vertex corrections are 
    small according to Migdal's theorem.
}    
    \label{fig:pcf_vertex_approx}
\end{figure}

The standard ME formalism probes superconductivity via an analysis of equations governing the normal and anomalous parts of the electron self energy. The normal self-energy, diagrammatically depicted in Fig.~\ref{fig:sigma_ph}, is  
 
\begin{equation}
    \label{eq:sigma}
        {\Sigma}_{\bk\omega} =  {\Sigma}_{\bk\omega}^\txe + \int_{\bq\nu} {G}_{\bk+\bq\omega-\nu}W^{\txph}_{\bk\omega,\bk+\bq\omega+\nu;\bq\nu}+ 
        O\left(\frac{\omega_\txD}{E_\txF}\right).
\end{equation} 
It consists of two terms with the first one, ${\Sigma}_{\bk\omega}^\txe$, coming entirely from the non-perturbative Coulomb interactions. 
The second term has the structure of the self-consistent Fock diagram
based on the dressed phonon-mediated interaction $W^\txph$ shown in Fig.~\ref{fig:wph_def}. By construction, $W^\txph$ incorporates all non-perturbative Coulomb effects such as screening and vertex corrections. 
Migdal’s theorem ensures that vertex corrections higher-order in $W^\txph$ are suppressed by $\omega_\txD/E_\txF$, justifying truncation at this level of accuracy~\cite{migdal1958}. Similarly, the anomalous (superconducting) self-energy 
has a Coulomb contribution, ${\Sigma}_{\bk\omega}^{\txa,\txe}$, 
and a self-consistent ``Fock-type'' contribution based on 
the phonon-mediated interaction
\begin{equation}
    \Sigma^{\txa}_{\bk\omega} = {\Sigma}_{\bk\omega}^{\txa,\txe} + \int_{\bq\nu} {F}_{\bk+\bq,\omega-\nu}W^{\txph}_{\bk\omega,-\bk-\omega;\bq\nu}+
O\left(\frac{\omega_\txD}{E_\txF}\right) .
\end{equation}
The equation for the anomalous self energy involves the anomalous (Gorkov) propagator,
$F_{\bk\omega}=\langle {\psi}^{\uparrow}_{\bk,\omega} {\psi}^{\downarrow}_{-\bk,-\omega}\rangle$,
which encodes precursor Cooper pairing fluctuations.

These two equations are in effect the standard Migdal-Eliashberg equations, expressed in the current notations. The transition temperature is typically found by linearizing the second equation. Analysis even of the linearized equation can be challenging for low $T_c$ superconductors where a dense computational mesh is required and for strong coupling situations, where extrapolation of the leading eigenvalues of the linearized gap equation is not straightforward. If the goal is limited to prediction of $T_\txc$, then following  Ref.~\cite{pengcheng}, it is more efficient to probe SC from the high-temperature normal state ($T \gg T_\txc$) by solving Bethe-Salpeter equation (BSE) for the anomalous vertex function defined from the linear response of the anomalous self energy to an applied pairing field.  Here, we apply this approach in the context of the electron-phonon SC by considering response to an infinitesimal symmetry-breaking pair-field 
term ($\eta^\txa \to 0$):
\begin{equation}
    S_e[\bar{\psi},\psi; \eta^\txa] \equiv S_e[\bar{\psi}, \psi] + \int_{\bk\omega} \left[\eta^\txa \bar{\psi}^{\uparrow}_{\bk,\omega}\bar{\psi}^{\downarrow}_{-\bk,-\omega} + {\rm H.c.} \right] .
\end{equation}
This symmetry-breaking term induces an anomalous (Gorkov) propagator,
$F_{\bk\omega}=\langle {\psi}^{\uparrow}_{\bk,\omega} {\psi}^{\downarrow}_{-\bk,-\omega}\rangle\propto \eta^\txa$,
which encodes precursor Cooper pairing fluctuations. The normal propagator, $G_{\bk\omega}$, remains unperturbed  to leading order in $\eta^\txa$.  
The linear response of the electron self-energy to $\eta^\txa$ is given by the anomalous vertex, $\Lambda_{\bk\omega}$, for details, see Appendix~\ref{sec:pcf_general_derivation}, which 
encodes the diverging Cooper susceptibility, if any, and obeys the Bethe-Salpeter equation (BSE),
\begin{equation}
    \Lambda_{\bk\omega}=\eta_{k\omega} +\int_{\bk'\omega'}\Gammapp_{\bk\omega;\bk'\omega'}G_{\bk'\omega'}G_{-\bk',-\omega'}\Lambda_{\bk'\omega'}.
    \label{eq:lambda_BSE}
\end{equation}
Here we have written $\eta^a=\bar{\eta}\eta_{k\omega}$ with $\bar{\eta}$ an infinitesimal. The source term $\eta_{k\omega}$ is a function of momentum and frequency defining the symmetry channel and the range over which pairing is probed. Since the critical temperature is defined by the singularity of the homogeneous BSE and is independent of the source profile (provided non-orthogonality to the critical mode) we will in many cases set $\eta_{k\omega}=1$.

The BSE kernel $\Gammapp=\frac{\delta \Sigma^\txa}{\delta F}$  is  the particle-particle irreducible four-point vertex function  with contributions from both the e-e interactions and the e-ph coupling. The subscript notation for $\Gammapp$ is a compact representation of the incoming and outgoing Cooper pair momentum-frequency indexes $(\bk,\omega; -\bk,-\omega)$ and $(\bk',\omega'; -\bk',-\omega')$, respectively.

Migdal’s theorem allows one to write the kernel as the sum of purely electronic particle-particle irreducible vertex ($\Gammapp^\txe$) and phonon-mediated interaction ($W^{\txph}$), as shown in Fig.~\ref{fig:pcf_vertex_approx},
\begin{equation}
\Gammapp = \Gammapp^\txe+W^{\txph}+O\left(\frac{\omega_\txD}{E_\txF}\right).
\label{eq:gammapp_approx}
\end{equation}
By combining the BSE with the Dyson equation for $G$ and evaluation of the effective phonon-mediated coupling $W^{\txph}$ one obtains a
closed set of self-consistent equations to be solved in the normal state.

\subsection{Precursory Cooper Flow}  
The anomalous vertex function, $\Lambda_{\bk\omega}$, is used
not only to predict $T_\txc$ but also to study pairing fluctuations in the normal state and the critical gap function $\Delta_{\bk\omega}$.
We denote the low-frequency limit of $\Lambda_{\bk\omega}$ averaged over the Fermi surface as $\Lambda_0$. It obeys a universal scaling relation known as the precursory Cooper flow
(PCF)~\cite{pengcheng}:
\begin{equation}
\label{eq:PCF_scaling}
\Lambda_0  = \frac{1}{1 + g \ln(\omega_\Lambda/T)} + \mathcal{O}(T),
\end{equation}
where both the dimensionless coupling constant $g$ and high-energy cutoff $\omega_\Lambda$ depend on the microscopic details of the system. For negative values of $g$, the vertex function diverges at $T_\txc = \omega_\Lambda e^{1/g}$, signaling the onset of Cooper instability. By computing $\Lambda_0$ at several low-temperature points above $T_\txc$ and extrapolating the data according to the PCF scaling, one can accurately predict $T_\txc$
{without the need to perform computationally challenging calculations at $T_\txc$. This is particularly advantageous when complex frequency-dependent interactions prevent standard linearized ME solutions from extrapolating reliably from higher temperatures~\cite{implicit,pengcheng}, which would otherwise force one to solve the equations directly at the critical point.}

As the temperature approaches $T_{\text{c}}$ from above, BSE solutions provide direct access to the superconducting gap function, $\Delta_{\bk\omega}$. In this limit, the diverging anomalous vertex $\Lambda_{\bk\omega}$ is proportional to the gap function, with $\Lambda_{\bk\omega} \sim \Delta_{\bk\omega}/(T - T_{\text{c}})$. Substituting this scaling into Eq.~(\ref{eq:lambda_BSE}), we observe that the diverging prefactor $(T - T_{\text{c}})^{-1}$ cancels out and BSE reduces to 
\begin{equation}  
\label{eq:delta1}
    \Delta_{\bk\omega} =  \int_{\bk'\omega'} \tilde{\Gamma}_{\bk\omega;\bk'\omega'} G_{\bk'\omega'} G_{-\bk',-\omega'} \Delta_{\bk'\omega'},  
\end{equation}  
which is identical to the linearized Migdal-Eliashberg (ME) gap equation. 
What favorably distinguishes PCF from the traditional approach based on the temperature dependence of the largest eigenvalue $h(T)$ in 
\begin{equation}  
   h(T)  \Delta_{\bk\omega} =  \int_{\bk'\omega'} \tilde{\Gamma}_{\bk\omega;\bk'\omega'} G_{\bk'\omega'} G_{-\bk',-\omega'} \Delta_{\bk'\omega'},  
\end{equation} 
is the precise scaling law (\ref{eq:PCF_scaling}). In a system with strong repulsive interactions, there is no simple way to reliably extrapolate $h(T)$ towards low temperature \cite{implicit}. 


\section{Downfolding the BSE}
\label{sec:downfolding}
\subsection{Theoretical Overview}

Eq.~\ref{eq:lambda_BSE}, Fig.~\ref{fig:pcf_vertex_approx} for the pairing kernel is exact but not in general tractable. 
The fundamental difficulty is that the BSE kernel, i.e., the two-particle irreducible vertex function $\Gammapp$, is not known in the general case. Widely used approximations, such as the RPA, neglect non-perturbative vertex corrections and are, thus, unreliable beyond the weakly interacting regime. Further, the full momentum and frequency dependence of $\Gammapp$  renders a solution of the BSE computationally intractable in the general case. Historically, these limitations have restricted applications of the BSE to simplified models and necessitated severe approximations.

Eliashberg~\cite{eliashberg1960interactions, eliashberg1961temperature} recognized that a discussion of superconductivity in conventional materials should focus on interactions near the Fermi surface and proposed a ``downfolding'' approximation based on effective interactions between low-energy electronic states that led to a more tractable frequency-only equation for the superconducting gap function. However, as discussed extensively above, to date the parameters (especially those relating to the electron-electron interactions) in the downfolded theory have typically been treated in  ways that are phenomenological or theoretically unjustified. 

The  Wilsonian renormalization scheme is a formally rigorous method for downfolding an interacting electron problem to obtain a low energy theory~\cite{RAINER1986371,Shankar94}. It formalizes Landau Fermi liquid theory by partitioning electron and phonon propagators into high-energy and low-energy contributions separated by an intermediate energy scale $\omega_\txD \ll \omega_\txc \ll E_F$.  
The high-energy contributions are then absorbed into effective couplings that govern the low-energy theory for the quasiparticle pair-field. However, a crucial complication arises in the standard applications of this scheme to the electron liquid:  the low-energy electrons, which are essential for dynamic screening of the long-range Coulomb interaction, are absent in the renormalization of the effective coupling because the separation is constructed in the particle-hole channel.  As a result, the Coulomb interaction remains unscreened below the energy scale $\omega_c$, leading to a singularity in the effective coupling that is awkward to handle. 

In this work, we introduce 
the energy scale separation in the two-electron channel rather than in the single-electron or electron-hole channels. This approach preserves key features of the screening while avoiding the subtle issues encountered in previous attempts \cite{RAINER1986371,Shankar94}. We decompose the pair propagator into low-energy (IR) coherent and high-energy (UV) incoherent components, separated at the energy scale $\omega_\txc$,
\begin{equation}
\label{eq:separation}
G_{\mathbf{k}, \omega}G_{-\mathbf{k}, -\omega} = \Pi_{BCS} +\phi_{\mathbf{k}\omega},
\end{equation}
with
\begin{equation}
\Pi_{BCS}=\frac{\left(z^{\mathrm{e}}\right)^2}{\left(\frac{\omega}{z_\omega^{\mathrm{ph}}}\right)^2+\varepsilon_{\mathbf{k}}^2} \Theta\left(\omega_{\mathrm{c}}-\left|\varepsilon_{\mathbf{k}}\right|\right)
\label{eq:PiBCS}
\end{equation} 
Here, $\mathbf{k},\omega$ is the momentum/frequency of the pair-field, $\phi$ denotes the incoherent contribution, an intrinsic property of the electron liquid with phonon contributions $\mathcal{O}\left(\frac{\omega_D}{E_F}\right)$ which remains regular across the Fermi surface at low temperature.
We have anticipated subsequent results by writing the coherent contribution in terms of $z^\txe$, the quasiparticle weight arising from e-e interactions which is independent of frequency and momentum on the scales of interest, and $z_\omega^\txph$, the frequency-dependent quasiparticle weight from e-ph interaction.
{$\varepsilon_{\mathbf{k}}$ is the linearized quasiparticle dispersion renormalized by e-e interactions.}
{Formally, the effective mass renormalization $m^*$ is fully incorporated into $\varepsilon_{\mathbf{k}}$ and the subsequent effective density of states $N_F^*$. Although established QMC and DiagMC results~\cite{hauleSingleparticleExcitationsUniform2022,holzmannStaticSelfEnergyEffective2023a} indicate that $m^*$ deviates only slightly from the bare mass in the density range of interest, we retain this distinction to separate mass renormalization from the spectral weight reduction $z^\txe$.}

In addition to the separation of the pair propagator $\Pi$ into coherent and incoherent parts, a consistent low-energy theory requires a corresponding treatment of the vertex $\Gammapp$. Formally, following Migdal's theorem, the vertex $\Gammapp$ can be decomposed into a phonon-mediated attraction $W^{\text{ph}}$ and an electron-electron (e-e) interaction $\tilde{\Gamma}^{\text{e}}$ [see Eq.~\eqref{eq:gammapp_approx}]. However, this decomposition of the kernel does not automatically guarantee the separability of the resulting Bethe-Salpeter equation. Specifically, the existence of cross terms of the form $\tilde{\Gamma}^{\text{e}}\cdot \phi \cdot W^{\text{ph}}$ in the renormalization process could potentially couple the two channels, thereby invalidating the assumption of a universal, independent e-e pseudopotential in the downfolded theory.

In metallic systems, the impact of these cross terms is particularly critical because $\tilde{\Gamma}^{\text{e}}$ must account for long-range, dynamically screened Coulomb interactions.  To quantify this effect, we utilize the fact that in the limit of vanishing momentum transfer and low frequency, the dynamical interaction rigorously follows the asymptotic form:
\begin{equation}
W_{\mathbf{k}-\mathbf{k}^{\prime}, \omega-\omega^{\prime}}^{\mathrm{s}}=\frac{4 \pi e^2}{\left|\mathbf{k}-\mathbf{k}^{\prime}\right|^2} \frac{\left(\omega-\omega^{\prime}\right)^2}{\left(\omega-\omega^{\prime}\right)^2+\omega_{\text{p}}^2}\, .
\end{equation}
This expression captures the exact infrared behavior of the interaction. While extending this asymptotic form to finite momenta and frequencies (effectively treating it as a plasmon-pole model) tends to overestimate the range of the interaction in phase space, it provides a strictly conservative estimate for the cross terms. As detailed in Appendix \ref{sec:app_eliashberg}, even with this overestimation, the resulting cross terms are suppressed by $\omega_{\text{c}}^2/\omega_{\text{p}}^2$, justifying the low-energy separation. Consequently, provided the plasmon frequency $\omega_{\text{p}}$ is sufficiently high compared to the cutoff $\omega_{\text{c}}$, the interaction channels remain effectively decoupled, justifying the low-energy separation.

We utilize this suppression to simplify the Bethe-Salpeter equation (BSE). We assume the external source $\eta_\omega$ is non-zero only at low frequencies ($|\omega| < \omega_{\text{c}}$) and partition the anomalous vertex $\Lambda$ into two frequency sectors: a low-energy component $\Lambda_L$ (where $|\omega| < \omega_{\text{c}}$) and a high-energy component $\Lambda_H$ (where $|\omega| > \omega_{\text{c}}$). Adopting a shorthand notation where integrals are implicit and the subscripts $L$ and $H$ denote the respective frequency ranges, we recast Eq.~\ref{eq:lambda_BSE} as:
\begin{eqnarray}
\Lambda_L&=&\eta_\omega+\tilde{\Gamma}_{LL}\Pi_{BCS}\Lambda_L+\tilde{\Gamma}_{LH}\phi\Lambda_H+O\left(\frac{\omega_{\text{c}}^2}{\omega_{\text{p}}^2}\right)
\nonumber
\\
\Lambda_H&=&\tilde{\Gamma}_{HL}\Pi_{BCS}\Lambda_L+\tilde{\Gamma}_{HH}\phi\Lambda_H+O\left(\frac{\omega_{\text{c}}^2}{\omega_{\text{p}}^2}\right).
\label{eq:BSE2}
\end{eqnarray}
By neglecting the suppressed cross terms and eliminating the high-energy sector $\Lambda_H$, we reduce the BSE to an effective description restricted to the low-energy window $|\omega| < \omega_{\text{c}}$:
\begin{equation}
\Lambda_{k\omega}=\eta_\omega+\sum_{k^\prime,\omega^\prime}\tilde{\Gamma}^{\omega_c}_{k\omega,k^\prime\omega^\prime}\Pi_{BCS}(k^\prime,\omega^\prime)\Lambda_{k^\prime,\omega^\prime}.
\label{eq:BSElow}
\end{equation}
The effective vertex $\tilde{\Gamma}^{\omega_c}$ appearing here is particle-particle irreducible only with respect to the coherent propagator $\Pi_{BCS}$ and incorporates all high-energy renormalization effects via the recursive relation:
\begin{equation}
\label{eq:effective_gamma}
    \tilde{\Gamma}^{\omega_c}_{\mathbf{k}\omega;\mathbf{k}'\omega'} = \Gammapp_{\mathbf{k}\omega;\mathbf{k}'\omega'} + \sum_{\mathbf{p}\nu}\Gammapp_{\mathbf{k}\omega;\mathbf{p}\nu}  \phi_{\mathbf{p}\nu} \tilde{\Gamma}^{\omega_c}_{\mathbf{p}\nu;\mathbf{k}'\omega'} \, .
\end{equation}

Combined with Migdal's theorem, we conclude that the effective coupling projected onto the low energy subspace has a separable form
\begin{equation}
    \tilde{\Gamma}^{\omega_\txc} = U^\txe +  W^{\txph} + O\left(\frac{\omega_\txD}{E_\txF}, \frac{\omega_\txc^2}{\omega_\txp^2}\right), 
\end{equation}
where the effective e-e interaction $U^\txe \equiv \tilde{\Gamma}^{\txe} + \tilde{\Gamma}^{\txe}\cdot \phi\cdot \tilde{\Gamma}^{\txe} + ...$ is the property of the electron liquid independent of phonon details. Although $\tilde{\Gamma}^{\txe}$ and $\phi$ separately are
typically not evaluated in simulations, the final result for
$U^\txe$ can be expressed through the conventional full 4-point vertex function $\Gamma^{\txe}$, as detailed in Appendix \ref{sec:app_eliashberg}, which can be evaluated in simulations.

Because
$W^{\txph}$ and $U^\txe$ as defined above are regular functions of momentum and frequency, we may now project all momenta onto the Fermi surface (in this work, we focus on the $s$-wave SC and isotropic Fermi surface---solely for clarity of presentation). Using the explicit form Eq.~\ref{eq:PiBCS}  of $\Pi_{BCS}$ and integrating over the magnitude of the momentum perpendicular to the Fermi surface we obtain a low-energy effective theory for the quasiparticle pair-field, described by a frequency-only downfolded BSE:
\begin{equation}
\label{eq:downfolded_BSE}
    \pcf_\omega=\eta_\omega+\pi T\sum_{|\omega'|<\omega_\txc}\left(\lambda_{\omega\omega'}-\psp_{\omega_\txc}\right) \frac{z^{\txph}_{\omega'}}{|\omega'|} \pcf_{\omega'},
\end{equation}
with corrections proportional to one of the three small parameters:
${\omega_\txD}/{E_\txF}$, ${\omega_\txc^2}/{\omega_\txp^2}$, or ${T}/{\omega_\txc}$.
By construction, the summation over Matsubara frequencies $\omega'$ is limited by the cutoff, $|\omega'|<\omega_\txc$. The symmetry-breaking 
term [see Eq.~(\ref{eq:rpcf}) for more details] can be set to unity,  $\eta_\omega = 1$,  for numerical convenience without affecting $T_\txc$ or the gap function. 

The BSE
equation depends only on Fermi-surface-averaged quantities. 
Specifically, $\pcf_\omega = \langle \pcf_{\bk_\txF\omega,-\bk_\txF-\omega; 0}\rangle_{\txFS}$ represents the frequency-dependent anomalous vertex function averaged over the Fermi surface. Similarly, 
$\lambda_{\omega\omega'}\equiv -\left(z^\txe\right)^2N_F^* \left<W^{\txph}_{\bk_\txF-\bk_\txF',\omega-\omega'} \right>_{\bk_\txF^{}\bk_\txF'}$ is the effective phonon-mediated interaction, where $N_\txF^*$ is the quasiparticle density of states. 

The frequency-dependent quasiparticle weight renormalization due to e-ph interactions, $z^{\txph}_{\omega}$, is  fully determined by $\lambda$ in the same way as in the standard Eliashberg formulation:
\begin{equation}
\label{eq:zph}
    \frac{1}{z^\txph_\omega}=1+\frac{\pi T}{\omega}\sum_{\omega'}\frac{\omega'}{|\omega'|}\lambda_{\omega\omega'}+O\left(\frac{\omega_\txD}{E_\txF}, \frac{\omega_\txc^2}{\omega_\txp^2}\right).
\end{equation}

Remarkably---despite singular momentum dependence at the microscopic level and complex dynamic screening, the projected effective e-e interaction reduces to a pseudopotential constant $\psp_{\omega_\txc}$. 

Repeating identically the derivation that lead to Eq.~(\ref{eq:delta1}) for the gap function at the critical point, we obtain its downfolded version as
\begin{equation}
\label{eq:delta2}
    \Delta_\omega=\pi T_\txc\sum_{|\omega'|<\omega_\txc}\left(\lambda_{\omega\omega'}-\mu^*\right) \frac{z^{\txph}_{\omega'}}{|\omega'|} \Delta_{\omega'}+O\left(\frac{\omega_\txD}{E_\txF}, \frac{\omega_*^2}{\omega_\txp^2}\right) \; ,
\end{equation}
where  the notation {$\mu^*=\mu_{\omega_\txc},\, \omega_*=\omega_\txc$} is used to present it in the familiar ME form.
{We note that in the low-frequency limit, the term $\lambda_{\omega\omega'} z^{\txph}_{\omega'}$ reduces to $\lambda/(1+\lambda)$.}

Eqs.~\eqref{eq:zph} and~\eqref{eq:delta2}  establishes the microscopic foundation for ME equation with 
precise definitions of the Coulomb pseudopotential and effective e-ph coupling in terms of electron/phonon propagators 
and vertex functions. 

\subsection{The pseudopotential}

$\psp_{\omega_\txc}$ is  the coherent pair propagator irreducible particle-particle coulomb interaction; it is defined with respect to an arbitrary renormalization scale $\omega_c$. Here we discuss its relation to quantities that may be calculated. We begin by noting that in a theory with only electron-electron interactions, solving Eq.~\ref{eq:delta2} with $\lambda=0$ and $z^{ph}=1$ shows that at a temperature $T$ the effective repulsion is
\begin{equation}
\label{eq:standard}
\gamma_T = \frac{\mu_{\omega_c}}{1 +\mu_{\omega_c} \ln \frac{\omega_\txc}{T}} \qquad \qquad (T \ll \omega_\txc) \, ,
\end{equation}
where the repulsion may also be defined in terms
of the Fermi-surface-averaged two-quasiparticle scattering amplitude
\begin{equation}
\label{eq:gammaT}
    \gamma_T \equiv z_\txe^2 N_\txF^* \left<\Gamma^\txe_4(\bk_F,\omega_0; \bk_F',\omega_0)\right>_{\bk_F,\bk_F'}\, ,
\end{equation}
with $\omega_0=\pi T$  the smallest Matsubara frequency. As we shall see, the right hand side of Eq.~\ref{eq:gammaT} can be calculated.

$\gamma_T$ is a physical quantity, and is therefore independent of the (arbitrarily chosen) separation scale $\omega_c$. This independence requires that the connection between $\psp_{\omega_\txc}$ defined at two different scales $\omega_c,\omega_c^\prime$ is the Bogoliubov-Tolmachev-Shirkov relation
\begin{equation}
\psp_{\omega_\txc} = \frac{\mu_{\omega_c^\prime}}{1 +\mu_{\omega_c^\prime}\ln \frac{{\omega_c^\prime}}{\omega_\txc}} \, .
\label{eq:mu_E_F}
\end{equation}
As a practical matter we may therefore compute  $\gamma_T$ at a convenient temperature, corresponding to a convenient $\omega_c$ and then scale to any desired separation scale. One physically meaningful choice is 
$E_F$, the physical scale below which coherent electronic quasiparticles exist. 
One normally interprets $\mu_{_{E_\txF}}$ as the pseudopotential at the Fermi energy, or, equivalently, ``bare''---free of Bogoliubov-Tolmachev-Shirkov renormalization---pseudopotential. 

\begin{figure}
    \centering
    \includegraphics[width=0.95\linewidth]{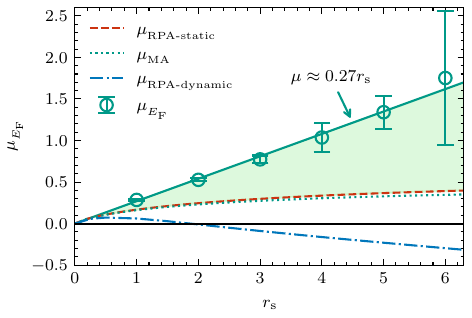}
    \caption{
    Dimensionless ``bare'' Coulomb pseudopotential, $\mu_{E_F}$, as a function of $r_\txs$ for the 3D UEG extracted from VDiagMC data for $\mu_{\omega_\txc}$ by inverting relation (\ref{eq:mu_E_F}); the solid line represents a linear fit to the VDiagMC data. The exact VDiagMC values corresponding to these integer $r_\txs$ points are listed in Table~\ref{tab:vdiagmc_data}. At $r_\txs> 0.5$, the VDiagMC results demonstrate a dramatic deviation from predictions of three standard approximations: static random phase approximation (RPA) $\mu_{\text{RPA-static}}$ (red dashed curve), $\mu_{\text{MA}}$ (green dashed curve) based on the Yukawa interaction with Thomas-Fermi screening momentum, and the dynamic RPA (see, e.g., Ref.~\cite{pengcheng}), $\mu_{\text{RPA-dynamic}}$. Note perfect agreement between all curves at $r_\txs \ll 1$.}
    \label{fig:uvsrs}
\end{figure}

Our results for $\mu_{_{E_\txF}}$ in UEG as a function of  $r_\txs$ are presented in Fig.~\ref{fig:uvsrs}; the corresponding numerical values are listed in Table~\ref{tab:vdiagmc_data}. We find $\mu_{_{E_\txF}}$ to be significantly larger (by a factor of three at $r_\txs=5$) compared to MA and static RPA estimates~\cite{morel_anderson, RPA_mu, RPA_mu2}. 
Even more profound is the disagreement with dynamic RPA that predicts negative $\mu_{\omega_\txc}$ values for $r_\txs>2$ \cite{takada1978plasmon, takada1993, rietschel, xiansheng, 23tao}. This outcome clearly demonstrates that the MA and RPA approximations are out of control at $r_\txs>0.5$, which one could have expected from the mere fact that at $r_\txs>0.5$ the results of static and dynamic RPA dramatically differ from each other. 

\begin{table}[b] 
    \begin{ruledtabular} 
    \begin{tabular}{ccccccc}
    $r_\txs$ & 1 & 2 & 3 & 4 & 5 & 6 \\
    \hline
    $\mu_{_{0.1E_F}}$ & 0.172(4) & 0.238(4) & 0.278(6) & 0.306(15) & 0.328(12) & 0.35(3) \\
    \hline
    $\mu_{_{E_F}}$ & 0.28(1) & 0.53(2) & 0.77(5) & 1.0(2) & 1.3(2) & 1.8(8) \\
    \end{tabular}
    \end{ruledtabular}
    \caption{
    VDiagMC results for the dimensionless Coulomb pseudopotential $\mu_{_{\omega_\txc}}$ for integer values of $r_\txs$ (see also Fig.~\ref{fig:uvsrs}). Values were computed at $\omega_\txc=0.1E_\txF$ and rescaled to $E_\txF$. Numbers in parentheses indicate the estimated systematic uncertainty in the last digit.
    }
    \label{tab:vdiagmc_data}
\end{table}

In addition to parameterizing the value of the Coulomb pseudopotential as a function of low frequency cutoff, Eq.~(\ref{eq:mu_E_F}), the bare pseudopotential $\mu_{_{E_\txF}}$ plays yet another role that is closely related to superconductivity. The low-temperature response of the normal-Fermi-liquid state to the uniform pair-creating perturbation, $\chi_0$, has the form (see, e.g., Ref.~\cite{pengcheng})
$\chi_0 \, \propto\, {z_{\rm e}^2\over m^*\mu_{_{E_\txF}}} $ for $\mu_{_{E_\txF}} >0, \; \omega_D \ll T \ll E_\txF, \omega_p).$
From our results for $\mu_{_{E_\txF}}$ we thus conclude that $\chi_0$ gets substantially suppressed with increasing $r_{\rm s}$.

\subsection{Validation of the Fermi-Surface Downfolding}
\label{sec:validation}
The controlled downfolding scheme relies on the condition that $\omega_\txc$ is much smaller than the characteristic energy scales of the electron Fermi liquid. In the absence of additional emergent collective excitations, the conditions $\omega_\txc/E_F \ll 1$ and $\omega_\txc/\omega_\txp \ll 1$ are the most relevant. In 3D electron gases, the ratio $\omega_\txc/\omega_\txp$ scales as $1/\sqrt{r_\txs}$. For most metals $r_\txs \gtrsim 1$ (even in high-pressure metallic hydrogen samples where $r_\txs \approx 0.9$), and one can safely estimate $\omega_\txc/\omega_\txp \lesssim 0.1$, indicating the robustness of the downfolding approximation for terrestrial and laboratory metals.

To validate this approximation for the systems under consideration, we investigate a toy model with the two-particle-irreducible vertex function, $\tilde{\Gamma}^\txe$, approximated by the dynamically screened Coulomb interaction of the uniform electron gas within the RPA, $W_\txRPA = v_q/(1 - v_q \Pi^0_{q\nu})$. For the effective phonon-mediated interaction, $W^{\txph}_{q\nu}$, we consider a model form from Ref.~\cite{23tao}:
\begin{equation}
\label{eq:elph}
W^{\rm ph}_{q\nu} = -\frac{g /N\txF}{1+{(q/2k_\txF)}^2}\frac{\omega_q^2}{\nu^2+\omega_q^2} ,
\end{equation}
with the phonon dispersion $\omega_q^2 = \omega_\txD^2(q/ k_\txF)^2 / (1 + (q/ k_\txF)^2)$ and a coupling strength of $g=0.4$. We consider a conventional metallic density corresponding to $r_\txs = 1.91916$ (similar to aluminum) and a small adiabatic ratio $\omega_\txD/E_F=0.005$. We then solve for the critical temperature $T_\txc$ using both the full, frequency-dependent BSE and the simplified, downfolded frequency-only BSE.

\begin{figure}
    \centering
    \includegraphics[width=1.0\linewidth]{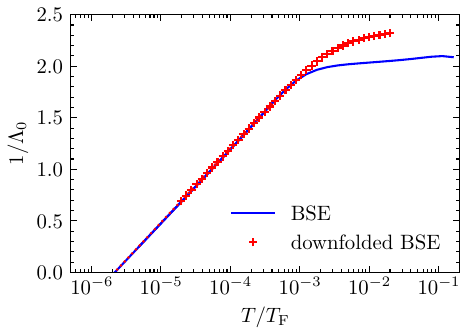}
    \caption{Comparison between the precursory Cooper flow solutions of the full and downfolded Bethe-Salpeter equations for a toy model with the two-particle-irreducible electron vertex function approximated by the Coulomb interaction screened by RPA polarization and a typical phonon-mediated interaction. The calculation is performed at moderate $r_\txs=1.91916$ case (representative of Al) with $T_\txc^{(\text{full})}/T_\txF=10^{-5.668}$ and $T_\txc^{(\text{approx})}/T_F=10^{-5.667}$ (difference $\sim0.2\%$), confirming the validity of the downfolding approximation. 
    }
    \label{fig:compareflow_zph}
\end{figure}

The comparison, shown in Fig.~\ref{fig:compareflow_zph}, demonstrates excellent agreement between the two methods. The calculated critical temperatures differ by a negligible $0.2\%$, confirming the quantitative accuracy of the downfolding approximation for conventional metals. When the approximation is valid, both solutions follow the same universal logarithmic scaling, aligning perfectly below the Debye frequency.

While the downfolding framework is clearly robust in this physical regime, its accuracy could be challenged in extreme cases where the separation of energy scales is less pronounced. For instance, downfolding can become questionable in systems with low electron densities where $E_\txF\sim\omega_\txD$ (a regime outside the scope of this discussion), as well as in dense systems with small $r_\txs$. This includes certain astrophysical objects, such as the interiors of white dwarfs (slowly evolving into black dwarfs). In these systems, electrons may become superconducting at sufficiently low temperatures~\cite{ginzburg_superconductivity_1968, trubnikov1968white}, and their $r_\txs$ values can be as small as 0.01. This leads to soft plasmon frequencies and potential problems with the accuracy of the downfolding approximation. Similar, and potentially more severe, problems are anticipated in two-dimensional electron gas (2DEG) systems~\cite{katsnelson2023screening}, provided the Coulomb interaction is not screened by a substrate. In that case, the plasmon mode remains gapless across all density regimes, a situation that warrants further careful investigation. A detailed analysis of such exotic regimes goes beyond the scope of the present work. 

\section{Coulomb Pseudopotential from the First Principles}
\label{sec:pseudopotential}
\subsection{Uniform Electron Gas}

In this section, we compute the Coulomb pseudopotential for the uniform electron gas (UEG). Now $S_\txe$ in Eq.~\eqref{eq:full_action} describes electrons on a uniform neutralizing background:
\begin{align}
    S_{\txUEG} &= \int_{\bk\omega\sigma} \bar{\psi}_{\bk\omega}^{\sigma}\left[-i\omega +\frac{\bk^2}{2m} -\mu \right]\psi_{\bk\omega}^{\sigma}\nonumber\\
            &+\frac{1}{2}\int_{\bk\omega\sigma,\bk'\omega\sigma',\bq\nu}\bar{\psi}_{\bk\omega}^{\sigma}\bar{\psi}_{\bk'\omega}^{\sigma'} v_\bq\psi_{\bk'-\bq,\omega'-\nu}^{\sigma'} \psi_{\bk+\bq\omega+\nu}^{\sigma}\, ,  
            \label{eq:UEG}
\end{align}
here $\bar{\psi}, \psi$ are Grassmann fields,
$v_\bq= {4\pi e^2}/{q^2}$ is the Coulomb potential, 
and $\mu$ is the chemical potential. 
The strength of interaction in UEG is characterized by the dimensionless Wigner-Seitz radius $r_\txs$, which measures the distance between electrons in units of the Bohr radius and is proportional to the ratio between the potential and kinetic energies. 
Below and in numerical calculations, we follow the atomic Rydberg units convention, setting $\hbar=2m=e^2/2=1$, while other quantities such as $E_\txF$ and $k_\txF$ vary with respect to $r_\txs$.

The UEG is a foundational model in condensed matter physics. 
Not only does it apply directly to simple metals with weak lattice potentials but also within local density approximation (LDA) it is used to formulate DFT. By treating inhomogeneous electron systems as locally uniform, LDA utilizes the knowledge of ground-state properties of UEG to capture complex exchange-correlation effects. The remarkable success of LDA-based DFT in describing a broad spectrum of materials highlights importance of UEG to our understanding of interacting electrons.

Despite the relative simplicity of the UEG model (compared to real materials) precise calculation of its Coulomb pseudopotential
presents a significant challenge. We are not aware of any calculation performed with controlled accuracy for $r_\txs>1$.   
The most challenging aspect is the evaluation of the four-point vertex function in the Cooper channel on the Fermi surface. In the strongly correlated regime, this quantity cannot be obtained by traditional ground-state methods such as variational and diffusion Monte Carlo~\cite{vmc1,vmc2,dmc1}. 

In this context, the recently developed method of Variational diagrammatic Monte Carlo~\cite{diagmc1,diagmc2,diagmc2010,detdiagmc,diagmclatest,diagmckozik,vdiagmc1,vdiagmc2} (VDiagMC) emerges as the right tool
to address the problem. Unlike the majority of quantum Monte Carlo techniques, VDiagMC is based on stochastic sampling of Feynman diagrams to high order
and provides direct access to the vertex function needed for evaluation
of $\mu^*$. The key principle of VDiagMC is optimization of the starting point to set up the diagrammatic expansion order-by-order.
As a result, one obtains accurate converged answers even beyond the 
weak-coupling regime. 

Within the VDiagMC framework, we treat the UEG action $S_{\txUEG}$ as that of the ``renormalized'' Yukawa Fermi gas with counter-terms. 
The bare chemical potential $\mu$ and Coulomb interaction $v$ are expressed as power series in terms of renormalized parameters: 
$\mu \equiv \mu_\txR + \delta \mu_1 \xi + \delta \mu_2 \xi^2 + ...$, 
$v(\bq) = 4\pi e^2/\bq^2 \equiv 4\pi e^2/(\bq^2+\lambda_\txR)
+ \delta v_1 \cdot \xi + \delta v_2 \cdot \xi^2+\dots $,
where $\xi$ is an auxiliary parameter used to track the
expansion order; physical result corresponding to $\xi=1$. Here $\mu_\txR$ is the renormalized 
chemical potential set to the Fermi energy in the low-temperature limit
in such a way that the tree-level propagator corresponds to the physical electron density. This is achieved by requiring that counterterms $\delta \mu_i$ cancel all self-energy corrections to the Fermi energy 
at order $i$. Similarly, the Coulomb interaction $v(\bq)$ is written as Yukawa interaction, $v_\txR(\bq) \equiv 4\pi e^2/(\bq^2+\lambda_\txR)$, and a series of counterterms $\delta v_i \equiv v_\txR^{i+1}\left(\frac{\lambda_\txR}{4\pi e^2}\right)^{i}$.
The screening parameter, $\lambda_\txR$, is optimized to improve
convergence. With these redefinitions, we perform perturbative 
expansions of physical quantities in powers of $\xi$, 
effectively removing Coulomb divergences and large expansion parameters
and significantly improving convergence and, thus, final accuracy. 

As detailed in Ref.~\cite{hou_feynman_2024}, the efficiency 
of VDiagMC is greatly enhanced by representing high-order Feynman diagrams of vertex functions as computational graphs. This approach leverages the structure of Dyson-Schwinger and Parquet equations to express diagrams in a compressed form with a fractal structure of tensor operations, significantly reducing computational redundancy. Notably, this computational graph representation allows for highly efficient implementation of field-theoretic renormalization schemes of the bare parameters using Taylor-mode automatic differentiation algorithms~\cite{Taylor1,Taylor2,Taylor3}, reducing the computational cost of evaluating the renormalized set of 
diagrams from exponential to sub-exponential scaling.

Building upon these techniques, we developed a Feynman diagram compiler that generates, optimizes, and converts the renormalized diagrammatic series into a compressed computational graph across various
platforms using machine learning frameworks. 
This enables the implementation of high-dimensional Monte Carlo integration algorithms with state-of-the-art importance sampling techniques, such as the VEGAS~\cite{vegas,vegasenhanced} adaptive algorithm and more advanced methods employing normalizing flow neural networks~\cite{normalization_flow}. Our codes support both the conventional CPU clusters and GPU platforms, allowing for scalable and efficient computations. 
In practice, we can reach sufficiently high expansion orders to accurately determine various quasiparticle properties, including density, spin susceptibility, and effective mass. This combination of computational graph representation, Taylor-mode automatic differentiation, and modern machine learning frameworks provides 
a robust and efficient AI Tech Stack for quantum field theory. 

\subsection{Coulomb Pseudopotential: Homotopic expansion}
\label{sec:expansion}
\begin{figure}
    \centering
    \includegraphics[width=0.7\linewidth]{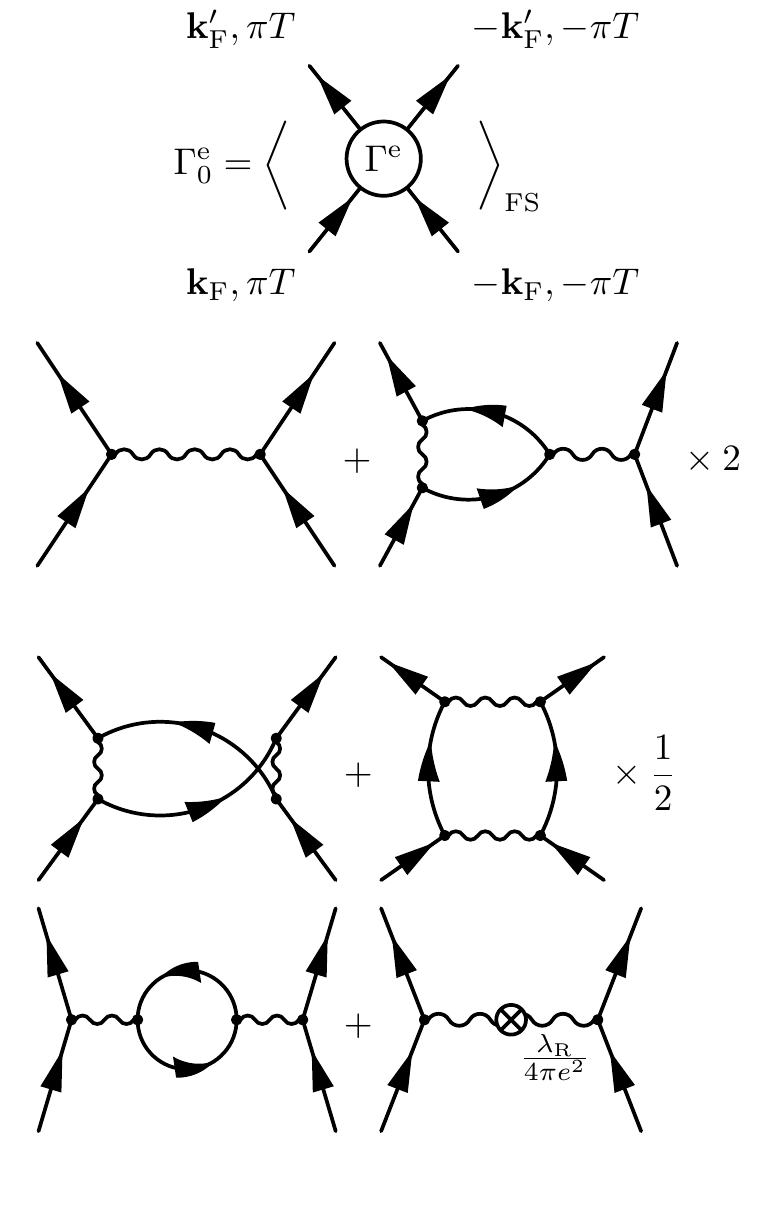}
    \caption{Diagrammatic contributions to the 4-point vertex 
    at the first and second order. The Coulomb interaction is re-expanded starting from Yukawa interaction with the screening parameter $\lambda_\txR$, resulting in a power series of counterterms based on $\lambda_\txR$ (see text).}
    \label{fig:psp_diags}
\end{figure}

\begin{figure*}
    \centering
    \includegraphics[width=0.80\linewidth]{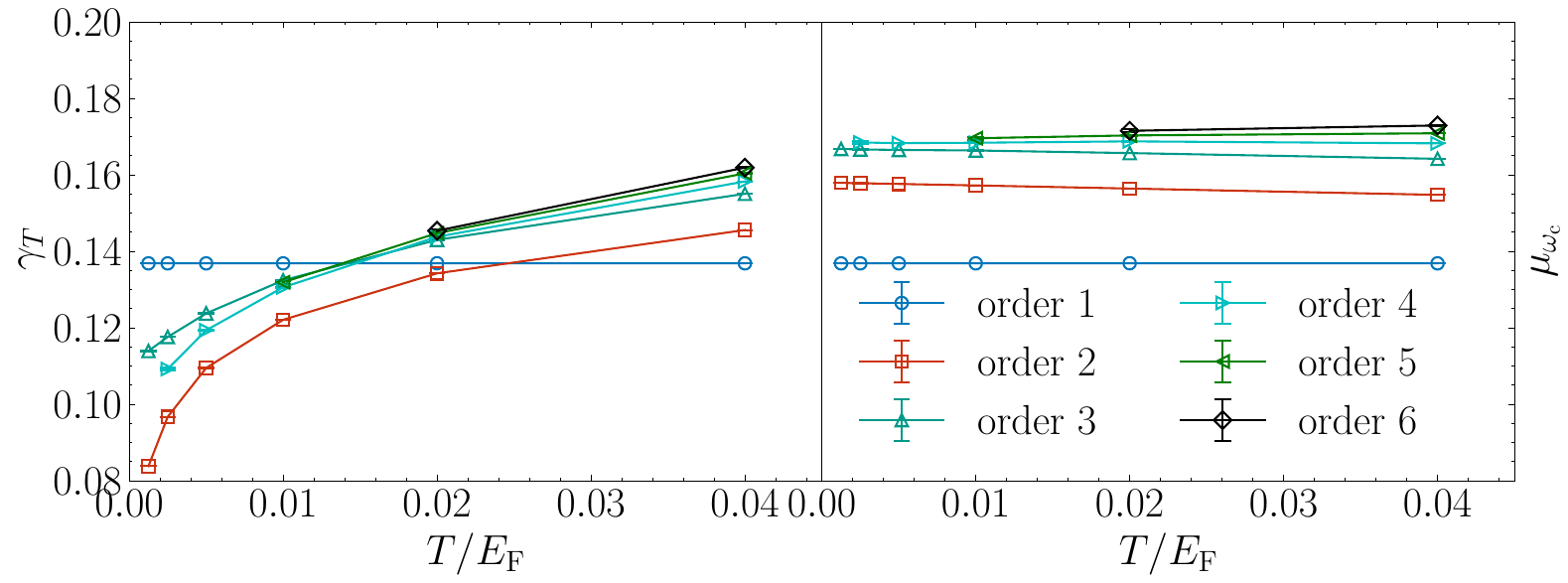}
    \caption{
    Temperature dependence of the partial sums of the series for $\gamma_T$ (left panel) and corresponding results for Coulomb pseudopotential $\mu_{\omega_c}$ 
    (right panel), for a given energy scale separation parameter 
    $\omega_\txc = 0.1 E_\txF$,  
    $r_\txs = 1$ and $\lambda_\txR = 3.5 k_\txF^2$. 
    Temperature dependence of 
    $\gamma_T$ is dominated by the logarithmic divergence characteristic of the Cooper channel, with contributions at order $N$ being proportional to $[\ln (E_\txF/T)]^N$. In contrast, $N$-th order results for $\mu_{\omega_c}$ saturate to a constant at low temperature. 
    The renormalized series for $\mu_{\omega_\txc}$ appear to converge to a well-defined value in the $T \to 0$ limit, thereby enabling reliable estimation of $\mu_{E_F}$.
    }
    \label{fig:lnT_m_renorm}
\end{figure*}

With VDiagMC we construct and compute high-order power series for $\psp_{\omega_\txc}$ for a certain computationally optimal value of $\omega_\txc$.
The calculation starts with evaluation of the self-energy and four-point vertex functions within the renormalized UEG 
approach described above. Subsequently, these quantities are used to extract the quasiparticle residue, $z^\txe(\xi)$, effective mass, $m_\txe^*(\xi)$, and the two-electron scattering amplitude, $\Gamma^\txe (\xi )$, as power series in $\xi$. 
Finally, we combine the results to obtain the power series for $\gamma_T$,
\begin{equation}
\label{eq:gamma_T_series}
    \gamma_T(\xi) \equiv \left[z^\txe(\xi)\right]^2 \! \frac{m_\txe^*(\xi)}{m}N_\txF\Gamma^\txe(\xi)\! \equiv \! \gamma_T^{(0)} + \gamma_T^{(1)}\xi +\gamma_T^{(2)}\xi^2+... \, .
\end{equation}

At low temperature, the convergence of the series at the physical value $\xi=1$ becomes problematic due to the $(\ln T)^N$ scaling of the $N$-th order term arising from nested particle-particle bubbles in the diagrammatic expansion. The solution to the convergence problem comes with a homotopic trick (cf.~Ref.~\cite{homotopic_expansions}). Assuming analyticity of $\gamma_T(\xi)$ as a function of $\xi$, we want to construct a matching temperature-independent analytic function $\mu_{\omega_c}(\xi)$ such that, on the one hand, it could be analytically expressed in terms of $\gamma_T(\xi)$ and, on the other hand, would feature the homotopic requirement that $\mu_{\omega_c}(\xi=1)$ be equal to the value of $\mu_{\omega_c}$ defined by Eq.~(\ref{eq:standard}). The series (\ref{eq:gamma_T_series}) will then be readily converted into the homotopic expansion
\begin{equation}
    \mu_{\omega_\txc}(\xi) = \mu_{\omega_\txc}^{(0)} + \mu_{\omega_\txc}^{(1)}\xi +\mu_{\omega_{\txc}}^{(2)}\xi^2 + \dots \, ,
    \label{eq:psp_series}
\end{equation}
with temperature-independent coefficients, thereby providing a natural cure for the low-$T$ convergence problem.

The guiding principle for constructing the desired homotopy is the above-mentioned $(\ln T)^N$ scaling of the $N$-th order term of the expansion (\ref{eq:psp_series}). The relation between $\mu_{\omega_\txc}(\xi)$ and $\gamma_T(\xi)$ has to generate corresponding counterterms. This suggests that we define
\begin{equation}
    \mu_{\omega_\txc}(\xi) \equiv \frac{\gamma_T(\xi)}{1-\gamma_T(\xi)\xi\ln(\omega_\txc /T) } \, .
    \label{eq:psp_resum}
\end{equation}
Expansion in powers of $\xi$ then yields  $\mu_{\omega_\txc}^{(0)} = \gamma_T^{(0)}$, $\mu_{\omega_\txc}^{(1)} = \gamma_T^{(1)} + [\gamma_T^{(0)}]^2\ln(\omega{_\txc}/T)$, etc.

The low-temperature behavior of the partial sums of the series (\ref{eq:gamma_T_series}) at $\xi=1$ is shown in 
Fig.~\ref{fig:lnT_m_renorm} (left panel). It reveals a logarithmic
divergence below $0.01E_F$.
In the right panel of Fig.~\ref{fig:lnT_m_renorm}, we present
results for partial sums of the series (\ref{eq:psp_series}) at $\xi=1$
(after incorporating the frequency cutoff shift induced by mass renormalization, see Appendix \ref{sssec:omegashift} for details). 
As anticipated, the $\mu_{\omega_\txc}$ series does not exhibit terms with
divergent behavior as $T\to 0$ and quickly converges thus allowing us to reliably extract the value of $\mu_{E_F}$.

The mathematical justification for this resummation protocol, including the underlying assumptions regarding analyticity and the details of the conformal map technique, is provided in Appendix~\ref{sec:app:math_status}.


\section{Electron-Phonon Coupling from Band Theory}
\label{sec:electronphonon}
Accurate determination of the phonon-mediated electron-electron attraction
quantified by the dimensionless coupling constant $\lambda$ is fundamental to understanding and predicting phonon-mediated SC. 
Current theories of conventional superconductivity are based on defining electron-phonon coupling by using density functional theory to estimate the response of the electronic states to changes in atomic positions. This response is efficiently implemented in the  density functional perturbation theory (DFPT) computational method~\cite{dfpt_rev,EPW1,EPW2,giustino_2017}. Currently, DFPT underlies most \emph{ab initio} predictions of conventional superconductivity, yet its accuracy in correlated systems remains untested. Moreover, precise benchmarks help establish semi-phenomenological
pathways for systematically correcting DFPT results, potentially 
extending its applicability to strongly correlated superconductors. In this section we address the fundamental question: under what conditions and with what accuracy does the EFT framework agree with the results of DFPT~\cite{dfpt_rev,EPW1,EPW2,giustino_2017} ?

Within the downfolded ME framework, $\lambda$ 
is defined by the Fermi-surface average of the square of the ratio of the physical electron-phonon coupling $g_\kappa(k,q)$ and the physical  phonon frequency $\omega_{\kappa,q}$ following combination (below $|\bk |=k_F,\,  |\bk ' | = | \bk +\bq |=k_F$), 
\begin{equation}
\label{eq:def_lambda}
\lambda = N_\txF \sum_{\kappa} \left\langle \frac{g_\kappa^2(\bk, \bq) }
          {\omega^2_{\kappa, \bq} }   \right\rangle_{\txFS} \, .
\end{equation}
Note that the physical coupling is corrected from the bare coupling $g^0_{\kappa\bq}$ by  electronic screening parameterized by the dielectric function $\epsilon_q$ and vertex corrections $\Gamma^e_3$, as well as
the quasiparticle residue $z^\txe$:
\begin{equation}
\label{eq:eft_g}
g_{\kappa}(\bk, \bq) \equiv g^{(0)}_{\kappa\bq} \frac{z^\txe}{\epsilon_\bq}\Gamma_3^\txe(\bk,\bq).
\end{equation}
Here, the combination $z^\txe \Gamma_3^\txe(\bk,\bq)$ can be interpreted as the quasiparticle vertex correction.
From now on in the main text  we omit the phonon branch and reciprocal lattice vector indexes for simplicity of presentation; a complete description is presented in the Appendix~\ref{app:epi}.

In the small $\bq$ limit, the bare coupling $g^{(0)}_\bq$ associated with longitudinal fluctuations diverges as $qv_q$, whereas the coupling to transverse modes is generally regular and vanishes in the free-electron limit, becoming non-zero only when non-zero reciprocal lattice vectors are considered.

Regardless of this asymptotic behavior, the bare interaction is screened and renormalized by the electronic response over the full kinematic range relevant to superconductivity. Given that $\bk$ and $\bk + \bq$ both reside on the Fermi surface, the typical momentum transfer in Eq.~(\ref{eq:def_lambda}) is not small, covering $|\bq | \in (0,2k_\txF)$. Therefore, one cannot rely on the properties of the long-wavelength limit when evaluating the effective interaction strength. In general, $g$ depends on both incoming and outgoing momenta.

In the framework of DFPT, the effective electron-phonon coupling is obtained by considering the linear response of the Kohn-Sham (KS) potential to ionic displacement, $\delta V^{\rm KS}_{\bq} = \delta V^{\rm ion}_{\bq} +v_{\bq} \delta n_{\bq}+f_{\rm xc} \delta n_{\bq}$. This expression sums the contributions from the change in ionic potential, the electrostatic potential arising from the electron density distortion, and the exchange-correlation potential $f_{\rm xc}$ (within LDA for simplicity). Using the linear density response $\delta n_{\bq}= \chi_0^e(\bq) \delta V^{\rm KS}_{\bq}$, where $\chi_0^e(\bq)$ is the Lindhard function of the Kohn-Sham orbitals, we arrive at the DFPT ansatz for the electron-phonon coupling:
\begin{equation}
\label{eq:KS_g}
g^{\txKS}(\bq) = \frac{g^{(0)}_{\bq}}{1-(v_{\bq}+f_{\txxc})\chi_0^\txe(\bq)} \, .
\end{equation}
The resulting quantity $g^{\txKS}(\bq)$ in Eq.~\eqref{eq:KS_g} should be understood as the screened ionic potential, and therefore depends only on the transferred momentum $\bq$. In standard \emph{ab initio} DFPT calculations for real materials, one then converts this potential into band-resolved electron–phonon matrix elements by projecting it onto the Kohn–Sham Bloch states, $\bra{\bk+\bq} \delta V^{\rm KS}_{\bq} \ket{\bk}$, which introduces an explicit dependence on the incoming momentum $\bk$ (and band indices). In the present work, however, our goal is to benchmark the screening of the underlying potential itself against the field-theoretic result. Since the EFT electron–phonon vertex is defined without any additional orbital projection, the natural DFPT object to compare with is the scalar function $g^{\txKS}(\bq)$. For the homogeneous electron gas, where the Bloch projection is trivial, we can therefore match $g^{\txKS}(\bq)$ to the EFT vertex $g(\bk,\bq)$ in Eq. \ref{eq:eft_g}; as shown below, the residual $\bk$-dependence of $g(\bk,\bq)$ in this regime is numerically weak and can be safely neglected.

\begin{figure}[htbp]
    \centering
    \includegraphics[width=0.9\linewidth]{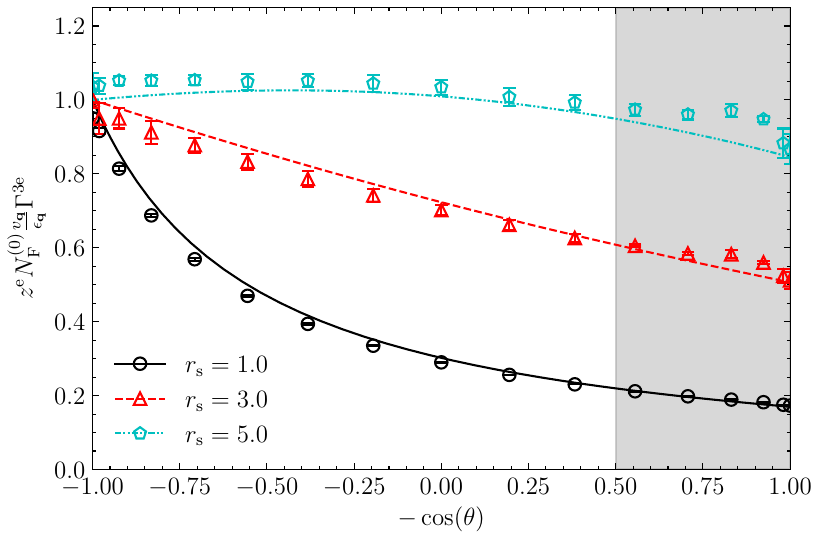}

    \caption{Comparison between the angle-resolved e-ph vertex correction in the uniform electron gas from variational diagrammatic Monte Carlo (points) and density functional perturbation theory (lines). The data are shown for $ z^{\text{qp}}N_\txF W^{\text{qp}}_{\mathbf{q}}\Gamma_3^{\text{qp}}(\mathbf{k},\mathbf{q})$ (VDiagMC data points), where $W^{\text{qp}}_{\mathbf{q}} = N_\txF v_{\mathbf{q}}/\epsilon_{\mathbf{q}}$ is the dimensionless screened Coulomb interaction.  
    The DFPT ansatz of $ z^{\text{qp}}N_\txF W^{\text{qp}}_{\mathbf{q}}\Gamma_3^{\text{qp}}(\mathbf{k},\mathbf{q})$, $\frac{N_\txF^{(0)} v_{\mathbf{q}}}{1 - (v_{\mathbf{q}} + f_{\text{xc}})\chi_0^e(\mathbf{q})}$ (lines), is based on 
    the Lindhard function, $\chi_0^e(\mathbf{q})$, and exchange-correlation kernel in the local density approximation, $f_{\text{xc}}$.
    Excellent agreement is observed for all values of $r_\txs$ and angles 
    $\theta$ between the incoming ($\mathbf{k}$) and outgoing ($\mathbf{k}+\mathbf{q}$) electron momenta on the Fermi surface, except 
    for the challenging backscattering region, $\theta \approx \pi$, where the diagrammatic series become sensitive to the logarithmic divergence in the Cooper channel.    
    }
 \label{fig:ver3angle}
\end{figure}

Using VDiagMC calculations for UEG at $r_s \in [1,5]$ we find that despite large interaction corrections to $z^{\txe}$,  $\epsilon_\bq$, and $\Gamma_3^\txe(\bq)$ separately, their product involves remarkable cancellation of interaction effects and the final result is very accurately
approximated by
\begin{equation}
z^\txe\frac{v_\bq}{\epsilon_\bq}\Gamma^\txe_3(\bk; \bq) \approx \frac{v_{\bq}}{1-(v_{\bq}+f_{xc})\chi_0^e(\bq)},
\label{eq:epi_ansatz}
\end{equation}
as demonstrated in Fig.~\ref{fig:ver3angle}. 
Diagrammatically, the electron-electron contribution to the quasiparticle weight $z^{\txe}$ is effectively cancelled by the renormalization of the electron-phonon vertex described by $\Gamma^\txe_3$. While this cancellation is exact in the long-wavelength limit ($q\rightarrow0$), our numerical results demonstrate that it holds with remarkable accuracy throughout the relevant range of momentum transfers between states on the Fermi surface ($|q| \le 2k_F$). This observation justifies our utilization of the DFPT-derived electron-phonon interaction within the present effective field theory framework.
Thus, DFPT and EFT for UEG produce nearly identical results
for UEG for all values of $r_\txs \le 5$:
\begin{equation}
g(\bk, \bq) \approx g^{\txKS}(\bq).
\label{eq:DFPTagreement}
\end{equation}

To obtain $\lambda$, the effective e-ph coupling must be combined with the physical phonon spectrum and density of states, $N_\txF$. 
This is where the two methods are radically different:
EFT is based on the quasiparticle density of states, 
while DFPT uses the band-structure density of states (bare one for UEG), $N^{(0)}_\txF$.
The two differ by the ratio of the quasiparticle mass to the electron mass.  Recent high-precision calculations for UEG \cite{vdiagmc2, markus_2023}
find this ratio to be very close to unity (with sub-percent accuracy at $r_\txs =5$). Therefore, the two densities of states end up nearly identical despite strong correlations.

Based on benchmark results for effective e-ph coupling and quasiparticle mass, we conclude that DFPT approximations for $\lambda$ are fully justified in simple metals such as Li, Na, and K. However, for strongly correlated systems
with complex band structure that cannot be approximated by UEG, 
such as semi-core electrons in transition metals, the accuracy of  
DFPT approximations remains an open question requiring further investigation 
by developing the corresponding  EFT strategy. Understanding DFPT limitations
and possible improvements in evaluating $\lambda$ is crucial for accurately predicting SC properties of a wider class of materials.

\begin{figure}
    \centering
    \includegraphics[width=0.9\linewidth]{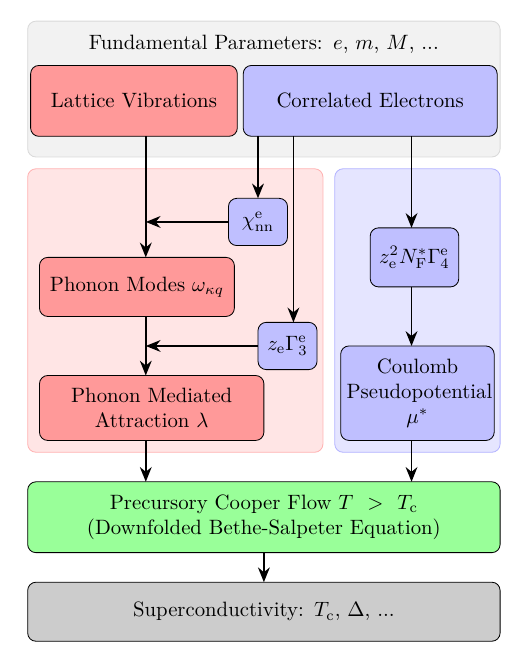}
    \caption{Proposed ab initio framework for electron-phonon SC beyond the weak correlation limit. Many-electron properties (blue boxes) such as the quasiparticle density of states $N_F^*$, quasiparticle weight, $z_e$, density-density correlation function, as well as the three-point ($\Gamma_3^e$) and four-point ($\Gamma_4^e$) vertex functions on the Fermi surface, are computed using high-order variational Diagrammatic Monte Carlo (or any other unbiased method). These quantities are subsequently used as 
    inputs for determining the phonon-mediated attraction (red box) and the Coulomb pseudopotential within a many-electron formalism. Specifically, $\mu^*$ is deduced from the quasiparticle scattering 
    in the Cooper-channel averaged over the Fermi surface. In radical departure from the standard ME framework, the critical temperature and gap function are calculated using the precursory Cooper flow of the anomalous vertex in the normal phase, enabling predictions of low-temperature SC and quantum phase transitions that are impossible to obtain within the conventional ME technique.  
}    
    \label{fig:flowchart}
\end{figure}

\section{Conventional Superconductors}
\label{sec:realmaterial}

We describe the \textit{ab initio} flowchart illustrated in Fig.~\ref{fig:flowchart} to predict SC from a microscopic model within the downfolded BSE theory, detailing a systematic approach to compute the superconducting transition temperature and related properties. This framework is broadly applicable to e-ph superconductors in generic correlated materials (e.g., simple metals, transition metals, and Hubbard-type tight-binding models), provided the adiabatic approximation holds (small $m/M$ and $\omega_\txD/E_\txF$ ratio), and the plasmon mode $\omega_p \sim E_F$. Here we consider clean metals, although generalization to include disorder is straightforward. The calculation proceeds along two interconnected directions: the correlated-electron part (blue boxes) and the phonon-related part (red boxes). All calculations start from the fundamental system parameters.

The correlated-electron part focuses on computing three crucial quantities:
(i) the density-density correlation function, $\chi^\txe(\bq)$,
(ii) the quasiparticle vertex correction, screened by the dielectric function, $z^\txe \Gamma_3^\txe(\bk, \bq)/\epsilon_\bq$, where $\bk$ is the incoming electron momentum, and $\bq$ is the momentum transfer, and (iii) the static quasiparticle e-e scattering amplitude averaged over the Fermi surface, $z_\txe^2 N_\txF^* \langle \Gamma^\txe_4 \rangle_\txFS$.
The last quantity is crucial for extracting the Coulomb pseudopotential $\mu^*$. Quantities (i) and (ii) are required for determining the effective phonon-mediated attraction in the presence of strong electron correlations.

The output of the correlated electrons flowchart, namely $\chi^\txe$ and $z^\txe \Gamma_3^\txe$, is fed into the phonon part of the flowchart. The effective mass density is used to compute the phonon dispersion renormalization, while $z^\txe \Gamma_3^\txe$ accounts for screening and renormalization of e-ph interaction between the quasiparticles dressed by e-e correlations. Finally, the two flowchart parts are combined by substituting $\lambda$ and $\mu^*$ as inputs to the downfolded ME (or BSE) equation. By solving this equation one obtains $T_\txc$, the frequency-dependent anomalous vertex correction, $\Lambda(\omega)$, and the frequency-dependent superconducting gap function, $\Delta(\omega)$, at the critical point. This completes the \textit{ab initio} procedure to predict superconducting properties from normal state calculations.


We apply this framework to revisit e-ph SC in simple metals where properties of conduction electrons can be approximated by UEG. Existing predictions of $T_\txc$ based on the downfolded ME theory or derived approximations, such as the McMillan or Allen-Dynes formulas, result in order-of-magnitude uncertainties for simple metals with sub-Kelvin $T_\txc$, depending on the choice of the phenomenological Coulomb pseudopotential~\cite{morel_anderson}.

Our theory is based on two assumptions:
(i) core and conduction bands are well separated by large gaps,
(ii) lattice potential experienced by the conduction electrons is weak/smooth allowing us to neglect Umklapp scattering processes.
Under these assumptions, it is possible to use our calculated Coulomb pseudopotential results for UEG corresponding to the material value of $r_s$. The same approach cannot be applied to materials exhibiting strong lattice potentials (e.g., Be), significant contributions from core electrons (e.g., Fe, Cu), strong spin-orbit coupling (e.g., Ta), flat-bands near the Fermi surface (e.g., Ca), or high-$T_c$ metals (with $T_\txc \gg 1 K$, e.g., Pb) where effective Coulomb interactions may change their sign.
Our investigation is, thus, focused on Li, Na, K, Mg, Al, and Zn---they are representative elements from the alkali, alkaline earth, and post-transition metal groups. Given remarkable agreement demonstrated in Fig.~\ref{fig:ver3angle}, our strategy is to use Eq.~(\ref{eq:DFPTagreement}) for the effective e-ph attraction from DFPT in combination with our Coulomb pseudopotential data to compute $T_\txc$ for several simple metals. This analysis showcases the effectiveness of the EFT approach to $\mu^*$ through higher accuracy of $T_\txc$ predictions and new insights into the prospects of low-$T_c$ superconductivity.

\subsection{Methods}

\setlength{\tabcolsep}{2pt} 
\begin{table*}[t]
    \centering
    \caption{First-Principle Calculation Results of Selected Simple Metals}
\begin{tabular}{|l|r|r|r|r|r|r|r|r|r|r|}
 & \(T_\txF(10^{3}\text{K})\) & \(\omega_{\log}\)\footnotemark[1](K) & \(\lambda_\text{prev}\)\footnotemark[2] & \(\lambda\) & \(r_\txs\)\footnotemark[3] & \(m_\txb\) & \(\psp\)\textsuperscript{*} & \(T_\text{EFT}\)(K) & \(T_\text{exp}\)(K) & $T_{\psp_{_\text{MA}}}$\footnotemark[6](K)\\
Li(9R) & 40 & 242\footnotemark[4] & 0.34\footnotemark[4] & 0.34\footnotemark[4] & 3.25\footnotemark[5] & 1.75\footnotemark[5] & 0.18 & \(5\times 10^{-3}\) & \(4\times 10^{-4}\) & 0.35\\
Li(hcp) & 41 & 243 & 0.33\footnotemark[4] & 0.37 & 3.19 & 1.4 & 0.17 & 0.03 &  \(4\times 10^{-4}\) & 0.64\\
Na & 42 & 127 & 0.181 & 0.2 & 3.96 & 1.0 & 0.15 & \(2\times 10^{-13}\) &  & $6\times 10^{-5}$\\
K & 26 & 85 & 0.132 & 0.11 & 4.86 & 1.0 & 0.16 & No SC &  & $10^{-120}$\\
Mg & 80 & 269 & 0.237 & 0.24 & 2.66 & 1.02 & 0.14 & \(5\times 10^{-5}\) & & 0.007\\
Al & 130 & 320 & 0.402 & 0.44 & 2.07 & 1.05 & 0.13 & 0.96 & 1.2& 1.9 \\
Zn & 121 & 111 & 0.508 & 0.502 & 2.90 & 1.0 & 0.12 & 0.874 & 0.875 & 1.37
\end{tabular}
    \footnotetext[1]{The $\log$-averaged frequency is calculated following Ref.~\cite{allen1975}.}
    \footnotetext[2]{Previous results for e-ph coupling 
    $\lambda_\text{prev}$ (mostly from Ref.\cite{kawamura_benchmark_2020} except for lithium) are shown for comparison with current $\lambda$ estimates. }
    \footnotetext[3]{Wigner-Seitz radius $r_\txs$ computed from the lattice constant and conduction electron properties from DFT calculations.}
    \footnotetext[4]{Electron-phonon coupling in lithium for 9R and hcp structures \cite{liu_structural_1999}. We computed the hcp case, but also adapted results from literature for the 9R case.}
    \footnotetext[5]{Data for $r_\txs$ and $m_\txb$ in lithium were 
    adapted from Ref.\cite{richardson_effective_1997}.}
    \footnotetext[6]{$T_{\psp_{_\text{MA}}}$ is the McMillan's formula~\cite{mcmillan1968}
    prediction using standard $\psp^*=0.1$ value.}

    \label{tab:simple_metals}
\end{table*}

To obtain band structures and phonon spectra we employed Quantum Espresso (QE)~\cite{QE1,QE2,QE3} package. In all calculations the optimized norm-conserving Vanderbilt (ONCV) potential~\cite{ONCV1, ONCV2} and Perdew-Burke-Ernzerhof (PBE) exchange-correlation functional~\cite{perdew_generalized_1996} were used.
Metallic crystals considered in this work form either FCC, or BCC, or HCP Bravais lattices. 
{To investigate pressure effects, the lattice constants (specifically for Al) were determined via a fitted equation of state from Ref.~\cite{Al_EOS}; a detailed discussion is provided in the Appendix~\ref{sec:app:al_pressure}.}
For the self-consistent-field (SCF) calculation, 
the energy cutoff was set to 90 Ry, the momentum-grid had $24 \times 24 \times 24$, $24 \times 24 \times 24$, $24 \times 24 \times 12$ points  for the FCC, BCC, and HCP structures, respectively. The error on the converged total energy was less than 0.001 Ry per atom.
Phonons were calculated using the DFPT approach of Ref.~\cite{dfpt_rev}, which was implemented within the QE package.
A coarse $\mathbf{q}$-grid for the FCC, BCC, and HCP lattices
contained $6\times 6 \times 6$, $6\times 6 \times 6$, and $6\times 6 \times 3$ points, respectively, and the acoustic sum rule was used to eliminate the small imaginary frequency at the $\mathbf{\Gamma}$ point.

To get the e-ph coupling constant $\lambda$, we used the EPW package~\cite{EPW1, EPW2}, which evaluates e-ph interaction for
Wannier functions, and subsequent Fourier interpolation to obtain the
e-ph interaction defined on an arbitrary $\mathbf{k},\mathbf{q}$ grid. 
Wannier functions were generated by the Wannier90 package~\cite{Wann1, Wannier_Vanderbilt_1997, Wannier_Vanderbilt_2001} using coarse 
$\mathbf{k}$-grids for the FCC, BCC, and HCP lattices with $12\times 12 \times 12$, $12\times 12 \times 12$, and $12\times 12 \times 6$ points, respectively.
The Wannier projectors for  Li, Na, K, Mg, Al were set to $s$ and $p$ orbitals, and the ones for Zn are $s$, $p$, $d$ orbitals.
To get a converged result for $\lambda$, we set a fine $\mathbf{k}$ ($\mathbf{q}$) grid for the FCC, BCC, and HCP lattices with $60\times 60 \times 60$, $60\times 60 \times 60$, and  $60\times 60 \times 30$ points, respectively.

To decide what Coulomb pseudopotential to use, we fit the band structure of considered metals with the UEG model. The UEG density is set to be that of conduction
electrons, and bare mass in the UEG model is extracted from the 
curvature of the band dispersion relation at the $\Gamma$ point.
This band mass, $m_\txb$, effectively rescales the $r_s$ parameter 
$r_\txs \to (m_\txb / m) r_\txs$. Effective mass renormalization in UEG 
is very small, and we assume that the same is true in our case, i.e.
$m_\txb$ can be interpreted as the true quasiparticle effective mass. 
The Fermi energy $E_\txF$ is determined by the difference between the calculated Fermi energy and energy at the $\Gamma$ point. Next, we 
interpolate pre-computed UEG results for Coulomb pseudopotential at rescaled value of $r_s$, see Fig.~\ref{fig:uvsrs},  to obtain the effective pseudopotential at the Fermi energy, $\mu_{E_F}$ of Eq.~(\ref{eq:mu_E_F}).

\subsection{Results}

\begin{figure}
    \centering
    \includegraphics[width=0.95\linewidth]{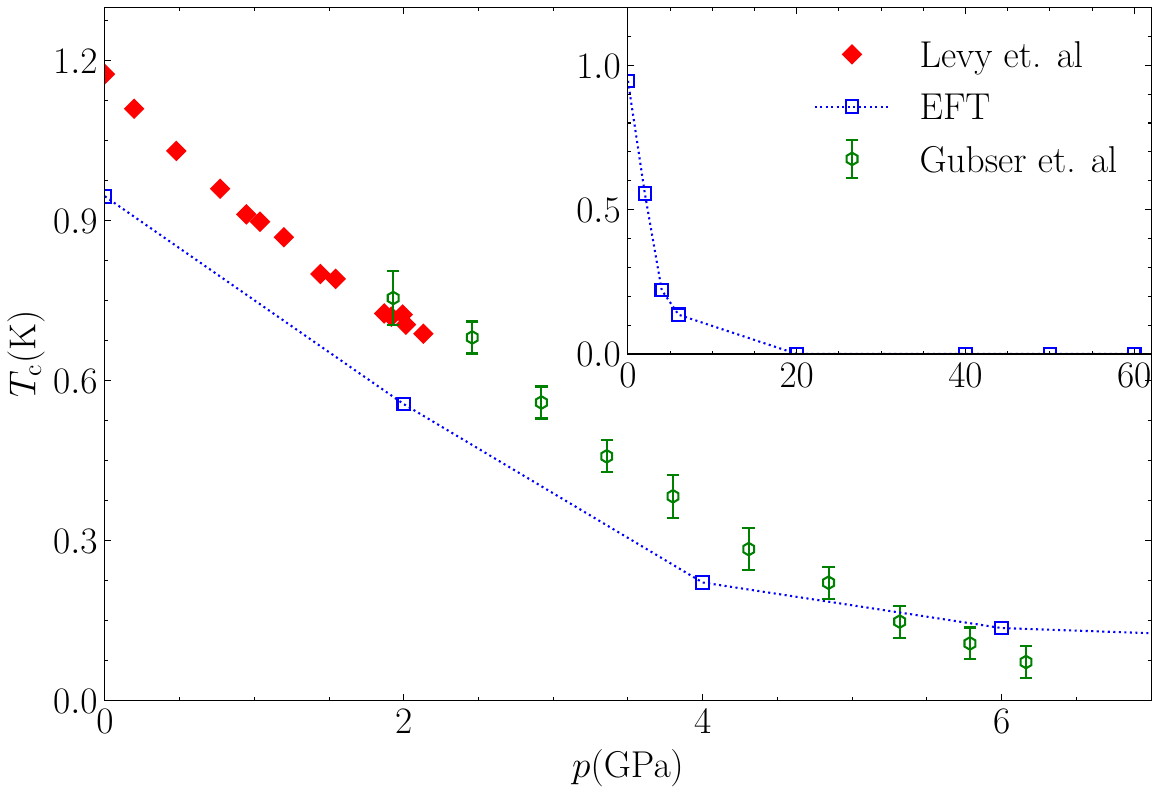}
    \caption{
    The pressure dependence of the superconducting critical temperature in aluminum. The squares are our theoretical results; the lines are guides to the eye. Experimental data from Levy et al.~\cite{levy_can_1964} and Gubser et al.~\cite{gubser_high-pressure_1975} are plotted as diamond and circular markers, respectively. 
    }
    \label{fig:mat_al}
\end{figure}

In Fig.~\ref{fig:mat_al}, we compare the experimental data for the pressure-dependent $T_\txc$ in aluminum \cite{levy_can_1964,gubser_high-pressure_1975}  with the corresponding results of our approach (up to the highest reported pressure). We see that our approach accurately captures the experimental trend, showing a clear decrease in $T_\txc$ as pressure increases from ambient conditions to $6$ GPa. 

\begin{figure}
    \centering
    \includegraphics[width=0.95\linewidth]{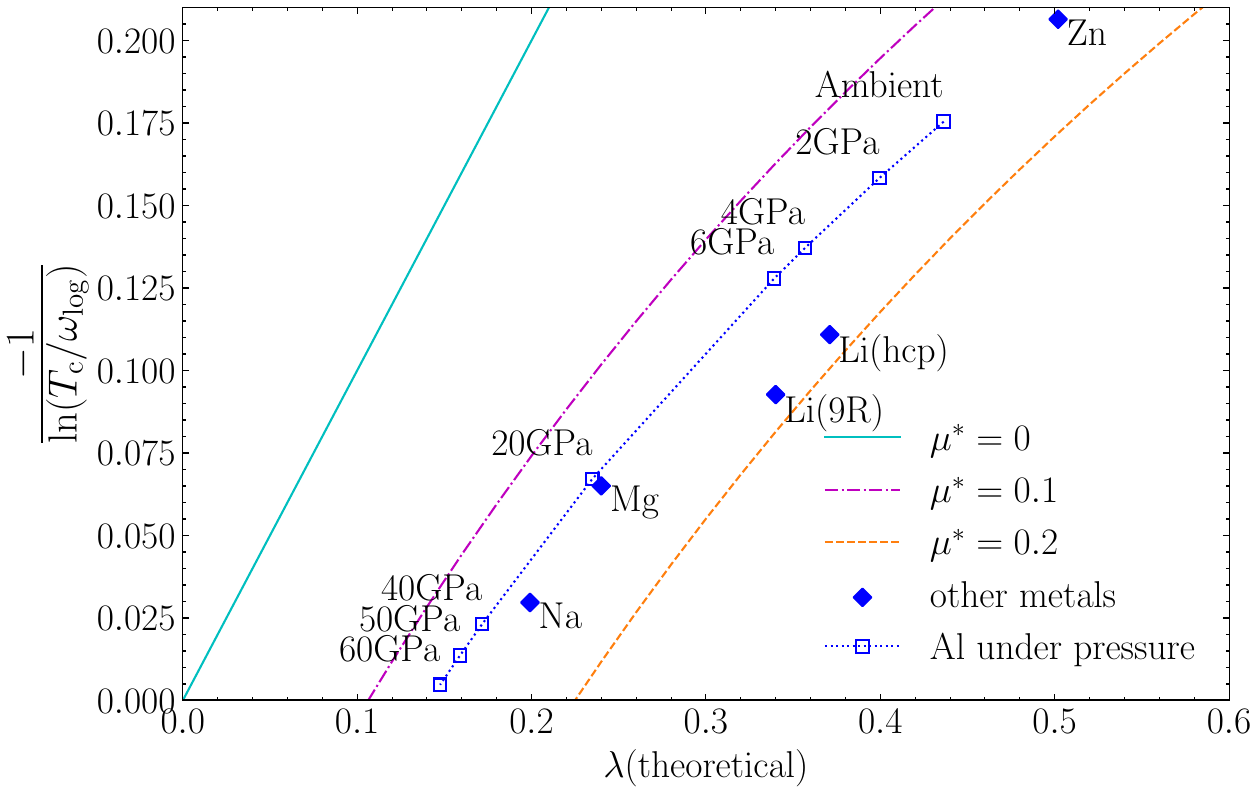}
    \caption{
    Effective BCS coupling strength, plotted as $-1/\ln(T_\txc/\omega_{\text{log}})$, for simple metals. The e-ph couplings were taken from the DFPT calculations, while the pseudopotentials were extracted from the VDiagMC results. In addition to the ambient pressure results for Li, Na, Mg, Al, and Zn, we also computed the pressure dependence of $T_\txc$ in Al. Potassium is not shown, as our calculations predict it to be non-superconducting.
    We also show predictions based on McMillan's formula for 
    different choices of $\psp^*$: $0$ (cyan line),  $0.1$ (magenta line) $0.2$ (orange line). We observe that our results for simple metals fall in between the conventional choices for $\mu^*$. Notably, $\psp^*$ for two Li 
    structures is larger than for other metals, resulting in $T_\txc$ values much lower than estimated based on $\mu \leq 0.15$. Detailed values for $T_\txc$ and $\omega_{\text{log}}$ are listed in Table~I.}
    \label{fig:mat_main}
\end{figure}

Although similar benchmarks using the SCDFT method~\cite{profeta_superconductivity_2006} also show good agreement with the experiment, the treatment of Coulomb repulsion within SCDFT involves uncontrolled approximations with respect to vertex corrections, screening dynamics, and quasiparticle weight. 
In contrast, our approach is based on a rigorous microscopic foundation and clear understanding of how one should proceed in cases significantly different from UEG. 
We performed calculations beyond the current experimental pressure limit of $6$~GPa; 
{see the inset of Fig.~\ref{fig:mat_al} and Fig.~\ref{fig:mat_main}}.
Assuming no structural transitions,  we predict that SC in Al vanishes at pressure $~60$GPa. 
{Notably, even at $20$~GPa, $T_\txc$ is already suppressed below $1$~mK, rendering it undetectable with current experimental techniques.}

With the same numeric protocol, we calculated $T_c$ for other simple metals; the results are presented in  Fig.~\ref{fig:mat_main} and TABLE~\ref{tab:simple_metals}.
Given $T_c / \omega_\text{log} \propto \exp(-1/g)$ scaling, we choose to plot all data using $y=1/\ln (\omega_\text{log} /T\txc)$ for the vertical axis, which may be considered as an effective attractive coupling in the Cooper channel (if $T_\txc$ is finite) for a given $\psp^*$ and $\lambda$
(here $\omega_\text{log}$ is the $\log$-averaged phonon frequency calculated following Ref.~\cite{allen1975}).

The solid points in Fig.~\ref{fig:mat_main} and values given in Table I are our predictions. We see that  for Na and Mg the predicted transition temperatures are so low that they cannot yet be
probed with existing experimental techniques. For Li, our first-principles approach is significantly lower than existing estimates for $T_c$, but still
remains about an order of magnitude higher than the reported experimental value, in part due to the existing controversy with respect to the 
lattice structure at extremely low temperatures.

We compare our results with those obtained from standard theories using a phenomenological $\mu^\star$. The differences between the cyan, magenta-dot-dashed and orange-dashed lines in Fig ~\ref{fig:mat_main} show that uncertainties in the $\mu^*$ values have a huge effect on the predicted $T_c$  of these low-$T_c$  materials, demonstrating that the phenomenological approach has no predictive power for low-temperature superconductors. We further see that  some superconductors have ultra-low $T_c$ not because $\lambda$ is vanishingly small, but rather because the e-ph attraction is nearly balanced out by the e-e repulsion.
On a practical note, precise values of $\mu^*$ are particularly relevant for metals with $T_\txc$ in the $10 \text{mK}$ to $1 \text{K}$ range, which is of interest to the superconducting electronics industry and thus highlight the need for a more thorough treatment of e-e scattering in real materials.

Our results suggest that Mg, Na, and K are close to the 
critical point between states with and without s-wave SC.
(We do not use the term ``quantum critical point'' here because SC in some other symmetry channel will preempt it.) 
Although these metals remain in the normal state as the temperature is lowered (e.g., well below $1K$), they enter a quantum critical regime  characterized by a diverging pair-field susceptibility $\chi \sim \ln (T)$~\cite{pengcheng} without any fine-tuning.  
This signal is worth verifying in future experiments.


\section{Conclusion}

We developed a rigorous {\it ab initio} framework, Fig.~\ref{fig:flowchart}, to obtain downfolded ME theory on the Fermi surface for correlated electrons, which resolves long-standing ambiguities in the treatment of Coulomb repulsion. By systematically integrating out high-energy electronic degrees of freedom using field-theoretical renormalization techniques, we arrive at frequency-only ME equations with all parameters precisely linked to the underlying microscopic Hamiltonian: the Coulomb pseudopotential $\mu^*$, determined by the two-quasiparticle scattering amplitude renormalized by the
Bogoliubov-Tolmachev-Shirkov logarithm~\cite{bogoliubov59,tolmachev1961logarithmic}, the electron-phonon coupling $\lambda$ screened by the quasiparticle vertex correction,
and the quasiparticle density of states.
Our approach eliminates the need for phenomenological---and thus uncontrolled---treatment of key parameters and establishes protocols for computing 
$\mu^*$ and $\lambda$ from the first principles.

For $\mu^*$, high-order VDiagMC calculations for UEG reveal that pseudopotential values at the Fermi energy scale are significantly
larger than estimates based on Yukawa or static-RPA screening. 
While UEG is barely a prototypical model featuring strong e-e correlations, its density-dependent $\mu^*$ can be used to parameterize pseudopotentials
in real materials. For $\lambda$, we confirm the accuracy of DFPT by comparing its predictions to our many-body vertex-corrected e-ph coupling, finding excellent agreement for UEG. By combining these results 
with observation that effective mass renormalization in the 
Coulomb system is negligibly small, we arrive at fitting-free 
predictions for $T_c$ in simple metals with an order-of-magnitude 
improvement for sub-Kelvin superconductors. In particular, we
predict that s-wave superconductivity in aluminum is suppressed to zero
at $60$ GPa, and that pair-susceptibility in Mg and Na should exhibit 
quantum critical scaling below $10$ K. These are testable predictions
following from the interplay between the Coulomb repulsion and phonon-mediated attraction.
The accuracy of the above-mentioned results rests on the power of precursory Cooper flow of the anomalous vertex correction to predict ultra-low $T_\txc$ and critical points from normal state calculations.

The \emph{ab initio} framework of Fig.~\ref{fig:flowchart} applies not only to simple metals but to any other correlated material
with well-defined quasiparticles at the Fermi surface and
large separation between the phonon and electron energy scales $\omega_\txD \ll E_\txF, \omega_\txp $.
However, implementing high-order diagrammatics and renormalization,
as it was done for UEG, is a major technical challenge for lattice
systems with complex band structure lacking spherical symmetry. For example, in transition metals more than one band is involved in
screening~\cite{schrieffer_1969} and Umklapp processes cannot be neglected. {In cases including two dimensional systems the particle-particle channel Coulomb interaction is  non-local (not well approximated by a structureless interaction and as noted by Simonato, Katsnelson and Rosner \cite{Simonato23} a more careful treatment is required. }
If the energy-scale separation condition, on which the downfolding approximation rests, is not satisfied, e.g. at extreme densities relevant to astrophysical objects such as white and black dwarf stars~\cite{ginzburg_superconductivity_1968}, {or perhaps in flat-band systems and 2D materials where the $\sqrt{q}$ plasmon dispersion creates challenges to the energy scale separation arguments} there may be no real alternative to working with full BSE.

Our findings highlight the crucial role of e-e 
correlations in the two-particle vertex function; they 
cannot be quantified by the single-electron exchange-correlation potential within the DFT approximation~\cite{kohn1964,kohn1965}. Modern numerical methods like VDiagMC enable accurate computation of these correlations, providing key insights into the topic of paring despite strong repulsive forces, and opening new possibilities for understanding other emergent phenomena, such as accurate calculation of many-body forces in metals, with direct implications for more reliable molecular dynamics simulations.
In conclusion, by providing a complete and predictive first-principles framework for incorporating e-e correlations into the existing machinery
of predicting superconductivity from e-ph interaction, 
this work represents a step forward in our ability to predict superconductivity starting from properties of the microscopic Hamiltonian
without uncontrolled approximations.


\begin{acknowledgments}
We thank S. Das Sarma and A. Chubukov for valuable discussions. 
K. Chen and X. Cai are supported by the National Natural Science Foundation of China under Grants No. 12474245 and No. 12447103, the National Key Research and Development Program of China under Grant No.2024YFA1408604, and the GHfund A (202407010637).
X. Cai and T. Wang acknowledge support from the National Science Foundation under Grant No. DMR-2335904. 
T. Zhang and S. Zhang acknowledge the support from National Key R\&D Project (Grant Nos. 2023YFA1407400 and 2024YFA1409200), and the National Natural Science Foundation of China (Grant Nos. 12374165 and 12447101).
A. Millis's work at Columbia University was supported by the Keele Foundation and the Columbia University Materials Research Science and Engineering Center (MRSEC) through NSF Grant DMR-2011738.
{N. Prokof'ev and B. Svistunov acknowledge support from the Simons Foundation (SFI-MPS-NFS-00006741-07, N.P. a B.S.) in the Simons Collaboration on New Frontiers in Superconductivity.}
The Flatiron Institute is a division of the simons foundation.
\end{acknowledgments}

\bibliography{ref.bib}

@article{bardeen_microscopic_1957,
  title = {Microscopic {{Theory}} of {{Superconductivity}}},
  author = {Bardeen, J. and Cooper, L. N. and Schrieffer, J. R.},
  year = {1957},
  month = apr,
  journal = {Physical Review},
  volume = {106},
  number = {1},
  pages = {162--164},
  publisher = {{American Physical Society}},
  doi = {10.1103/PhysRev.106.162},
}

@article{bardeen_theory_1957,
  title = {Theory of {{Superconductivity}}},
  author = {Bardeen, J. and Cooper, L. N. and Schrieffer, J. R.},
  year = {1957},
  month = dec,
  journal = {Physical Review},
  volume = {108},
  number = {5},
  pages = {1175--1204},
  publisher = {{American Physical Society}},
  doi = {10.1103/PhysRev.108.1175},
}

@article{pellegrini2024ab,
  title={Ab initio methods for superconductivity},
  author={Pellegrini, Camilla and Sanna, Antonio},
  journal={Nature Reviews Physics},
  volume={6},
  number={8},
  pages={509--523},
  year={2024},
  publisher={Nature Publishing Group UK London}
}

@article{cooper_bound_1956-1,
  title = {Bound {{Electron Pairs}} in a {{Degenerate Fermi Gas}}},
  author = {Cooper, Leon N.},
  year = {1956},
  month = nov,
  journal = {Physical Review},
  volume = {104},
  number = {4},
  pages = {1189--1190},
  publisher = {{American Physical Society}},
  doi = {10.1103/PhysRev.104.1189},
}

@article{migdal1958,
  title        = {Interaction between electrons and lattice vibrations in a normal metal},
  author       = {A.B. Migdal},
  year         = {1958},
  monthsup     = nov,
  journal      = {Sov.Phys.-JETP},
  volume       = {7},
  numbersup    = {17},
  pagesups     = {11163--11166},
  publishersup = {{American Physical Society}},
  doisup       = {10.1103/PhysRevB.46.11163}
}

@article{eliashberg1960interactions,
  title   = {Interactions between electrons and lattice vibrations in a superconductor},
  author  = {Eliashberg, GM},
  journal = {Sov. Phys. JETP},
  volume  = {11},
  number  = {3},
  pages   = {696--702},
  year    = {1960}
}

@article{eliashberg1961temperature,
  title   = {Temperature Green’s function for electrons in a superconductor},
  author  = {Eliashberg, GM},
  journal = {Sov. Phys. JETP},
  volume  = {12},
  number  = {5},
  pages   = {1000--1002},
  year    = {1961}
}

@incollection{irwin_transition-edge_2005,
  title = {Transition-{{Edge Sensors}}},
  booktitle = {Cryogenic {{Particle Detection}}},
  author = {Irwin, K.D. and Hilton, G.C.},
  editor = {Enss, Christian},
  year = {2005},
  series = {Topics in {{Applied Physics}}},
  pages = {63--150},
  publisher = {{Springer}},
  address = {{Berlin, Heidelberg}},
  doi = {10.1007/10933596_3},
  urldate = {2024-01-11},
  isbn = {978-3-540-31478-3},
  langid = {english},
}

@article{qubit_poisoning,
  title = {Quasiparticle Poisoning in Superconducting Quantum Computers},
  author = {Aumentado, Jos{\'e} and Catelani, Gianluigi and Serniak, Kyle},
  year = {2023},
  month = aug,
  journal = {Physics Today},
  volume = {76},
  number = {8},
  pages = {34--39},
  issn = {0031-9228},
  doi = {10.1063/PT.3.5291},
  urldate = {2024-12-23},
}

@article{scalapino1969electron,
  title={THE ELECTRON-PHONON INTERACTION AND STRONG-COUPLING SUPERCONDUCTORS},
  author={Scalapino, Douglas J},
  journal={Superconductivity: Part 1 (In Two Parts)},
  volume={1},
  pages={449},
  year={1969},
  publisher={Marcel Dekker, Inc.}
}

@article{gladstone2018superconductivity,
  title={SUPERCONDUCTIVITY IN THE TRANSITION METALS: THEORY AND EXPERIMENT},
  author={Gladstone, G and Jensen, MA and Schrieffer, JR},
  journal={Superconductivity: Part 2 (In Two Parts)},
  volume={2},
  pages={665},
  year={1969},
  publisher={Marcel Dekker, Inc.}
}

@article{morel_anderson,
  title = {Calculation of the Superconducting State Parameters with Retarded Electron-Phonon Interaction},
  author = {Morel, P. and Anderson, P. W.},
  journal = {Phys. Rev.},
  volume = {125},
  issue = {4},
  pages = {1263--1271},
  numpages = {0},
  year = {1962},
  month = {Feb},
  publisher = {American Physical Society},
  doi = {10.1103/PhysRev.125.1263},
  url = {https://link.aps.org/doi/10.1103/PhysRev.125.1263}
}

@incollection{RAINER1986371,
title = {Chapter 4: Principles of AB Initio Calculations of Superconducting Transition Temperatures},
editor = {D.F. Brewer},
series = {Progress in Low Temperature Physics},
publisher = {Elsevier},
volume = {10},
pages = {371-424},
year = {1986},
issn = {0079-6417},
doi = {https://doi.org/10.1016/S0079-6417(08)60024-4},
url = {https://www.sciencedirect.com/science/article/pii/S0079641708600244},
author = {Dierk Rainer}
}

@article{margine2016electron,
  title={Electron-phonon interaction and pairing mechanism in superconducting Ca-intercalated bilayer graphene},
  author={Margine, ER and Lambert, Henry and Giustino, Feliciano},
  journal={Scientific Reports},
  volume={6},
  number={1},
  pages={21414},
  year={2016},
  publisher={Nature Publishing Group UK London}
}

@article{RPA_mu,
  title = {Effect of Van Hove singularities on high-${T}_{\mathrm{c}}$ superconductivity in ${\mathrm{H}}_{3}\mathrm{S}$},
  author = {Sano, Wataru and Koretsune, Takashi and Tadano, Terumasa and Akashi, Ryosuke and Arita, Ryotaro},
  journal = {Phys. Rev. B},
  volume = {93},
  issue = {9},
  pages = {094525},
  numpages = {16},
  year = {2016},
  month = {Mar},
  publisher = {American Physical Society},
  doi = {10.1103/PhysRevB.93.094525},
  url = {https://link.aps.org/doi/10.1103/PhysRevB.93.094525}
}

@article{RPA_mu2,
  title = {Revisiting homogeneous electron gas in pursuit of properly normed ab initio Eliashberg theory},
  author = {Akashi, Ryosuke},
  journal = {Phys. Rev. B},
  volume = {105},
  issue = {10},
  pages = {104510},
  numpages = {11},
  year = {2022},
  month = {Mar},
  publisher = {American Physical Society},
  doi = {10.1103/PhysRevB.105.104510},
  url = {https://link.aps.org/doi/10.1103/PhysRevB.105.104510}
}

@article{xiansheng,
  title = {Superconductivity in the uniform electron gas: Irrelevance of the Kohn-Luttinger mechanism},
  author = {Cai, Xiansheng and Wang, Tao and Prokof'ev, Nikolay V. and Svistunov, Boris V. and Chen, Kun},
  journal = {Phys. Rev. B},
  volume = {106},
  issue = {22},
  pages = {L220502},
  numpages = {5},
  year = {2022},
  month = {Dec},
  publisher = {American Physical Society},
  doi = {10.1103/PhysRevB.106.L220502},
  url = {https://link.aps.org/doi/10.1103/PhysRevB.106.L220502}
}

@article{23tao,
  title = {Origin of the Coulomb pseudopotential},
  author = {Wang, Tao and Cai, Xiansheng and Chen, Kun and Svistunov, Boris V. and Prokof'ev, Nikolay V.},
  journal = {Phys. Rev. B},
  volume = {107},
  issue = {14},
  pages = {L140507},
  numpages = {5},
  year = {2023},
  month = {Apr},
  publisher = {American Physical Society},
  doi = {10.1103/PhysRevB.107.L140507},
  url = {https://link.aps.org/doi/10.1103/PhysRevB.107.L140507}
}

@article{pengcheng,
  title = {Precursory Cooper flow in ultralow-temperature superconductors},
  author = {Hou, Pengcheng and Cai, Xiansheng and Wang, Tao and Deng, Youjin and Prokof'ev, Nikolay V. and Svistunov, Boris V. and Chen, Kun},
  journal = {Phys. Rev. Res.},
  volume = {6},
  issue = {1},
  pages = {013099},
  numpages = {10},
  year = {2024},
  month = {Jan},
  publisher = {American Physical Society},
  doi = {10.1103/PhysRevResearch.6.013099},
  url = {https://link.aps.org/doi/10.1103/PhysRevResearch.6.013099}
}

@misc{hou_feynman_2024,
  title = {Feynman {{Diagrams}} as {{Computational Graphs}}},
  author = {Hou, Pengcheng and Wang, Tao and Cerkoney, Daniel and Cai, Xiansheng and Li, Zhiyi and Deng, Youjin and Wang, Lei and Chen, Kun},
  year = {2024},
  month = feb,
  number = {arXiv:2403.18840},
  eprint = {2403.18840},
  primaryclass = {cond-mat, physics:hep-ph, physics:hep-th, physics:physics},
  publisher = {arXiv},
  urldate = {2024-03-29},
  archiveprefix = {arXiv},
}

@article{ashcroft1,
  title = {Effective electron-electron interactions and the theory of superconductivity},
  author = {Richardson, C. F. and Ashcroft, N. W.},
  journal = {Phys. Rev. B},
  volume = {55},
  issue = {22},
  pages = {15130--15145},
  numpages = {0},
  year = {1997},
  month = {Jun},
  publisher = {American Physical Society},
  doi = {10.1103/PhysRevB.55.15130},
  url = {https://link.aps.org/doi/10.1103/PhysRevB.55.15130}
}

@article{ashcroft2,
  title = {High Temperature Superconductivity in Metallic Hydrogen: Electron-Electron Enhancements},
  author = {Richardson, C. F. and Ashcroft, N. W.},
  journal = {Phys. Rev. Lett.},
  volume = {78},
  issue = {1},
  pages = {118--121},
  numpages = {0},
  year = {1997},
  month = {Jan},
  publisher = {American Physical Society},
  doi = {10.1103/PhysRevLett.78.118},
  url = {https://link.aps.org/doi/10.1103/PhysRevLett.78.118}
}

@article{hubbard_mu,
  title = {Retardation effects and the Coulomb pseudopotential in the theory of superconductivity},
  author = {Bauer, Johannes and Han, Jong E. and Gunnarsson, Olle},
  journal = {Phys. Rev. B},
  volume = {87},
  issue = {5},
  pages = {054507},
  numpages = {21},
  year = {2013},
  month = {Feb},
  publisher = {American Physical Society},
  doi = {10.1103/PhysRevB.87.054507},
  url = {https://link.aps.org/doi/10.1103/PhysRevB.87.054507}
}

@article{liu_structural_1999,
  title = {Structural Phase Stability and Electron-Phonon Coupling in Lithium},
  author = {Liu, Amy Y. and Quong, Andrew A. and Freericks, J. K. and Nicol, E. J. and Jones, Emily C.},
  year = {1999},
  month = feb,
  journal = {Physical Review B},
  volume = {59},
  number = {6},
  pages = {4028--4035},
  publisher = {American Physical Society},
  doi = {10.1103/PhysRevB.59.4028},
  urldate = {2024-03-22},
}

@article{richardson_effective_1997,
  title = {Effective Electron-Electron Interactions and the Theory of Superconductivity},
  author = {Richardson, C. F. and Ashcroft, N. W.},
  year = {1997},
  month = jun,
  journal = {Physical Review B},
  volume = {55},
  number = {22},
  pages = {15130--15145},
  issn = {0163-1829, 1095-3795},
  doi = {10.1103/PhysRevB.55.15130},
  urldate = {2020-08-27},
  langid = {english},
}

@article{tuoriniemi_superconductivity_2007,
  title = {Superconductivity in Lithium below 0.4 Millikelvin at Ambient Pressure},
  author = {Tuoriniemi, Juha and {Juntunen-Nurmilaukas}, Kirsi and Uusvuori, Johanna and Pentti, Elias and Salmela, Anssi and Sebedash, Alexander},
  year = {2007},
  month = may,
  journal = {Nature},
  volume = {447},
  number = {7141},
  pages = {187--189},
  publisher = {Nature Publishing Group},
  issn = {1476-4687},
  doi = {10.1038/nature05820},
  urldate = {2020-08-27},
  copyright = {2007 Nature Publishing Group},
  langid = {english},
}

@article{kawamura_benchmark_2020,
  title = {Benchmark of Density Functional Theory for Superconductors in Elemental Materials},
  author = {Kawamura, Mitsuaki and Hizume, Yuma and Ozaki, Taisuke},
  year = {2020},
  month = apr,
  journal = {Physical Review B},
  volume = {101},
  number = {13},
  pages = {134511},
  issn = {2469-9950, 2469-9969},
  doi = {10.1103/PhysRevB.101.134511},
  urldate = {2024-05-28},
  langid = {english},
}

@article{stefanucci_2023,
  title = {In and {{Out-of-Equilibrium Ab Initio Theory}} of {{Electrons}} and {{Phonons}}},
  author = {Stefanucci, Gianluca and {van Leeuwen}, Robert and Perfetto, Enrico},
  year = {2023},
  month = sep,
  journal = {Physical Review X},
  volume = {13},
  number = {3},
  pages = {031026},
  publisher = {American Physical Society},
  doi = {10.1103/PhysRevX.13.031026},
  urldate = {2023-09-11},
}

@article{QE1,
doi = {10.1088/0953-8984/21/39/395502},
url = {https://dx.doi.org/10.1088/0953-8984/21/39/395502},
year = {2009},
month = {sep},
publisher = {},
volume = {21},
number = {39},
pages = {395502},
author = {Paolo Giannozzi and Stefano Baroni and Nicola Bonini and Matteo Calandra and Roberto Car and Carlo Cavazzoni and Davide Ceresoli and Guido L Chiarotti and Matteo Cococcioni and Ismaila Dabo and Andrea Dal Corso and Stefano de Gironcoli and Stefano Fabris and Guido Fratesi and Ralph Gebauer and Uwe Gerstmann and Christos Gougoussis and Anton Kokalj and Michele Lazzeri and Layla Martin-Samos and Nicola Marzari and Francesco Mauri and Riccardo Mazzarello and Stefano Paolini and Alfredo Pasquarello and Lorenzo Paulatto and Carlo Sbraccia and Sandro Scandolo and Gabriele Sclauzero and Ari P Seitsonen and Alexander Smogunov and Paolo Umari and Renata M Wentzcovitch},
title = {QUANTUM ESPRESSO: a modular and open-source software project for quantum
simulations of materials},
journal = {Journal of Physics: Condensed Matter}
}

@article{QE2,
doi = {10.1088/1361-648X/aa8f79},
url = {https://dx.doi.org/10.1088/1361-648X/aa8f79},
year = {2017},
month = {oct},
publisher = {IOP Publishing},
volume = {29},
number = {46},
pages = {465901},
author = {P Giannozzi and O Andreussi and T Brumme and O Bunau and M Buongiorno Nardelli and M Calandra and R Car and C Cavazzoni and D Ceresoli and M Cococcioni and N Colonna and I Carnimeo and A Dal Corso and S de Gironcoli and P Delugas and R A DiStasio and A Ferretti and A Floris and G Fratesi and G Fugallo and R Gebauer and U Gerstmann and F Giustino and T Gorni and J Jia and M Kawamura and H-Y Ko and A Kokalj and E Küçükbenli and M Lazzeri and M Marsili and N Marzari and F Mauri and N L Nguyen and H-V Nguyen and A Otero-de-la-Roza and L Paulatto and S Poncé and D Rocca and R Sabatini and B Santra and M Schlipf and A P Seitsonen and A Smogunov and I Timrov and T Thonhauser and P Umari and N Vast and X Wu and S Baroni},
title = {Advanced capabilities for materials modelling with Quantum ESPRESSO},
journal = {Journal of Physics: Condensed Matter}
}

@article{QE3,
    author = {Giannozzi, Paolo and Baseggio, Oscar and BonfĂ , Pietro and Brunato, Davide and Car, Roberto and Carnimeo, Ivan and Cavazzoni, Carlo and de Gironcoli, Stefano and Delugas, Pietro and Ferrari Ruffino, Fabrizio and Ferretti, Andrea and Marzari, Nicola and Timrov, Iurii and Urru, Andrea and Baroni, Stefano},
    title = "{Quantum ESPRESSO toward the exascale}",
    journal = {The Journal of Chemical Physics},
    volume = {152},
    number = {15},
    pages = {154105},
    year = {2020},
    month = {04},
    issn = {0021-9606},
    doi = {10.1063/5.0005082},
    url = {https://doi.org/10.1063/5.0005082},
}

@article{ONCV1,
  title = {Optimized norm-conserving Vanderbilt pseudopotentials},
  author = {Hamann, D. R.},
  journal = {Phys. Rev. B},
  volume = {88},
  issue = {8},
  pages = {085117},
  numpages = {10},
  year = {2013},
  month = {Aug},
  publisher = {American Physical Society},
  doi = {10.1103/PhysRevB.88.085117},
  url = {https://link.aps.org/doi/10.1103/PhysRevB.88.085117}
}

@article{ONCV2,
title = {Optimization algorithm for the generation of ONCV pseudopotentials},
journal = {Computer Physics Communications},
volume = {196},
pages = {36-44},
year = {2015},
issn = {0010-4655},
doi = {https://doi.org/10.1016/j.cpc.2015.05.011},
url = {https://www.sciencedirect.com/science/article/pii/S0010465515001897},
author = {Martin Schlipf and FranÃ§ois Gygi}
}

@article{perdew_generalized_1996,
  title={Generalized gradient approximation made simple},
  author={Perdew, John P and Burke, Kieron and Ernzerhof, Matthias},
  journal={Phys. Rev. Lett.},
  volume={77},
  number={18},
  pages={3865},
  year={1996},
  publisher={APS},
  url = {https://link.aps.org/doi/10.1103/PhysRevLett.77.3865}
}

@article{dfpt_rev,
  title = {Phonons and related crystal properties from density-functional perturbation theory},
  author = {Baroni, Stefano and de Gironcoli, Stefano and Dal Corso, Andrea and Giannozzi, Paolo},
  journal = {Rev. Mod. Phys.},
  volume = {73},
  issue = {2},
  pages = {515--562},
  numpages = {0},
  year = {2001},
  month = {Jul},
  publisher = {American Physical Society},
  doi = {10.1103/RevModPhys.73.515},
  url = {https://link.aps.org/doi/10.1103/RevModPhys.73.515}
}

@Article{EPW1,
     Title   = {Electron-phonon interaction using Wannier functions},
     Author  = {F. Giustino and M. L. Cohen and S. G. Louie},
     Journal = {Phys. Rev. B},
     Year    = {2007},
     Volume  = {76},
     Pages   = {165108},
     Doi     = {10.1103/PhysRevB.76.165108}
   }

@Article{EPW2,
     Title   = {EPW: Electron–phonon coupling, transport and superconducting properties using maximally localized Wannier functions},
     Author  = {S. Ponc\'e and E.R. Margine and C. Verdi and F. Giustino},
     Journal = {Computer Physics Communications},
     Year    = {2016},
     Volume  = {209},
     Pages   = {116 - 133},
     Doi     = {https://doi.org/10.1016/j.cpc.2016.07.028}
   }

@article{AEPW,
  title = {Anisotropic {{Migdal-Eliashberg}} Theory Using {{Wannier}} Functions},
  author = {Margine, E. R. and Giustino, F.},
  year = {2013},
  month = jan,
  journal = {Physical Review B},
  volume = {87},
  number = {2},
  pages = {024505},
  issn = {1098-0121, 1550-235X},
  doi = {10.1103/PhysRevB.87.024505},
  urldate = {2022-08-29},
  langid = {english},
}

@article{Wann1,
doi = {10.1088/1361-648X/ab51ff},
url = {https://dx.doi.org/10.1088/1361-648X/ab51ff},
year = {2020},
month = {jan},
publisher = {IOP Publishing},
volume = {32},
number = {16},
pages = {165902},
author = {Giovanni Pizzi and Valerio Vitale and Ryotaro Arita and Stefan Blügel and Frank Freimuth and Guillaume Géranton and Marco Gibertini and Dominik Gresch and Charles Johnson and Takashi Koretsune and Julen Ibañez-Azpiroz and Hyungjun Lee and Jae-Mo Lihm and Daniel Marchand and Antimo Marrazzo and Yuriy Mokrousov and Jamal I Mustafa and Yoshiro Nohara and Yusuke Nomura and Lorenzo Paulatto and Samuel Poncé and Thomas Ponweiser and Junfeng Qiao and Florian Thöle and Stepan S Tsirkin and Małgorzata Wierzbowska and Nicola Marzari and David Vanderbilt and Ivo Souza and Arash A Mostofi and Jonathan R Yates},
title = {Wannier90 as a community code: new features and applications},
journal = {Journal of Physics: Condensed Matter}
}

@article{Wannier_Vanderbilt_1997,
  title = {Maximally localized generalized Wannier functions for composite energy bands},
  author = {Marzari, Nicola and Vanderbilt, David},
  journal = {Phys. Rev. B},
  volume = {56},
  issue = {20},
  pages = {12847--12865},
  numpages = {0},
  year = {1997},
  month = {Nov},
  publisher = {American Physical Society},
  doi = {10.1103/PhysRevB.56.12847},
  url = {https://link.aps.org/doi/10.1103/PhysRevB.56.12847}
}

@article{Wannier_Vanderbilt_2001,
  title = {Maximally localized Wannier functions for entangled energy bands},
  author = {Souza, Ivo and Marzari, Nicola and Vanderbilt, David},
  journal = {Phys. Rev. B},
  volume = {65},
  issue = {3},
  pages = {035109},
  numpages = {13},
  year = {2001},
  month = {Dec},
  publisher = {American Physical Society},
  doi = {10.1103/PhysRevB.65.035109},
  url = {https://link.aps.org/doi/10.1103/PhysRevB.65.035109}
}

@article{giustino_2017,
  title = {Electron-Phonon Interactions from First Principles},
  author = {Giustino, Feliciano},
  year = {2017},
  month = feb,
  journal = {Reviews of Modern Physics},
  volume = {89},
  number = {1},
  pages = {015003},
  issn = {0034-6861, 1539-0756},
  doi = {10.1103/RevModPhys.89.015003},
  urldate = {2022-11-07},
  langid = {english},
}

@article{rietschel,
  title = {Role of electron Coulomb interaction in superconductivity},
  author = {Rietschel, H. and Sham, L. J.},
  journal = {Phys. Rev. B},
  volume = {28},
  issue = {9},
  pages = {5100--5108},
  numpages = {0},
  year = {1983},
  month = {Nov},
  publisher = {American Physical Society},
  doi = {10.1103/PhysRevB.28.5100},
  url = {https://link.aps.org/doi/10.1103/PhysRevB.28.5100}
}

@article{takada1978plasmon,
  title={Plasmon mechanism of superconductivity in two-and three-dimensional electron systems},
  author={Takada, Yasutami},
  journal={Journal of the Physical Society of Japan},
  volume={45},
  number={3},
  pages={786--794},
  year={1978},
  publisher={The Physical Society of Japan}
}

@article{takada1993,
  title = {s- and p-wave pairings in the dilute electron gas: Superconductivity mediated by the Coulomb hole in the vicinity of the Wigner-crystal phase},
  author = {Takada, Yasutami},
  journal = {Phys. Rev. B},
  volume = {47},
  issue = {9},
  pages = {5202--5211},
  numpages = {0},
  year = {1993},
  month = {Mar},
  publisher = {American Physical Society},
  doi = {10.1103/PhysRevB.47.5202},
  url = {https://link.aps.org/doi/10.1103/PhysRevB.47.5202}
}

@article{dielectric1,
  title = {Microscopic {{Theory}} of {{Force Constants}} in the {{Adiabatic Approximation}}},
  author = {Pick, Robert M. and Cohen, Morrel H. and Martin, Richard M.},
  year = {1970},
  month = jan,
  journal = {Physical Review B},
  volume = {1},
  number = {2},
  pages = {910--920},
  issn = {0556-2805},
  doi = {10.1103/PhysRevB.1.910},
  urldate = {2023-11-16},
  langid = {english},
}

@article{dielectric2,
  title = {Self-Consistent-Screening Calculation of Interatomic Force Constants and Phonon Dispersion Curves from First Principles},
  author = {Quong, Andrew A. and Klein, Barry M.},
  year = {1992},
  month = nov,
  journal = {Physical Review B},
  volume = {46},
  number = {17},
  pages = {10734--10737},
  issn = {0163-1829, 1095-3795},
  doi = {10.1103/PhysRevB.46.10734},
  urldate = {2023-11-16},
  langid = {english},
}

@misc{polchinski_effective_1999,
  title = {Effective {{Field Theory}} and the {{Fermi Surface}}},
  author = {Polchinski, Joseph},
  year = {1999},
  month = jun,
  number = {arXiv:hep-th/9210046},
  eprint = {hep-th/9210046},
  publisher = {arXiv},
  urldate = {2023-07-20},
  archiveprefix = {arXiv},
  langid = {english},
}

@article{kohn1964,
  title = {Inhomogeneous {{Electron Gas}}},
  author = {Hohenberg, P. and Kohn, W.},
  year = {1964},
  month = nov,
  journal = {Physical Review},
  volume = {136},
  number = {3B},
  pages = {B864-B871},
  publisher = {American Physical Society},
  doi = {10.1103/PhysRev.136.B864},
  urldate = {2024-10-12},
}

@article{kohn1965,
  title = {Self-{{Consistent Equations Including Exchange}} and {{Correlation Effects}}},
  author = {Kohn, W. and Sham, L. J.},
  year = {1965},
  month = nov,
  journal = {Physical Review},
  volume = {140},
  number = {4A},
  pages = {A1133-A1138},
  publisher = {American Physical Society},
  doi = {10.1103/PhysRev.140.A1133},
  urldate = {2024-10-12},
}

@article{mcmillan1968,
  title = {Transition Temperature of Strong-Coupled Superconductors},
  author = {McMillan, W. L.},
  journal = {Phys. Rev.},
  volume = {167},
  issue = {2},
  pages = {331--344},
  numpages = {0},
  year = {1968},
  month = {Mar},
  publisher = {American Physical Society},
  doi = {10.1103/PhysRev.167.331},
  url = {https://link.aps.org/doi/10.1103/PhysRev.167.331}
}

@article{allen1975,
  title = {Transition temperature of strong-coupled superconductors reanalyzed},
  author = {Allen, P. B. and Dynes, R. C.},
  journal = {Phys. Rev. B},
  volume = {12},
  issue = {3},
  pages = {905--922},
  numpages = {0},
  year = {1975},
  month = {Aug},
  publisher = {American Physical Society},
  doi = {10.1103/PhysRevB.12.905},
  url = {https://link.aps.org/doi/10.1103/PhysRevB.12.905}
}

@book{bogoliubov59,
  title={A new method in the theory of superconductivity},
  author={Bogoliubov, Nikolay N and Tolmachov, Vladimir Veniaminovic and Shirkov, DV},
  year={1959},
  publisher={Consultants Bureau, New York}
}

@inproceedings{tolmachev1961logarithmic,
  title={Logarithmic criterion for superconductivity},
  author={Tolmachev, Vladimir Veniaminovich},
  booktitle={Doklady Akademii Nauk},
  volume={140},
  number={3},
  pages={563--566},
  year={1961},
  organization={Russian Academy of Sciences}
}

@article{scdft-eliashberg,
author = {Sanna ,Antonio and Flores-Livas ,Jos\'{e} A. and Davydov ,Arkadiy and Profeta ,Gianni and Dewhurst ,Kay and Sharma ,Sangeeta and Gross ,E. K. U.},
title = {Ab initio Eliashberg Theory: Making Genuine Predictions of Superconducting Features},
journal = {Journal of the Physical Society of Japan},
volume = {87},
number = {4},
pages = {041012},
year = {2018},
doi = {10.7566/JPSJ.87.041012},
URL = {https://doi.org/10.7566/JPSJ.87.041012},
eprint = {https://doi.org/10.7566/JPSJ.87.041012}
}

@article{SCDFT1,
  title = {Ab Initio Theory of Superconductivity. {{I}}. {{Density}} Functional Formalism and Approximate Functionals},
  author = {L{\"u}ders, M. and Marques, M. A. L. and Lathiotakis, N. N. and Floris, A. and Profeta, G. and Fast, L. and Continenza, A. and Massidda, S. and Gross, E. K. U.},
  year = {2005},
  month = jul,
  journal = {Physical Review B},
  volume = {72},
  number = {2},
  pages = {024545},
  publisher = {American Physical Society},
  doi = {10.1103/PhysRevB.72.024545},
  urldate = {2022-05-19},
}

@article{SCDFT2,
  title = {Ab Initio Theory of Superconductivity. {{II}}. {{Application}} to Elemental Metals},
  author = {Marques, M. A. L. and L{\"u}ders, M. and Lathiotakis, N. N. and Profeta, G. and Floris, A. and Fast, L. and Continenza, A. and Gross, E. K. U. and Massidda, S.},
  year = {2005},
  month = jul,
  journal = {Physical Review B},
  volume = {72},
  number = {2},
  pages = {024546},
  publisher = {American Physical Society},
  doi = {10.1103/PhysRevB.72.024546},
  urldate = {2022-05-19},
}

@article{vdiagmc1,
  title = {A Combined Variational and Diagrammatic Quantum {{Monte Carlo}} Approach to the Many-Electron Problem},
  author = {Chen, Kun and Haule, Kristjan},
  year = {2019},
  month = dec,
  journal = {Nature Communications},
  volume = {10},
  number = {1},
  pages = {3725},
  issn = {2041-1723},
  doi = {10.1038/s41467-019-11708-6},
  urldate = {2020-07-08},
  langid = {english},
}

@article{vdiagmc2,
  title = {Single-Particle Excitations in the Uniform Electron Gas by Diagrammatic {{Monte Carlo}}},
  author = {Haule, Kristjan and Chen, Kun},
  year = {2022},
  month = feb,
  journal = {Scientific Reports},
  volume = {12},
  number = {1},
  pages = {2294},
  publisher = {Nature Publishing Group},
  issn = {2045-2322},
  doi = {10.1038/s41598-022-06188-6},
  urldate = {2023-11-06},
  copyright = {2022 The Author(s)},
  langid = {english},
}

@article{diagmc2,
  title = {Fermi-Polaron Problem: {{Diagrammatic Monte Carlo}} Method for Divergent Sign-Alternating Series},
  shorttitle = {Fermi-Polaron Problem},
  author = {Prokof'ev, Nikolay and Svistunov, Boris},
  year = {2008},
  month = jan,
  journal = {Physical Review B},
  volume = {77},
  number = {2},
  pages = {020408},
  publisher = {American Physical Society},
  doi = {10.1103/PhysRevB.77.020408},
  urldate = {2023-12-04},
}

@article{diagmc1,
  title = {Polaron {{Problem}} by {{Diagrammatic Quantum Monte Carlo}}},
  author = {Prokof'ev, Nikolai V. and Svistunov, Boris V.},
  year = {1998},
  month = sep,
  journal = {Physical Review Letters},
  volume = {81},
  number = {12},
  pages = {2514--2517},
  publisher = {American Physical Society},
  doi = {10.1103/PhysRevLett.81.2514},
  urldate = {2023-12-04},
}

@article{diagmckozik,
  title = {Combinatorial Summation of {{Feynman}} Diagrams},
  author = {Kozik, Evgeny},
  year = {2024},
  month = sep,
  journal = {Nature Communications},
  volume = {15},
  number = {1},
  pages = {7916},
  publisher = {Nature Publishing Group},
  issn = {2041-1723},
  doi = {10.1038/s41467-024-52000-6},
  urldate = {2025-02-08},
  copyright = {2024 The Author(s)},
  langid = {english},
}

@article{detdiagmc,
  title = {Determinant {{Diagrammatic Monte Carlo Algorithm}} in the {{Thermodynamic Limit}}},
  author = {Rossi, Riccardo},
  year = {2017},
  month = jul,
  journal = {Physical Review Letters},
  volume = {119},
  number = {4},
  pages = {045701},
  publisher = {American Physical Society},
  doi = {10.1103/PhysRevLett.119.045701},
  urldate = {2020-10-30},
}

@article{diagmclatest,
  title = {Diagrammatic {{Monte Carlo}} Scheme for Dielectric Losses in Metals},
  author = {Tupitsyn, I. S. and Prokof'ev, N. V.},
  year = {2025},
  month = jan,
  journal = {Physical Review B},
  volume = {111},
  number = {4},
  pages = {L041106},
  publisher = {American Physical Society},
  doi = {10.1103/PhysRevB.111.L041106},
  urldate = {2025-02-08},
}

@article{diagmc2010,
  title = {Diagrammatic {{Monte Carlo}}},
  author = {Van Houcke, Kris and Kozik, Evgeny and Prokof'ev, N. and Svistunov, B.},
  year = {2010},
  month = jan,
  journal = {Physics Procedia},
  series = {Computer {{Simulations Studies}} in {{Condensed Matter Physics XXI}}},
  volume = {6},
  pages = {95--105},
  issn = {1875-3892},
  doi = {10.1016/j.phpro.2010.09.034},
  urldate = {2023-12-04},
}

@article{vegasenhanced,
  title = {Adaptive {{Multidimensional Integration}}: {{VEGAS Enhanced}}},
  shorttitle = {Adaptive {{Multidimensional Integration}}},
  author = {Lepage, G. Peter},
  year = {2021},
  month = aug,
  journal = {Journal of Computational Physics},
  volume = {439},
  eprint = {2009.05112},
  primaryclass = {hep-ph, physics:physics},
  pages = {110386},
  issn = {00219991},
  doi = {10.1016/j.jcp.2021.110386},
  urldate = {2022-07-16},
  archiveprefix = {arXiv},
}

@article{markus_2023,
  title = {Static Self-Energy and Effective Mass of the Homogeneous Electron Gas from Quantum Monte Carlo Calculations},
  author = {Holzmann, Markus and Calcavecchia, Francesco and Ceperley, David M. and Olevano, Valerio},
  journal = {Phys. Rev. Lett.},
  volume = {131},
  issue = {18},
  pages = {186501},
  numpages = {6},
  year = {2023},
  month = {Nov},
  publisher = {American Physical Society},
  doi = {10.1103/PhysRevLett.131.186501},
  url = {https://link.aps.org/doi/10.1103/PhysRevLett.131.186501}
}

@article{vegas,
  title = {A New Algorithm for Adaptive Multidimensional Integration},
  author = {Peter Lepage, G},
  year = {1978},
  month = may,
  journal = {Journal of Computational Physics},
  volume = {27},
  number = {2},
  pages = {192--203},
  issn = {0021-9991},
  doi = {10.1016/0021-9991(78)90004-9},
  urldate = {2024-12-25},
}

@incollection{schrieffer_1969,
  title = {Superconductivity in the Transition Metals: Theory and Experiment},
  booktitle = {Superconductivity},
  author = {{G. Gladstone and M. A. Jensen and J. R. Schrieffer}},
  editor = {R. D. Parks},
  year = {1969},
  publisher = {Marcel Dekker, Inc.},
  address = {New York},
  volume = {2},
  pages = {665--816},
}

@article{ginzburg_superconductivity_1968,
  title = {Superconductivity in {{White Dwarfs}} and {{Pulsars}}},
  author = {Ginzburg, V. L. and Kirzhniz, D. A.},
  year = {1968},
  month = oct,
  journal = {Nature},
  volume = {220},
  number = {5163},
  pages = {148--149},
  publisher = {Nature Publishing Group},
  issn = {1476-4687},
  doi = {10.1038/220148b0},
  urldate = {2024-12-25},
  copyright = {1968 Springer Nature Limited},
  langid = {english}
}

@article{trubnikov1968white,
  title={Are white dwarfs superconductors?},
  author={Trubnikov, BA},
  journal={Zh. Eksp. Teor. Fiz},
  volume={55},
  pages={1893--1902},
  year={1968}
}

@article{kostrzewa_anomalously_2018,
  title = {Anomalously High Value of {{Coulomb}} Pseudopotential for the {{H5S2}} Superconductor},
  author = {Kostrzewa, Ma{\l}gorzata and Szcz{\k e}{\'s}niak, Rados{\l}aw and Kalaga, Joanna K. and Wrona, Izabela A.},
  year = {2018},
  month = aug,
  journal = {Scientific Reports},
  volume = {8},
  number = {1},
  pages = {11957},
  publisher = {Nature Publishing Group},
  issn = {2045-2322},
  doi = {10.1038/s41598-018-30391-z},
  urldate = {2024-11-22},
  copyright = {2018 The Author(s)},
  langid = {english},
}

@article{RevModPhys.62.1027,
  title = {Properties of boson-exchange superconductors},
  author = {Carbotte, J. P.},
  journal = {Rev. Mod. Phys.},
  volume = {62},
  issue = {4},
  pages = {1027--1157},
  numpages = {0},
  year = {1990},
  month = {Oct},
  publisher = {American Physical Society},
  doi = {10.1103/RevModPhys.62.1027},
  url = {https://link.aps.org/doi/10.1103/RevModPhys.62.1027}
}

@article{PhysRevB.84.020508,
  title = {Vibrational spectrum and electron-phonon coupling of doped solid picene from first principles},
  author = {Subedi, Alaska and Boeri, Lilia},
  journal = {Phys. Rev. B},
  volume = {84},
  issue = {2},
  pages = {020508},
  numpages = {5},
  year = {2011},
  month = {Jul},
  publisher = {American Physical Society},
  doi = {10.1103/PhysRevB.84.020508},
  url = {https://link.aps.org/doi/10.1103/PhysRevB.84.020508}
}

@article{zfactor_2022,
  title = {Revisiting homogeneous electron gas in pursuit of properly normed ab initio Eliashberg theory},
  author = {Akashi, Ryosuke},
  journal = {Phys. Rev. B},
  volume = {105},
  issue = {10},
  pages = {104510},
  numpages = {11},
  year = {2022},
  month = {Mar},
  publisher = {American Physical Society},
  doi = {10.1103/PhysRevB.105.104510},
  url = {https://link.aps.org/doi/10.1103/PhysRevB.105.104510}
}

@article{katsnelson2023screening,
  title={Screening induced crossover between phonon-and plasmon-mediated pairing in layered superconductors},
  author={Katsnelson, Mikhail I and Millis, Andrew J and R{\"o}sner, Malte and others},
  journal={2D Materials},
  volume={10},
  number={4},
  pages={045031},
  year={2023},
  publisher={IOP Publishing}
}

@article{vmc2,
  title     = {Monte {{Carlo}} Simulation of a Many-Fermion Study},
  author    = {Ceperley, D. and Chester, G. V. and Kalos, M. H.},
  year      = {1977},
  month     = oct,
  journal   = {Physical Review B},
  volume    = {16},
  number    = {7},
  pages     = {3081--3099},
  publisher = {American Physical Society},
  doi       = {10.1103/PhysRevB.16.3081},
  urldate   = {2024-11-11},
}

@article{vmc1,
  title     = {Ground {{State}} of {{Liquid}} $\mathrm{He}^4$},
  author    = {McMillan, W. L.},
  year      = {1965},
  month     = apr,
  journal   = {Physical Review},
  volume    = {138},
  number    = {2A},
  pages     = {A442-A451},
  publisher = {American Physical Society},
  doi       = {10.1103/PhysRev.138.A442},
  urldate   = {2024-11-11},
}

@article{dmc1,
  title   = {Diffusion {{Quantum Monte Carlo}}},
  author  = {Reynolds, Peter J. and Tobochnik, Jan and Gould, Harvey},
  year    = {1990},
  month   = nov,
  journal = {Computer in Physics},
  volume  = {4},
  number  = {6},
  pages   = {662--668},
  issn    = {0894-1866},
  doi     = {10.1063/1.4822960},
  urldate = {2024-11-11},
}

@inproceedings{Taylor1,
title={Taylor-Mode Automatic Differentiation for Higher-Order Derivatives in {JAX}},
author={Jesse Bettencourt and Matthew J. Johnson and David Duvenaud},
booktitle={Program Transformations for ML Workshop at NeurIPS 2019},
year={2019},
url={https://openreview.net/forum?id=SkxEF3FNPH}
}

@book{Taylor2,
  title={Evaluating derivatives: principles and techniques of algorithmic differentiation},
  author = {Griewank, Andreas and Walther, Andrea},
  year = {2008},
  month = jan,
  doi = {10.1137/1.9780898717761},
  urldate = {2023-12-07},
  isbn = {978-0-89871-659-7},
  publisher={SIAM}
}

@phdthesis{Taylor3,
  title={Higher-Order Automatic Differentiation and Its Applications},
  author={Tan, Songchen},
  year={2023},
  school={Massachusetts Institute of Technology}
}

@article{normalization_flow,
  title = {Normalizing Flows for Microscopic Many-Body Calculations: An Application to the Nuclear Equation of State},
  author = {Brady, Jack and Wen, Pengsheng and Holt, Jeremy W.},
  journal = {Phys. Rev. Lett.},
  volume = {127},
  issue = {6},
  pages = {062701},
  numpages = {6},
  year = {2021},
  month = {Aug},
  publisher = {American Physical Society},
  doi = {10.1103/PhysRevLett.127.062701},
  url = {https://link.aps.org/doi/10.1103/PhysRevLett.127.062701}
}

@article{bipolaron1,
  title = {Breakdown of the Migdal-Eliashberg theory: A determinant quantum Monte Carlo study},
  author = {Esterlis, I. and Nosarzewski, B. and Huang, E. W. and Moritz, B. and Devereaux, T. P. and Scalapino, D. J. and Kivelson, S. A.},
  journal = {Phys. Rev. B},
  volume = {97},
  issue = {14},
  pages = {140501},
  numpages = {5},
  year = {2018},
  month = {Apr},
  publisher = {American Physical Society},
  doi = {10.1103/PhysRevB.97.140501},
  url = {https://link.aps.org/doi/10.1103/PhysRevB.97.140501}
}

@article{implicit,
  title = {Implicit renormalization approach to the problem of Cooper instability},
  author = {Chubukov, Andrey and Prokof'ev, Nikolay V. and Svistunov, Boris V.},
  journal = {Phys. Rev. B},
  volume = {100},
  issue = {6},
  pages = {064513},
  numpages = {11},
  year = {2019},
  month = {Aug},
  publisher = {American Physical Society},
  doi = {10.1103/PhysRevB.100.064513},
  url = {https://link.aps.org/doi/10.1103/PhysRevB.100.064513}
}

@article{bipolaron2,
  title={Eliashberg theory of phonon-mediated superconductivity—When it is valid and how it breaks down},
  author={Chubukov, Andrey V and Abanov, Artem and Esterlis, Ilya and Kivelson, Steven A},
  journal={Annals of Physics},
  volume={417},
  pages={168190},
  year={2020},
  publisher={Elsevier}
}

@article{bipolaron3,
  title = {Bipolaronic High-Temperature Superconductivity},
  author = {Zhang, C. and Sous, J. and Reichman, D. R. and Berciu, M. and Millis, A. J. and Prokof'ev, N. V. and Svistunov, B. V.},
  journal = {Phys. Rev. X},
  volume = {13},
  issue = {1},
  pages = {011010},
  numpages = {19},
  year = {2023},
  month = {Jan},
  publisher = {American Physical Society},
  doi = {10.1103/PhysRevX.13.011010},
  url = {https://link.aps.org/doi/10.1103/PhysRevX.13.011010}
}

@article{gubser_high-pressure_1975,
  title = {High-{{Pressure Effects}} on the {{Superconducting Transition Temperature}} of {{Aluminum}}},
  author = {Gubser, D. U. and Webb, A. W.},
  year = {1975},
  month = jul,
  journal = {Physical Review Letters},
  volume = {35},
  number = {2},
  pages = {104--107},
  issn = {0031-9007},
  doi = {10.1103/PhysRevLett.35.104},
  urldate = {2024-05-28},
  copyright = {http://link.aps.org/licenses/aps-default-license},
  langid = {english},
}

@article{levy_can_1964,
  title = {Can Pressure Destroy Superconductivity in Aluminum?},
  author = {Levy, M. and Olsen, J. L.},
  year = {1964},
  month = may,
  journal = {Solid State Communications},
  volume = {2},
  number = {5},
  pages = {137--139},
  issn = {0038-1098},
  doi = {10.1016/0038-1098(64)90401-6},
  urldate = {2024-06-20},
}

@article{profeta_superconductivity_2006,
  title = {Superconductivity in {{Lithium}}, {{Potassium}}, and {{Aluminum}} under {{Extreme Pressure}}: {{A First-Principles Study}}},
  shorttitle = {Superconductivity in {{Lithium}}, {{Potassium}}, and {{Aluminum}} under {{Extreme Pressure}}},
  author = {Profeta, G. and Franchini, C. and Lathiotakis, N. N. and Floris, A. and Sanna, A. and Marques, M. A. L. and L{\"u}ders, M. and Massidda, S. and Gross, E. K. U. and Continenza, A.},
  year = {2006},
  month = feb,
  journal = {Physical Review Letters},
  volume = {96},
  number = {4},
  pages = {047003},
  issn = {0031-9007, 1079-7114},
  doi = {10.1103/PhysRevLett.96.047003},
  urldate = {2024-05-20},
  copyright = {http://link.aps.org/licenses/aps-default-license},
  langid = {english},
}

@article{dftdmft1,
  title = {First-Principles Calculations of the Electronic Structure and Spectra of Strongly Correlated Systems: Dynamical Mean-Field Theory},
  shorttitle = {First-Principles Calculations of the Electronic Structure and Spectra of Strongly Correlated Systems},
  author = {Anisimov, V. I. and Poteryaev, A. I. and Korotin, M. A. and Anokhin, A. O. and Kotliar, G.},
  year = {1997},
  month = sep,
  journal = {Journal of Physics: Condensed Matter},
  volume = {9},
  number = {35},
  pages = {7359},
  issn = {0953-8984},
  doi = {10.1088/0953-8984/9/35/010},
  urldate = {2025-02-10},
  langid = {english},
}

@article{dftdmft4,
  title = {Electrodynamics of Correlated Electron Materials},
  author = {Basov, D. N. and Averitt, Richard D. and {van der Marel}, Dirk and Dressel, Martin and Haule, Kristjan},
  year = {2011},
  month = jun,
  journal = {Reviews of Modern Physics},
  volume = {83},
  number = {2},
  pages = {471--541},
  publisher = {American Physical Society},
  doi = {10.1103/RevModPhys.83.471},
  urldate = {2025-02-10},
}

@article{dftdmft3,
  title = {Dynamical Mean-Field Theory within the Full-Potential Methods: {{Electronic}} Structure of \$\{{\textbackslash}text\{\vphantom{\}\}}{{CeIrIn}}\vphantom\{\}\vphantom\{\}\_\{5\}\$, \$\{{\textbackslash}text\{\vphantom{\}\}}{{CeCoIn}}\vphantom\{\}\vphantom\{\}\_\{5\}\$, and \$\{{\textbackslash}text\{\vphantom{\}\}}{{CeRhIn}}\vphantom\{\}\vphantom\{\}\_\{5\}\$},
  shorttitle = {Dynamical Mean-Field Theory within the Full-Potential Methods},
  author = {Haule, Kristjan and Yee, Chuck-Hou and Kim, Kyoo},
  year = {2010},
  month = may,
  journal = {Physical Review B},
  volume = {81},
  number = {19},
  pages = {195107},
  publisher = {American Physical Society},
  doi = {10.1103/PhysRevB.81.195107},
  urldate = {2025-02-10},
}

@article{dmft_sc2,
  title = {Superconductivity and Antiferromagnetism in \$\{{\textbackslash}mathrm\{\vphantom{\}\}}{{NdNiO}}\vphantom\{\}\vphantom\{\}\_\{2\}\$ and \$\{{\textbackslash}mathrm\{\vphantom{\}\}}{{CaCuO}}\vphantom\{\}\vphantom\{\}\_\{2\}\$: {{A}} Cluster {{DMFT}} Study},
  shorttitle = {Superconductivity and Antiferromagnetism in \$\{{\textbackslash}mathrm\{\vphantom{\}\}}{{NdNiO}}\vphantom\{\}\vphantom\{\}\_\{2\}\$ and \$\{{\textbackslash}mathrm\{\vphantom{\}\}}{{CaCuO}}\vphantom\{\}\vphantom\{\}\_\{2\}\$},
  author = {Karp, Jonathan and Hampel, Alexander and Millis, Andrew J.},
  year = {2022},
  month = may,
  journal = {Physical Review B},
  volume = {105},
  number = {20},
  pages = {205131},
  publisher = {American Physical Society},
  doi = {10.1103/PhysRevB.105.205131},
  urldate = {2025-02-10},
}

@article{dmft_sc1,
  title = {Efficient Lattice Dynamics Calculations for Correlated Materials with \${\textbackslash}mathrm\{\vphantom\}{{DFT}}\vphantom\{\}+{\textbackslash}mathrm\{\vphantom\}{{DMFT}}\vphantom\{\}\$},
  author = {Ko{\c c}er, Can P. and Haule, Kristjan and Pascut, G. Lucian and Monserrat, Bartomeu},
  year = {2020},
  month = dec,
  journal = {Physical Review B},
  volume = {102},
  number = {24},
  pages = {245104},
  publisher = {American Physical Society},
  doi = {10.1103/PhysRevB.102.245104},
  urldate = {2025-02-10},
}

@article{dftdmft2,
  title = {Electronic Structure Calculations with Dynamical Mean-Field Theory},
  author = {Kotliar, G. and Savrasov, S. Y. and Haule, K. and Oudovenko, V. S. and Parcollet, O. and Marianetti, C. A.},
  year = {2006},
  month = aug,
  journal = {Reviews of Modern Physics},
  volume = {78},
  number = {3},
  pages = {865--951},
  publisher = {American Physical Society},
  doi = {10.1103/RevModPhys.78.865},
  urldate = {2025-02-10},
}

@article{dftdmft5,
  title = {Computing Total Energies in Complex Materials Using Charge Self-Consistent {{DFT}} + {{DMFT}}},
  author = {Park, Hyowon and Millis, Andrew J. and Marianetti, Chris A.},
  year = {2014},
  month = dec,
  journal = {Physical Review B},
  volume = {90},
  number = {23},
  pages = {235103},
  publisher = {American Physical Society},
  doi = {10.1103/PhysRevB.90.235103},
  urldate = {2025-02-10},
}

@article{dftdmftreview,
  title = {Applications of {{DFT}} + {{DMFT}} in {{Materials Science}}},
  author = {Paul, Arpita and Birol, Turan},
  year = {2019},
  month = jul,
  journal = {Annual Review of Materials Research},
  volume = {49},
  number = {Volume 49, 2019},
  pages = {31--52},
  publisher = {Annual Reviews},
  issn = {1531-7331, 1545-4118},
  doi = {10.1146/annurev-matsci-070218-121825},
  urldate = {2025-02-10},
  langid = {english},
}

@article{homotopic_expansions,
  title = {Homotopic Action: A Pathway to Convergent Diagrammatic Theories},
  author = {Kim, Aaram J. and Prokof'ev, Nikolay V. and Svistunov, Boris V. and Kozik, Evgeny},
  year = {2021},
  month = June,
  journal = {Phys. Rev. Lett.},
  volume = {126},
  pages = {257001},
  publisher = {Americam Physical Society},
  doi = {10.1103/PhysRevLett.126.25700},
  url = {https://journals.aps.org/prl/abstract/10.1103/PhysRevLett.126.257001},
 
}

@article{frohlich1950,
  title = {Theory of the {{Superconducting State}}. {{I}}. {{The Ground State}} at the {{Absolute Zero}} of {{Temperature}}},
  author = {Fr{\"o}hlich, H.},
  year = 1950,
  journal = {Physical Review},
  volume = {79},
  number = {5},
  pages = {845--856},
  doi = {10.1103/PhysRev.79.845},
}

@article{maxwell1950,
  title = {Isotope {{Effect}} in the {{Superconductivity}} of {{Mercury}}},
  author = {Maxwell, Emanuel},
  year = 1950,
  month = may,
  journal = {Physical Review},
  volume = {78},
  number = {4},
  pages = {477--477},
  publisher = {American Physical Society},
  doi = {10.1103/PhysRev.78.477},
  urldate = {2025-11-26},
}

@article{reynolds1950,
  title = {Superconductivity of {{Isotopes}} of {{Mercury}}},
  author = {Reynolds, C. A. and Serin, B. and Wright, W. H. and Nesbitt, L. B.},
  year = 1950,
  month = may,
  journal = {Physical Review},
  volume = {78},
  number = {4},
  pages = {487--487},
  publisher = {American Physical Society},
  doi = {10.1103/PhysRev.78.487},
  urldate = {2025-11-26},
}

@misc{dassarma2025,
  title = {Conventional and Practical Metallic Superconductivity Arising from Repulsive {{Coulomb}} Coupling},
  author = {Sarma, Sankar Das and Sau, Jay D. and Tu, Yi-Ting},
  year = 2025,
  month = nov,
  number = {arXiv:2511.00625},
  eprint = {2511.00625},
  primaryclass = {cond-mat},
  publisher = {arXiv},
  doi = {10.48550/arXiv.2511.00625},
  urldate = {2025-11-26},
  archiveprefix = {arXiv},
}

@article{veld23,
  title = {Screening Induced Crossover between Phonon- and Plasmon-Mediated Pairing in Layered Superconductors},
  author = {{in't Veld}, Y and Katsnelson, M I and Millis, A J and R{\"o}sner, M},
  year = 2023,
  month = sep,
  journal = {2D Materials},
  volume = {10},
  number = {4},
  pages = {045031},
  publisher = {IOP Publishing},
  issn = {2053-1583},
  doi = {10.1088/2053-1583/acf944},
  urldate = {2025-11-26},
  langid = {english},
}

@article{Shankar94,
  title = {Renormalization-group approach to interacting fermions},
  author = {Shankar, R.},
  journal = {Rev. Mod. Phys.},
  volume = {66},
  issue = {1},
  pages = {129--192},
  numpages = {0},
  year = {1994},
  month = {Jan},
  publisher = {American Physical Society},
  doi = {10.1103/RevModPhys.66.129},
  url = {https://link.aps.org/doi/10.1103/RevModPhys.66.129}
}

@article{Simonato23,
  title = {Revised Tolmachev-Morel-Anderson pseudopotential for layered conventional superconductors with nonlocal Coulomb interaction},
  author = {Simonato, M. and Katsnelson, M. I. and R\"osner, M.},
  journal = {Phys. Rev. B},
  volume = {108},
  issue = {6},
  pages = {064513},
  numpages = {10},
  year = {2023},
  month = {Aug},
  publisher = {American Physical Society},
  doi = {10.1103/PhysRevB.108.064513},
  url = {https://link.aps.org/doi/10.1103/PhysRevB.108.064513}
}

@article{Al_EOS,
  title = {Aluminum under high pressure. I. Equation of state},
  author = {Friedli, Carlos and Ashcroft, N. W.},
  journal = {Phys. Rev. B},
  volume = {12},
  issue = {12},
  pages = {5552--5559},
  numpages = {0},
  year = {1975},
  month = {Dec},
  publisher = {American Physical Society},
  doi = {10.1103/PhysRevB.12.5552},
  url = {https://link.aps.org/doi/10.1103/PhysRevB.12.5552}
}

@article{hauleSingleparticleExcitationsUniform2022,
  title = {Single-Particle Excitations in the Uniform Electron Gas by Diagrammatic {{Monte Carlo}}},
  author = {Haule, Kristjan and Chen, Kun},
  year = 2022,
  month = feb,
  journal = {Scientific Reports},
  volume = {12},
  number = {1},
  pages = {2294},
  publisher = {Nature Publishing Group},
  issn = {2045-2322},
  doi = {10.1038/s41598-022-06188-6},
  urldate = {2023-11-06},
  copyright = {2022 The Author(s)},
  langid = {english},
}

@article{holzmannStaticSelfEnergyEffective2023a,
  title = {Static {{Self-Energy}} and {{Effective Mass}} of the {{Homogeneous Electron Gas}} from {{Quantum Monte Carlo Calculations}}},
  author = {Holzmann, Markus and Calcavecchia, Francesco and Ceperley, David M. and Olevano, Valerio},
  year = 2023,
  month = nov,
  journal = {Physical Review Letters},
  volume = {131},
  number = {18},
  pages = {186501},
  publisher = {American Physical Society},
  doi = {10.1103/PhysRevLett.131.186501},
  urldate = {2025-12-02},
}

\appendix
\begin{appendices}

\section{Field theoretical approach to the electron-phonon problem}
\label{sec:app_epi}
In this section, we employ the principles of effective field theory (EFT) to investigate the electron-phonon problem. Starting from a generic electron-ion model, we integrate out the ion vibration degrees of freedom to derive an EFT describing the coupled dynamics of electrons and phonons. Our derivation is distinguished by three key features that, to our knowledge, have not been collectively addressed in the existing literature on the electron-phonon problem.

First, we do not approximate the electron-electron interactions using an exchange-correlation potential, as is commonly done in density functional theory (DFT) calculations\cite{giustino_2017}. This allows our derivation to remain applicable to strongly coupled materials. The only assumption we make is that the ions form a rigid lattice structure, implying that the vibration amplitude is small compared to the lattice spacing.

Second, our derivation is applicable to generic electron-ion models with realistic lattice setups, making it suitable for ab-initio calculations of real materials. This generality ensures that our EFT can be used to study a wide range of systems, from simple metals to complex multi-component materials.

Third, we express the electron-phonon action in terms of the physical phonon dispersion using a renormalization scheme. This approach circumvents the need to introduce unphysical and potentially ill-defined bare phonon degrees of freedom, which can lead to complications in the interpretation of the results.

The EFT framework we develop for the electron-phonon problem provides a solid foundation for systematically investigating the dynamics of electrons below the Fermi scale using field-theoretical techniques. This approach is particularly well-suited for studying the pairing mechanism of superconductivity, as it allows for a consistent treatment of the electron-electron and electron-phonon interactions, while also taking into account the realistic lattice structure of the material.

\subsection*{Conventions}
To simplify notations, in appendices we adopted the following conventions.
We used the integral and summation notions interchangeably. Unless otherwise claimed, the integral of a discrete variable were summation over the variable without any prefactors. 
Special cases were listed below.
We interchangeably wrote integral and summation of momentums of electrons and phonons in the Brillouin zone, 
so that $\int_{\bq}=\int_{\text{BZ}}\frac{\text{d}\bq}{\Omega_\text{BZ}}=\frac{1}{N}\sum_{\bq}$, 
where $\Omega_\text{BZ}$ is the volume of Brillouin zone.
For G-vector $\bG_m$ and position in a unit cell $\Delta\br$, we have $\int_{\Delta\br}=\int_{\Omega_\text{Cell}}{\text{d}(\Delta\br)}$, where $\Omega_\text{Cell}$ represents the volume of a unit cell, and $\int_{\bG_m}=\frac{1}{{\Omega_\text{Cell}}}\sum_{m}$.
For imaginary time $\tau$ and Matsubara frequency $\omega_n$, 
we have $\int_\tau=\int_{0}^{\beta}\text{d}\tau$ and $\int_{\omega_n}=T\sum_n$, 
where $T=\frac{1}{\beta}$ is the temperature.

\subsection{Electron-ion Model}

In this subsection, we present a concise review of the electron-ion problem from first principles, following a derivation adapted from Ref.~\cite{stefanucci_2023}. We consider a solid-state system composed of electrons and ions, where the ions form a rigid lattice with equilibrium positions $\bR^0=( \bR^0_1, \ldots, \bR^0_{N_n} )$. Due to quantum and thermal fluctuations, the ions oscillate around their equilibrium positions with small amplitudes $\bu=( \bu_1, \ldots, \bu_{N_n} )$. At a given time, the ions are located at coordinates $\bR \equiv \bR^0 + \bu$. The electrons and ions are governed by the full Hamiltonian:
\begin{equation}
  \hat{H} = \hat{H}_\text{e}+\hat{H}_\text{n}+\hat{H}_\text{en}.
\end{equation}
where $\hat{H}_\text{e}$, $\hat{H}_\text{n}$, and $\hat{H}_\text{en}$ represent the electron, ion, and electron-ion interaction terms, respectively.  

The ion Hamiltonian, $\hat{H}_\text{n}$, consists of the kinetic energy of the ions and the potential energy due to the Coulomb interaction between the ions:
\begin{equation}
  \hat{H}_\txn = \sum_{i=1}^{N_n}\frac{\hat{\bP}_i^2}{2M}+E_{\txnn}( \bu ),
\end{equation}
where $\hat{\bP}_i$ are the ion momentum operators, $M_i$ are the ion masses, and $E_{nn}$ is the ion potential energy, 
\begin{equation}
  E_{\txnn}( \bu ) \equiv \frac{1}{2}\sum_{i\neq j}^{N_n}Z_iZ_jv(\bR_i-\bR_j),
\end{equation} 
with $Z_i$ being the atomic number of the $i$-th ion, and $v(\bR_i-\bR_j) \equiv e^2/|\bR_i-\bR_j|$ representing the Coulomb repulsion between ions.

The electron part of the Hamiltonian, $\hat{H}_e$, is given by
\begin{align*}
  \hat{H}_\txe = &\int_{\br\sigma} \hat{\psi}^{\dagger}_{\br\sigma}\left[ -\frac{\nabla^2}{2m} +V^0_{\br} \right]\hat{\psi}_{\br\sigma}  \\
            &+\frac{1}{2}\int_{\substack{\br\br'\\ \sigma\sigma'}} \hat{\psi}^{\dagger}_{\br\sigma}\hat{\psi}^{\dagger}_{\br'\sigma'}v(\br-\br')\hat{\psi}_{\br'\sigma'}\hat{\psi}_{\br\sigma},
\end{align*}
where $\hat{\psi}_{\br\sigma}$ is the field operator for electrons at the position $\br$ with the spin index $\sigma$, $m$ is the electron mass, and $v(\br-\br')$ is the electron-electron Coulomb repulsion. The electrons moving in an inhomogeneous background potential $V^0_{\br}$, which is a static potential generated by the ions
\begin{equation}
  V^{\bu}_{\br} \equiv -\sum_{j=1}^{N_n}Z_j\cdot v(\bR_j-\br),
\end{equation}
at their equilibrium position $\bu = 0$.

The vibration of the ions around the equilibrium position induces an electron-ion interaction term, $\hat{H}_{e-n}$, 
\begin{equation}
  \hat{H}_{\txen} = \int_\br \hat{n}_\br \left(V^{\bu}_{\br}-V^0_{\br}\right).
\end{equation}
where $\hat{n}_\br=\sum_{\sigma}\hat{\psi}_{\br\sigma}^\dagger\hat{\psi}_{\br\sigma}$ is the electron density operator..

To study the finite temperature superconductivity, we consider a solid at a given temperature $T$. The central quantity of interest is the partition function,
\begin{equation}
    Z=\mathrm{Tr} e^{-\beta \hat{H}},
\end{equation}
where $\beta=1/k_BT$, $k_B$ is the Boltzmann constant, and $\hat{H}$ is the Hamiltonian of the system. We work in the canonical ensemble with fixed ion and electron numbers. 

To derive the action formulation of the electron-ion model, we represent the partition function with a path integral over the ionic and electronic degrees of freedom,
\begin{equation}
\label{eq:electron-ion-Z}
    Z = \int_{\bu} e^{-S_{\txn}[\bu]}  \int_{\bar{\psi}, \psi}  e^{-S_{\txe}[\bar{\psi}, \psi]}e^{- S_{\txen}[\bu,\bar{\psi}, \psi]},
\end{equation}
where $S_{\txn}$, $S_{\txe}$, and $S_{\txen}$ are the ion, electron, and electron-ion actions, respectively. The path integrals are performed over the ionic positions $\bu_\tau$ and the Grassmann fields $\psi_{\br\tau}^{\sigma}$ and $\bar{\psi}_{\br\tau}^{\sigma}$, which represent the electrons. The imaginary time $\tau$ runs from $0$ to $\beta$, and the fields satisfy periodic (anti-periodic) boundary conditions for bosons (fermions).

The electron Lagrangian, $S_\txe$, describes the dynamics of the electrons and is given by 
\begin{align}
\label{eq:electron_model}
    S_{\txe} =& \int_{\br\sigma\tau} \bar{\psi}_{\br\tau}^{\sigma}\left[\frac{\partial}{\partial \tau} -\frac{\nabla^2}{2m} +V^0_\br \right]\psi_{\br\tau}^{\sigma}\nonumber\\
            &+\frac{1}{2}\int_{\br\sigma,\br'\sigma',\tau}\bar{\psi}_{\br\tau}^{\sigma}\bar{\psi}_{\br'\tau}^{\sigma'} v(\br-\br')\psi_{\br'\tau}^{\sigma'} \psi_{\br\tau}^{\sigma}.   
\end{align}

The ion action, $S_\txn$, describes the dynamics of the ions and is given by
\begin{equation}
    S_{\txn} = \int_\tau \left[\sum_{i=1}^{N_n}\frac{M_i}{2}\left(\frac{\partial \bu_{i\tau}}{\partial\tau}\right)^2 + E_{\txnn}\left(\bu_\tau\right)\right],
\end{equation}
where the imaginary-time integration $\int_\tau \equiv \int_0^\beta d\tau$ with $\beta$ the inverse temperature.

The electron-ion Lagrangian, $S_{\txen}$, describes the interaction between the electrons and ions and is given by
\begin{equation}
\label{eq:en_action}
    S_{\txen} = \int_{\br\sigma\tau} \bar{\psi}_{\br\tau}^{\sigma}\psi_{\br\tau}^{\sigma} \left(V^{\bu_\tau}_{\br} -V^0_{ \br}\right),
\end{equation}
where $V^{\bu_\tau}_{\br}$ is the ionic potential with the ions at positions $\bu_\tau$, and $V^0_{ \br}$ is the background ionic potential with the ions at their equilibrium positions.

By expressing the partition function as a path integral over the ionic and electronic degrees of freedom, we can systematically investigate the electron-phonon problem with the standard field-theoretical techniques, as explained in the next subsection.

\subsection{Electron-Phonon Action}

In this subsection, we reinterpret the ion vibration degrees of freedom as propagating phonon modes and derive an effective field theory (EFT) for the electron-phonon problem based on the electron-ion action. Our approach takes advantage of the small vibration amplitude of the ions compared to the lattice constant, which allows us to systematically expand the electron-ion action in powers of the dimensionless parameter $u/a \sim \left(m/M\right)^{1/4}$, where $m$ is the electron mass, $M$ is the mass of the lightest ion species and $a$ is the characteristic lattice spacing.

We begin by expanding the ion potential energy, $E_{\txnn}(\bu_\tau)$, up to second order in the ion displacement $\bu$:
\begin{equation}
E_{\txnn}(\bu_\tau) = E_{\txnn}(0) +  \frac{1}{2}\sum_{ij}\bu_{i\tau} \cdot  \frac{\partial^2 E_{\txnn}(\bu\rightarrow 0)}{\partial{\bu_i}\partial{\bu_j}}   \cdot \bu_{j\tau}+O(u^4),
\end{equation}
where the odd terms vanish due to symmetry. Similarly, we expand the ion potential in the electron-ion coupling, $V^{\bu_\tau}_{\br}$, up to second order in $\bu$:
\begin{equation}
    V^{\bu_\tau}_{\br} =V^0_{ \br} +\sum_{i} \bg^0_{i}(\br) \cdot \bu_{i\tau} +   \frac{1}{2}\sum_{ij}\bu_{i\tau} \cdot \frac{\partial^2 V^{\bu\rightarrow 0}_{\br}}{\partial{\bu_i}\partial{\bu_j}} \cdot \bu_{j\tau}+O(u^3)
\end{equation}
where $\bg^0_{i} \equiv (g^0_{ix}, g^0_{iy}, g^0_{iz})$ is the bare electron-phonon coupling, defined as:
 \begin{equation}
   g^0_{i\alpha}\left( \br \right)\equiv  \frac{\partial V^{\bu\rightarrow 0}_\br}{\partial u_{i\alpha}}= Z_i\frac{\partial}{\partial r_\alpha}v( \br-\bR_i^0 ).
   \label{eq:epi_bare}
\end{equation}

Note that we retain the linear term in the expansion of $V^{\bu_\tau}_{\br}$ because the electron density in Eq.\eqref{eq:en_action} is odd in $\bu$. There is no need to go beyond $O(u^2)$ because, when the ion vibration degrees of freedom are integrated out, the electron-ion coupling always contributes in pairs.

Applying the above approximations to the electron-ion action, then collecting terms of the same order in $\bu$, we derive the electron-phonon action:
\begin{equation}
\label{eq:ep0_action}
    S = S_\txe + S^0_{\txph} +S_{\txeph} + O\left(\frac{m}{M}\right)
\end{equation}

where the $O(u)$ term is the electron-phonon coupling:
\begin{equation}
\label{eq:eph_action}
    S_{\txeph} = \sum_{i\alpha} \int_{\br\sigma\tau} g^0_{i\alpha}(\br) \bar{\psi}_{\br\tau}^{\sigma}\psi_{\br\tau}^{\sigma} u_{i\alpha\tau}
\end{equation}

and the $O(u^2)$ is identified as the bare phonon action,
\begin{equation}
    S_{\txph}^0 = \int_\tau \left[\sum_{i\alpha}\frac{M_i}{2}\left(\frac{\partial u_{i\alpha\tau}}{\partial\tau}\right)^2 - \frac{1}{2}\sum_{ij\alpha\beta} u_{i\alpha\tau} K_{i\alpha;j\beta} u_{j\beta\tau} \right],
\end{equation}

Here, $K_{i\alpha;j\beta}$ is the bare elastic tensor, given by: 
\begin{equation}
  K_{i\alpha;j\beta} \equiv \frac{\partial^2 \left[E_{\txnn}\left( \bu \rightarrow 0\right) +\int_\br n^0_\br V^{\bu\rightarrow 0}_{\br}\right]}{\partial u_{i\alpha}\partial u_{j\beta}},
\end{equation}
where the Greek letters $\alpha$ and $\beta$ labels the dimension of vector $\bu$. 

It is important to note that bare elastic tensor $K$, which appears in the original formulation of the phonon action, is not directly measurable and may not even be a well-defined quantity in real materials. This is because the bare elastic tensor does not take into account the renormalization effects arising from the electron-phonon interaction and the electron correlations in the system. To avoid working with the unphysical bare parameters, it makes sense to re-express $K$ with an effective elastic tensor $K^{\txeff}$, which is a physically measurable quantity directly related to the phonon dispersion $\omega_{\kappa\bq}$ and can be probed by inelastic neutron scattering experiments. It is a linear response function with respect to the deformation of the ion positions:
\begin{align}
    K^{\txeff}_{i\alpha; j\beta} &\equiv \left< u_{i\alpha} u_{j\beta}\right> \\
    &= K_{i\alpha; j\beta}+\int_{\br_1\br_2} g^0_{i\alpha}\left( \br_1 \right)
  \chi^\txe_{\br_1\br_2} g^0_{j\beta}\left( \br_2 \right),
\end{align}
where the first term is the contribution from the direct ion-ion Coulomb interaction, and the second term is the contribution mediated by the many electron system, with $\chi^\txe$ the static electron density-density correlation function,
\begin{equation}
    \chi^\txe_{\br_1\br_2} =\langle \hat{n}_{\br_1}\hat{n}_{\br_2}\rangle_{\txe}-\langle \hat{n}_{\br_1}\rangle_\txe \langle\hat{n}_{\br_2}\rangle_\txe,
\end{equation}
The averages are taken with respect to the many-electron action in Eq.\eqref{eq:electron_model}, assuming all ions are fixed at the equilibrium positions. 

Re-expressing the bare quantity with the renormalized quantity using $K = K_{\txeff}+g^0\cdot \chi^\txe\cdot g^0$, we replace the bare phonon action with an effective phonon action $S_{\txph}$ and a counterterm $S_{\txCT}$.

The effective phonon action $S_{\txph}$, with their dispersion directly related to measurable quantities such as the neutron scattering amplitude, is given by:
\begin{equation}
    S_{\txph} = \int_\tau \left[\sum_{i\alpha}\frac{M_i}{2}\left(\frac{\partial u_{i\alpha\tau}}{\partial\tau}\right)^2 - \frac{1}{2}\sum_{ij\alpha\beta} u_{i\alpha\tau} K^{\txeff}_{i\alpha;j\beta} u_{j\beta\tau} \right]
\end{equation}

The phonon action can be further simplified by diagonalizing the effective elastic tensor $K^{\txeff}$ using proper eigenvectors $e^\kappa_{s,\alpha}(\bq)$, where $i=(\bn,s)$ with $\bn$ being a lattice vector, $s$ labels different ions in a unit cell, and $\kappa$ labels the branches of vibration modes, and $\bq$ is the reciprocal lattice momentum. 
The $i$-th atom position is then $\bR_i = \bn + \delta\br_s$, with $\delta\br_s$ sublattice position of the atom.
By inserting the following transformation:
\begin{equation}
  {u}_{\bn s,\alpha}=\frac{1}{{M_s}}\int_{\bq}e^{i\bq\cdot\bn}\sum_{\kappa}e_{s,\alpha}^\kappa\left( \bq \right){u}_{\kappa\bq},
\end{equation}
into the phonon action, we obtain:
\begin{align}
\label{appeq:phonon_action}
    S_{\txph}[\bu] &= \frac{1}{2}\sum_{\kappa}\int_{\bq\tau}\left[\left(\frac{\partial u_{\kappa\bq\tau}}{\partial\tau}\right)^2 -  \omega_{\kappa\bq}^2 u_{\kappa\bq\tau}^2\right] \\
   & = \frac{1}{2}\sum_{\kappa}\int_{\bq\nu} D^{-1}_\kappa(\bq, \nu) \left|u_{\kappa\bq\nu}\right|^2
\end{align}
where eigenvalues of effective elastic tensor $\omega_{\kappa
\bq}$ the physical dispersion of the phonon mode.
Function $D_\kappa(\bq, \nu)$ is the Fourier transform of the phonon propagator $D_\kappa(\bq, \tau-\tau')\equiv -\langle TU_{\kappa\bq\tau}U_{\kappa\bq\tau'}\rangle$ into the Matsubara-frequency representation: 
\begin{equation}
D_\kappa(\bq, \nu)=-\frac{1}{\nu^2+\omega_{\kappa\bq}^2}
\end{equation}

In the effective phonon action, the renormalized phonon dispersion has resumed the electron correlations. To avoid double-counting, a counterterm is required to compensate the resummation:
\begin{equation}
\label{appeq:ct_action}
    S_{\txCT} = -\frac{1}{2}\sum_{\kappa}\int_{\bq\tau}  g^0_{\kappa\bq}(\br_1)\chi_{\br_1\br_2}^\txe g^0_{\kappa\bq}(\br_2 )u_{\kappa\bq\tau}^2,
\end{equation}
where the electron-phonon coupling in the reciprocal lattice momentum representation is given by:
\begin{equation}
    g^0_{\kappa\bq}(\br)=\sum_{\bn s\alpha}e^{i\bq\cdot \bn}e_{s,\alpha}^\kappa\left( \bq \right)g^0_{\bn s\alpha}(\br).
    \label{eq:epi_bare_fourier}
\end{equation}

Substituting the bare phonon action $S^0_{eph}$ in Eq.\eqref{eq:ep0_action} with the renormalized action $S_{ph} + S_{CT}$, we derive the effective field theory for the electron-phonon problem by integrating out the ion vibration degrees of freedom:
\begin{equation}
    S=S_\txe+S_{\txph}+S_{\txeph}+S_{\txCT}+O\left(\frac{m}{M}\right)
\end{equation}

In summary, by expanding the electron-ion action in powers of the ion displacement and expressing the bare quantities in terms of physical observables, we have derived an effective field theory for the electron-phonon problem. This EFT captures the essential physics of the coupled electron-phonon system, with the phonon dispersion directly related to measurable quantities. The inclusion of counterterms ensures that the double-counting of electron correlation effects is systematically removed, providing a well-defined framework for studying the properties of electron-phonon systems in real materials.

\subsection{Electron-phonon interaction}
\label{app:epi}

Now we return to the electron part of the action with the effective phonon action defined.
The interaction term, then, becomes
\begin{equation}
  S_{\txeph} = \sum_{\kappa\sigma} \int_{\br\bq\tau} \bar{\psi}_{\br\tau}^{\sigma}\psi_{\br\tau}^{\sigma}g^0_{\kappa\bq}(\br)u_{\kappa\bq\tau}.
\end{equation}
The Eq.\eqref{eq:epi_bare} shows that the electron-phonon interaction $g_{\kappa\bq}(\br)$ has the following property:
\begin{equation}
    g^0_{\kappa\bq}(\br+\bn)=e^{i\bq\cdot\bn}g^0_{\kappa\bq}(\br),
\end{equation}
when $\bn$ is a lattice vector. This property comes from the discrete translational invariance of the lattice.

Then, we switch the electron part to the momentum space by defining
\begin{equation}
    \psi_{\br}^{\sigma} = \int_{\bk\bG_m}e^{i\bk\bn}e^{i\bG_m\Delta\br} \psi_{\bk}^{m\sigma},
\end{equation}
with $\br=\bn+\Delta\br$, where $\bn$ is a lattice vector and $\Delta\br$ is within a unit cell of volume $\Omega_\text{Cell}$. The momentum $\bk$ is defined within the first Brillouin zone, and $\bG_m$ is the G-vector. 
Plugging this into the interaction Lagrangian leads to
\begin{align}
  S_{\txeph} =& \sum_{\sigma\kappa}\int\limits_{\substack{
    \bk\bq\tau\\
    \bG_m\bG_m'
  }} g^0_{mm'\kappa}(\bq) \bar{\psi}_{\bk+\bq\tau}^{m\sigma}\psi_{\bk\tau}^{m'\sigma}u_{\kappa\bq\tau},
\end{align}
with 
\begin{equation}
    g^0_{mm'\kappa}(\bq)=\int_{\Delta\br}e^{-i(\bG_m-\bG_{m'})\Delta\br}g_{\bq\kappa}(\Delta\br).
\end{equation}

The electron part of the action is now
\begin{align}
    S_{\txe} &= \int\limits_{\substack{\bk\tau\\ \bG_m\bG_m'}} \sum_{\sigma} \bar{\psi}_{\bk\tau}^{ m\sigma}\left[\left(\frac{\partial}{\partial \tau} -\frac{\left(\bk+\bG_m\right)^2}{2m}\right)\delta_{mm'} +V_{mm'}\right]\psi_{\bk\tau}^{{m'}\sigma}\\
            &+\frac{1}{2}\int\limits_{\substack{\bk\bk'\bq\tau\\ \bG_m\bG_m'}}\sum_{\sigma\sigma'n}\bar{\psi}_{\bk\tau}^{m\sigma}\bar{\psi}_{\bk'\tau}^{{m'}\sigma'} v_{\bq+\bG_n}\psi_{\bk'-\bq\tau}^{{m'-n}\sigma'}\psi_{\bk+\bq\tau}^{{m+n}\sigma},
\end{align}
with
\begin{equation}
    V_{mm'}=\int_{\Delta\br}e^{-i(\bG_m-\bG_{m'})\Delta\br}V(\Delta\br,\bR^0).
\end{equation}
We abbreviate the notation of G-vectors in subscripts of the field so that any arithmetic notation of indices should be understood as operation of corresponding G-vectors, i.e., $\psi^{m-n}=\psi\left({\bG_m-\bG_n}\right)$.
While the G-vector remains conserved during scattering processes due to Coulomb interactions and the bare electron-phonon interaction, the lattice potential $V_{mm'}$ may introduce an external source of G-vectors. Consequently, the Green's function for the electron could exhibit different incoming and outgoing G-vectors.

\begin{figure}
  \centering
  \includegraphics[width=1.0\linewidth]{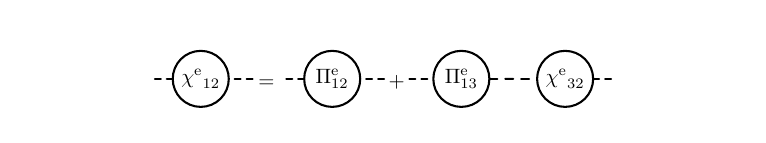}
  \caption{Electron's density-density correlation function $\chi^{\txe}$ from electron's polarization $\Pi^\txe$.
  } 
  \label{fig:chi_def}
\end{figure}

\subsection{Screening of the Electron-phonon Interaction}

While the interaction between bare electrons and phonons is described by the electron-phonon coupling $g^0$, the quasiparticle counterpart $g$ was screened by the electrons and dressed by the electron's 3-vertex. 
It turns out that this effect, together with the effect of quasiparticle renormalization factor $z^\txe$, resulted in a net effect that could be neglected for practical purposes. 
In this section we derived the quantity that described this effect, and in later sections we would show the numerical results.

\newcommand{\regepi}{g^\text{r}}

\begin{figure}
  \centering
  \includegraphics[width=1.0\linewidth]{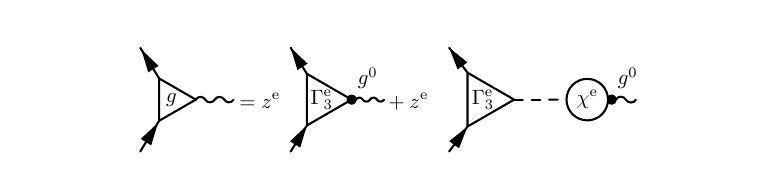}
  \caption{Electron-phonon interaction $g$ from bare electron-phonon coupling $g^0$. 
  } 
  \label{fig:epi_def}
\end{figure}

The bare electron-phonon interaction $g^0_{mm'\kappa}(\bq)$ generally depends on the structure of the lattice and has complex structures in momentum space. 
Specifically, the bare electron-phonon interaction defined in Eq.\eqref{eq:epi_bare} diverges at $\bq=0$. 
However, after the screening effect was taken into account, the quasiparticle counterpart $g$ would be regular.
The screening effect was shown in the definition of $g$ as illustrated in Fig.\ref{fig:epi_def}.

To simplify the discussion below, we split the lattice dependent part and the Coulomb singular part by writing $g^0_{mm'\kappa}(\bq)=\regepi_{mm'\kappa}(\bq)v_{\bq}$, such that $\regepi$ is a regular function at $\bq=0$.
By plugging this definition into the definition of $g$, we obtain $g(\bk\omega;\bq\nu)=\regepi z^\txe  W^\txe_{\bq\nu} \Gamma^\txe_3(\bk\omega;\bq\nu)$, where $\frac{v_\bq}{\varepsilon_{\bq\nu}}=W^\txe_{\bq\nu}=\frac{v_\bq}{1-v_\bq P_{\bq\nu}}$ is the screened Coulomb interaction between electrons, and $P_{\bq\nu}$ is the electron's polarization.
Thus we could see that the screening effect is encoded in the structure of $W\Gamma^\txe_3$. 
In Fig.\ref{fig:ver3angle} we show the numerical results of $z^\txe W\Gamma^\txe_3$ in UEG via VDMC, and in next section we discuss the connection between $g$ and the ones obtained from first-principle methods such as DFPT.

\subsection{Density Functional Perturbation Theory of the Electron-Phonon Interaction}
\label{app:DFPT}
We started by reviewing the derivation of the electron-phonon interaction within the framework of DFT. 
We followed the spirit of dielectric approach \cite{dielectric1, dielectric2}, which is theoretically equivalent to the DFPT as pointed out in Ref.\cite{EPW1}.
Starting from the effective potential experienced by an electron, we had the Kohn-Sham potential,
\begin{equation}
   V^{\txKS}\left(\mathbf{r} ;\left\{\mathbf{R}_0\right\}\right)=V^{\txion}\left(\mathbf{r} ;\left\{\mathbf{R}_0\right\}\right)+V^\txH\left(\mathbf{r} ;\left\{\mathbf{R}_0\right\}\right)+V^{\txxc}[\br; n(\left\{\bR_0\right\})],
\end{equation}
where $V^{\txion}$ is the potential from the ion lattice, $V^\txH$ is the Hartree energy, which is the electrostatic energy generated from $n(\br; \left\{\bR_0\right\})$.
We assumed that the exchange-correlation effects between electrons could be captured by a functional $V^{\txxc}[\br; n(\left\{\bR_0\right\})]$ depending on the electron density.
The dependence on $n$ could be non-local in general;
while for our purpose, it is sufficient to consider within LDA, 
where the potential at $\br$ is given by a functional $V^{\txxc}[n(\br; \left\{\bR_0\right\})]$ depending only on the local electron density $n(\br; \left\{\bR_0\right\})$ at $\br$. 
The density of electrons is in turn determined by an ionic potential imposed by the ions at the position $\bR^0$. 

Now consider a small displacement of ions $\delta\bR$,
the response of the potential would be
\begin{equation}
\delta V^{\txKS} \equiv V^{\txKS}\left(\mathbf{r} ;\left\{\mathbf{R}_0+\delta \mathbf{R}\right\}\right)-V^{\txKS}\left(\mathbf{r} ;\left\{\mathbf{R}_0\right\}\right),
\end{equation}
which was a sum of three components,
\begin{equation}
  \delta V^{\txion}=\sum_{i \alpha} g^0_{i \alpha} \delta R_{i \alpha},
\end{equation}
\begin{equation}
\delta V^\txH=\int_{\mathbf{r}^{\prime}} v\left(\mathbf{r}, \mathbf{r}^{\prime}\right) \delta n\left(\mathbf{r}^{\prime} ;\left\{\mathbf{R}_0\right\}\right),
\end{equation}
and
\begin{equation}
\delta V^{\txxc}=\int_{\mathbf{r}^{\prime}} f_{\txxc}\left(\mathbf{r}, \mathbf{r}^{\prime}\right) \delta n\left(\mathbf{r}^{\prime} ;\left\{\mathbf{R}_0\right\}\right),
\end{equation}
where $f_{\txxc}\left(\mathbf{r}, \mathbf{r}^{\prime}\right)=f_{\txxc}\left(\br\right)\delta_{\br-\br^\prime}=\left.\frac{\partial V^{\txxc}[n]}{\partial n}\right|_{n=n(\br)} \delta_{\br-\br^\prime}$ was the exchange correlation kernel in LDA.

The density deviation is proportional to the Kohn-Sham potential deviation,
\begin{equation}
    \delta n\left(\mathbf{r} ;\left\{\mathbf{R}_0\right\}\right)=\int_{\mathbf{r}^{\prime}} \chi^\txe_0\left(\mathbf{r}, \mathbf{r}^{\prime}\right) \delta V^{\txKS}\left(\mathbf{r}^{\prime} ;\left\{\mathbf{R}_0\right\}\right)
\end{equation}
where $\chi^\txe_0$ is the density-density correlation of free electrons. 
Without interaction, we have $\chi^\txe_0=P^{(0)}$, the polarization of free electrons, given by the Lindhard function.
We then symbolically have
\begin{equation}
\delta V^{\txKS}=\frac{1}{1-\left(v+f_{\txxc}\right) \chi^\txe_0}\delta V^{\txion},
\end{equation}
where both $v$ and $f_\txxc$ are functions of $\br,\br^\prime$, and their multiplication to $\chi^\txe_0$ and inverse correspond to convolution and inverse of functions. 
If we further assume that the density of electrons varies slowly in space, we have $f_{\txxc}\left(\mathbf{r}\right)\approx f_{\txxc}$, and then the Fourier transformed counterpart of this equation became algebraic.

Now the electron-phonon interaction,
\begin{equation}
\left. g^{\txKS}_{i \alpha}(\mathbf{r}) \equiv \frac{\partial V^{\txKS}(\mathbf{r} ;\{\mathbf{R}\})}{\partial R_{i \alpha}}\right|_{\mathbf{R}=\mathbf{R}_0},
\end{equation}
as defined in DFPT, 
could be expressed as 
\begin{equation}
\begin{aligned}
g^{\txKS}=& \frac{g^0}{1-\left(v+f_{\txxc}\right) \chi^\txe_0}
\end{aligned}
\end{equation}
where $g^0=v_\bq g^\txr$.
By observing that in LDA
\[\chi^\txe(\bq)=\frac{P_\bq}{1-v_\bq P_\bq}\approx\frac{\chi^\txe_0}{1-\left(v+f_{\txxc}\right) \chi^\txe_0} \, ,
\]
we have
\begin{equation}
  g^{\txKS}\approx g^\txr v_\bq \frac{\chi^\txe(\bq)}{\chi_0^\txe(\bq)}.
\end{equation}
Then we could see that the quality of this approximation was determined by the difference between $v_\bq\frac{\chi^\txe(\bq)}{\chi_0^\txe(\bq)}$ and $z^\txe W_\bq\Gamma^\txe_3(\bk;\bq)$ by comparing $g^{\txKS}$ with its many-body counterpart $g=g^\txr\left(z^\txe W^\txe \Gamma^\txe_3\right)$, as pointed out in Eq.\eqref{eq:epi_ansatz}.

\section{Pair-field BSE}
\label{sec:pcf_general_derivation}
This section details the derivation of the BSE for the anomalous vertex, which governs PCF. We employ the Nambu-Gor'kov formalism, defining the Nambu spinors as:
\begin{equation}
    \bar{\Psi}=\left(\bar{\psi}_{\uparrow},{\psi}_{\downarrow}\right), \quad \Psi=\binom{\psi_{\uparrow}}{\bar{\psi}_{\downarrow}}.
\end{equation}
We consider the system at $T>T_c$ with a source term,
\begin{equation}
    S[\nambu{\eta}] = S + \int_{12} \bar{\Psi}_1 \nambu{\eta}_{12} \Psi_2 + h.c.\quad,
    \label{eq:source_action}
\end{equation}
where the subscript $i$ abbreviates $(\br_i,\tau_i)$.
The source term matrix is
\begin{equation}
    \nambu{\eta}_{12} = \begin{bmatrix}
        \eta_{12} & \eta^a_{12} \\
        (\eta^a_{12})^* & -\eta_{21}
        \end{bmatrix},
\end{equation}
introducing a normal source term $\eta$ and an anomalous source term $\eta^a$.
These modify the bare electron propagator's inverse: $\nambu{g}^{-1}_{12}[\nambu{\eta}]=\text{diag}(g^{-1}_{12},-g^{-1}_{21})+\nambu{\eta}_{12}$.

The full Green's function is defined as
\begin{equation}
\nambu{G}_{12}\equiv -\langle\Psi_1\bar{\Psi}_2\rangle \equiv \begin{bmatrix}
    G_{12} & F_{12} \\
    F^*_{12} & -G_{21}
    \end{bmatrix},
\end{equation}
where $G$ is the normal and $F$ is the anomalous (Gorkov) Green's function.
These can be obtained by functional derivatives of the partition function:
\begin{equation}
    G_{12}[\eta,\eta^\txa] = \frac{\delta \ln Z[\eta,\eta^\txa]}{\delta \eta_{12}}, \quad F_{12}[\eta,\eta^\txa] =\frac{\delta \ln Z[\eta,\eta^\txa]}{\delta \eta^\txa_{12}}.
\end{equation}
We retain the source terms in the Green's functions, keeping zeroth order for $G$ and first order for $F$ through the derivation. The physical quantities are obtained by setting the sources to zero at the end.

The Nambu self-energy is defined as
$\nambu{\Sigma}_{12}=\nambu{G}^{-1}_{12}-\nambu{g}^{-1}_{12}$.
This yields a matrix equation that expands into two coupled Dyson's equations for the normal and anomalous Green's functions.
At $T>T_\txc$, these equations decouple in the leading order, resulting in
$\Sigma= G^{-1}-g^{-1} + O(\eta,\eta^\txa)$ 
and $\Sigma^\txa= G^{-1}FG^{-1}$.

We define the element-wise functional derivative of a Nambu matrix $\nambu{A}$ with respect to the Nambu source matrix $\nambu{\eta}$ as:
\begin{equation}
    \frac{\delta \nambu{A}}{\delta{\nambu{\eta}}_{12}}
    =
    \begin{bmatrix}
        \frac{\delta A}{\delta \eta_{12}} & \frac{\delta A^\txa}{\delta \eta^\txa_{12}}\\
        \frac{\delta \left(A^\txa\right)^*}{\delta \left(\eta^\txa_{12}\right)^*} & -\frac{\delta A}{\delta \eta_{21}}
    \end{bmatrix}.
\end{equation}
Applying this definition to the functional derivative of the inverse Green's function yields:
\begin{equation}
    \left.\frac{\delta\nambu{G}_{12}^{-1}}{\delta\nambu{\eta}_{33}}\right\vert_{\nambu{\eta}=0}
    =
    \begin{bmatrix}
        \Gamma^3_{12;3} & \Lambda_{12;3}\\
        \left(\Lambda_{12;3}\right)^* & \Gamma^3_{21;3}
    \end{bmatrix},
\end{equation}
where $\Gamma^3$ is the normal three-point vertex and
$\Lambda$ is the anomalous three-point vertex.

Now we derive the BSE for $\Lambda$.
Starting with the relation $\nambu{G}^{-1}_{12}=\nambu{g}^{-1}_{12}+\nambu{\Sigma}_{12}$ and using the functional derivative relation $\frac{\delta \nambu{G}}{\delta \nambu{\eta}} = -\nambu{G}\frac{\delta \nambu{G}^{-1}}{\delta \nambu{\eta}}\nambu{G}$, we obtain from the off-diagonal components:
\begin{align}
    \Lambda_{12;3} &= \delta_{13}\delta_{23}+\frac{\delta\Sigma^\txa_{12}}{\delta\eta^\txa_{33}}\nonumber\\
    &=\delta_{13}\delta_{23}+\int_{45}\frac{\delta \Sigma^\txa_{12}}{\delta F_{45}}G_{44'}G_{55'}\Lambda^a_{4'5';3}.
    \label{eq:lambda}
\end{align}
Here, the functional derivative of $\Sigma^\txa$ with respect to $F$ yields the particle-particle irreducible four-point vertex function $\Gammapp_{14;25}=\frac{\delta \Sigma^a_{12}}{\delta F_{45}}$.
This equation for $\Lambda$ gives the desired Bethe-Salpeter equation, solvable self-consistently given $G$ and $\Gammapp$.

\section{Finite Temperature Scaling: Generic BCS }
\label{app:pcf}

We started by writing the Eq.\eqref{eq:lambda} explicitly without phonons:
\begin{eqnarray}
  \label{eq:LR}
  \pcf_{\bk\omega}=1+\int _{\bk'\omega'} \Gammapp_{\bk^{}\omega^{};\bk'\omega'} \Pi_{\bk'\omega'} \pcf_{\bk'\omega'},
\end{eqnarray}
where we set the source term $\eta\equiv1$.
The vertex $\tilde{\Gamma}_{\bk^{}\omega^{};\bk'\omega'}$ is exempt from Cooper singularities, and connects to the full vertex function via the self-consistent equation:
\begin{equation}
\Gamma_{\bk^{}\omega^{};\bk'\omega'} = \tilde{\Gamma}_{\bk^{}\omega^{};\bk'\omega'} + \int_{\bk''\omega''} \tilde{\Gamma}_{\bk^{}\omega^{};\bk''\omega''}\Pi_{\bk''\omega''}\Gamma_{\bk''\omega'';\bk'\omega'},
\end{equation}
where the $\Pi_{\bk\omega}$ is the quasiparticle pair-field propagator. 

The source term $\eta_{\bk\omega}$ is a smooth function of $\bk\omega$ that we normally set to unity in real calculations, and $\Gamma_{\bk^{}\omega^{};\bk'\omega'}=\Gamma(\bk^{},\omega^{},-\bk^{} ,\omega^{};\bk',\omega',-\bk',\omega')$ is the
full four-point vertex function that can be derived from the connected
two-electron Green's function. 

As first illustrated in Ref.~\cite{pengcheng}, for emergent BCS superconductors and non-superconductors at the Fermi surface,
the linear response function follows a universal scaling relation, which we call
precursory Cooper flow (PCF): 
\begin{equation}
  \pcf_0(T)=\frac{1}{1+g \ln \left(\omega_\Lambda / T\right)}+\mathcal{O}(T),
  \label{eq:flow}
\end{equation}
Here the label $0$ represents the low-energy limit $|\bk|=k_\txF,\quad \omega = \pi T$, which we denote by $\bk\omega=0$.
Consequently, one can estimate the critical temperature $T_\txc$ as $T_\txc\equiv
\omega_\Lambda e^{-1/g}$ by computing $\pcf_0(T)$ at temperatures well above the
superconducting state. This methodology circumvents the need to perform
calculations within the challenging superconducting regime, offering a more
practical and robust predictive tool. The goal of this section is to 
derive this universal scaling and its prerequisite. 

\subsection{Singularity Analysis}
For a generic electron system, the BCS pairing originates from $\Pi_{\bk\omega}=G_{\bk\omega}G_{-\bk\omega}$,
where $G_{\bk\omega}$ is the electron Green's function 
\begin{align}
\label{G}
  G_{\bk\omega} = \frac{z}{-i \omega + \epsilon_{\bk}}+\text{reg}
\end{align}
Here $\varepsilon_{\bk}=\mathbf{v}_\txF^*\cdot (\bk-\bk_\txF)$ is the quasi-particle dispersion, $z$ is the fermi liquid renormalization factor, and $v_\txF^*$ is the renormalized fermi velocity.  The first term is the leading contribution of taylor expansion of the exact dispersion around Fermi surface. The corrections lead to a regular contribution that saturates to a constant at $\bk\omega =0$.Below we denote such terms as $\text{reg}$.

The same separation applies to  $\Pi_{\bk\omega}$
\begin{gather}
	\label{G2limit}
\Pi_{\bk\omega} = \frac{z^{2}}{\omega^2+\epsilon_{\bk}^2} + \text{reg}
\end{gather}  
It immediately follows that
\begin{gather}
\label{eq:G2}
\int_{\bk\omega} \, \Pi_{\bk\omega}= A \ln {T} + B + \mathcal O(T)
\end{gather}
where $A = \frac{z^{2}\pi}{2v_\txF}$ and $B$ are two constants, and the finite-temperature correction $\mathcal O(T)$ vanishes as power series of $T$. This low-energy singularity has been thoroughly studied in the BCS theory.

Another singular property comes from dynamical screened Coulomb interaction in the irreducible vertex $\Gammapp$. As revealed in the random phase approximation (RPA), only the static part of Coulomb interaction is fully screened. For any transfer frequency $\omega-\omega'\neq 0$, it remains long range with a plasmon contribution $W^s$  that diverges when transfer momentum $\bq=\bk - \bk'$ goes to zero. Therefore, we adopt the following separation of irreducible vertex
\begin{equation}
    \label{eq:Gammapp_sep}
    \Gammapp_{\bk^{}\omega^{};\bk'\omega'} = \Gammapp^\txr_{\bk^{}\omega^{};\bk'\omega'} +    W^\txs_{\bk^{}-\bk', \omega^{}-\omega'}
\end{equation}
Since  the parametrization of $W^s$ remains non-unique, provided it accurately captures the singularity, we choose the following simple form:
\begin{equation}
\label{eq:plasmonsing}
    W^\txs_{\bk^{}-\bk', \omega^{}-\omega'} =\frac{4\pi e^2}{\left|\bk^{}-\bk'\right|^2} \frac{(\omega^{}-\omega')^2}{(\omega^{}-\omega')^2+\omega_p^2}.
\end{equation}
Here $\omega_\txp$ is the characteristic plasma frequency. This singular contribution was ignored by the conventional treatment, where Coulomb pseudopotential only takes into account the static interaction. In the next section, we will demonstrate the Coulomb singularity does not change the universal temperature scaling in PCF.

\subsection{Temperature Scaling of the PCF}
Without loss of generality, we focus on the isotropic case, where all functions in the equations could be decomposed with spherical harmonics.
The results could be extended to the cases where anisotropy of the system is not
extreme.
If, on the other hand, the system shows drastic behavior, for instance some
singularity of dispersion at certain symmetry points on the Fermi surface, then
there is no guarantee the results persist, and the derivation should be re-examined carefully.

 To simplify the notation, we first introduce a symbolic expression of Eq.\eqref{eq:LR}
\begin{gather}
\label{eq:R_symbol}
\pcf = \frac{1}{1+\Gammapp \Pi}I
\end{gather}
Here $I_{\bk'\omega'} = 1$ is an uniform function of ${\bk'\omega'}$, and $\Gammapp \Pi$ should be considered as an operator that follows
\begin{gather}
	\Gammapp \Pi X = \int_{\bk' \omega'} \Gammapp_{\bk^{}\omega^{};\bk'\omega'} \Pi_{\bk'\omega'} X_{\bk'\omega'}
\end{gather}
 where $X_{\bk'\omega'}$ is the function that this operator acts on. Pay attention that (i)
 The operators do not commute with each other, thus it is important to keep
 their order the same as in the original equation; (ii) Any function of symbolic
 operators should be treated as a taylor expansion.
 For example $1/(1+X)$ in
 above equations represents a geometric series
 $\delta+\sum\limits_{n=1}^{\infty}X^n$, where the identity matrix $\delta$ is defined by 
\begin{gather}
	\delta = \delta_{\bk^{}-\bk',\omega^{}-\omega'} = \frac{1}{T} \delta_{\omega^{}-\omega'}\delta_{\bk^{}-\bk'},
\end{gather}
so that $\int_{\bk'\omega'}\, \delta_{\bk^{}\omega^{}-{\bk'\omega'}}X_{\bk'\omega'} = X_{\bk\omega}$.

For an emergent BCS system that satisfies Eq.\eqref{eq:LR}, $\Gammapp_{\bk^{}\omega^{};\bk'\omega'}$ must be free of
singularity at $\bk'\omega'=0$ that is strong enough to alter the $\ln T$ dependence.
In first order, this condition could be rigorously formulated as
\begin{gather}
\label{eq:oneloop}
	\int_{\bk'\omega'} \, \Gammapp_{\bk^{}\omega^{};{\bk'\omega'}}\Pi_{\bk'\omega'}=\tilde{g}_{\bk\omega} \ln T + \tilde{f}_{\bk\omega} + \mathcal O(T),
\end{gather} 
which guarantees that no singular term beyond $\ln T$ is generated.
In the following we show that as long as the temperature-independent functions $\tilde{g}_{\bk\omega}$ and $\tilde{f}_{\bk\omega}$ are regular in the $\bk\omega\to 0$ limit,
the singular $\ln T$ terms at higher order automatically follows a
simple geometric series.

To see this, we first consider the second order term 
\begin{gather}
\Gammapp \Pi\Gammapp \Pi = \int_{\bk'\omega'} \, \Gammapp_{\bk^{}\omega^{};\bk'\omega'} \Pi_{\bk'\omega'}(\tilde{g}_{\bk'\omega'} \ln T + \tilde{f}_{\bk'\omega'})+\mathcal{O}(T)
\end{gather}
 Since $\tilde{g}_{\bk'\omega'}$ and $\tilde{f}_{\bk'\omega'}$ are regular at $\bk'\omega'\rightarrow 0$ and independent of $\bk^{}\omega^{}$, they do not change the singular structure of $\Gammapp \Pi$, meaning that 
\begin{gather}
\int_{\bk'\omega'} \, \Gammapp_{\bk^{}\omega^{};\bk'\omega'} \Pi_{\bk'\omega'} \tilde{g}_{\bk'\omega'}=\tilde{g}^{\prime}_{\bk\omega} \ln T + \tilde{f}^{\prime}_{\bk\omega} + \mathcal O(T)
\end{gather}
where $\tilde{g}^{\prime}_{\bk\omega}$ and $\tilde{f}^{\prime}_{\bk\omega}$ are also regular at
$\bk'\omega'\rightarrow0$. Invoking the mathematical induction, we find that only one $\ln T$
singular contribution is generated when the order increases by one. 

The above observation enables us to determine whether the system is emergent BCS by
checking the validity of Eq.\eqref{eq:oneloop}. It also allows us to extract the renormalized
constants in Eq.\eqref{eq:flow} by defining 
\begin{gather}
	\label{eq:split}
	\Gammapp_{\bk^{}\omega^{};\bk'\omega'}\Pi_{\bk'\omega'}=\left[\tilde{g}_{\bk\omega} \ln T + \tilde{f}_{\bk\omega}\right] \delta_{\bk'\omega'} + \phi_{\bk^{}\omega^{};\bk'\omega'}
\end{gather}
where $ \int_{\bk'\omega'} \phi_{\bk^{}\omega^{};\bk'\omega'} = \mathcal O(T)$.
We start by resumming the regular terms of $\pcf_{\bk\omega}$ using the symbolical
expression. First, we insert Eq.\eqref{eq:split} into Eq.\eqref{eq:R_symbol}
(notice the operator nature of denominators): 
\begin{align}
	\pcf &=\frac{1}{1+\phi+(\tilde{g}\ln{T}+\tilde{f})\delta}I\nonumber\\ 
    &=\frac{1}{(1+\phi)\left(1+\frac{1}{1+\phi}(\tilde{g}\ln{T}+\tilde{f})\delta\right)}I\nonumber\\
  &=\frac{1}{1+\frac{1}{1+\phi}(\tilde{g}\ln{T}+\tilde{f})\delta}\frac{1}{1+\phi}I\nonumber\\
  &= \frac{1}{1+ \gamma (\tilde{g}\ln{T}+\tilde{f})\delta} \gamma I
  \label{eq:R_resum}  
\end{align}
where we have introduced
\begin{align}
\label{eq:gamma}
\gamma_{\bk^{}\omega; \bk'\omega'} =\delta_{\bk^{}\omega^{}-\bk'\omega'} -  \int_{\bk''\omega''} \, \phi_{\bk^{}\omega^{};\bk''\omega''}\gamma_{\bk''\omega'';\bk'\omega'}.
\end{align}
Returning to Eq.\eqref{eq:R_resum} with its explicit form, we find
\begin{align}
	\label{eq:LR_full}
	&\pcf_{\bk\omega}=1-\int_{\bk'\omega'} \gamma_{\bk^{}\omega^{};\bk'\omega'}\left(\tilde{g}_{\bk'\omega'}\ln {T}+ \tilde{f}_{\bk'\omega'}\right) \pcf_0 + \mathcal O(T)
\end{align}
where we have used $\int_{\bk'\omega'} \gamma_{\bk^{}\omega^{};\bk'\omega'} = 1+\mathcal O(T)$ that follows $ \int_{\bk'\omega'} \phi_{\bk^{}\omega^{};\bk'\omega'} = \mathcal O(T)$. Defining 
 \begin{equation}
\label{eq:R_final}
g_{\bk\omega} = -\int_{\bk'\omega'}\,  \gamma_{\bk^{}\omega^{};\bk'\omega'}\tilde{g}_{\bk'\omega'}, \quad  f_{\bk\omega} = -\int_{\bk'\omega'}\,  \gamma_{\bk^{}\omega^{};\bk'\omega'}\tilde{f}_{\bk'\omega'}
\end{equation}
we can solve the algebraic equation at $\bk\omega=0$, which immediately recovers the key
result Eq.\eqref{eq:flow} with $g = g_0$, $f = f_0$ and $\omega_\Lambda = e^{-\frac{f_0}{g_0}}$
for its coefficients. 

We also immediately restore expression of $\pcf_{\bk\omega}$:
\begin{align}
\label{eq:RK_final}
\pcf_{\bk\omega}&= \frac{1+(g_{\bk\omega} - g)\ln T +(f_{\bk\omega} -f)}{1 - g\ln {T} - f}\nonumber\\
	&= \frac{J_{\bk\omega}}{1+g \ln \frac{\omega_\Lambda}{T}} + 1 - \frac{g_{\bk\omega}}{g} + \mathcal O(T)\\
J_{\bk\omega} &= f_{\bk\omega} +(1-f)\frac{g_{\bk\omega}}{g}
\end{align}
This final form shows that $\pcf_{\bk\omega}$ is also regular at $\bk\omega=0$, as directly inherited from $g_{\bk\omega}$ and $f_{\bk\omega}$.

Following the same derivation, we can also establish the temperature scaling of full vertex $\Gamma$
\begin{align}
    \Gamma& = \frac{1}{1+\Gammapp \Pi}\Gammapp \nonumber\\
    &= \frac{1}{1+ \gamma (\tilde{g}\ln{T}+\tilde{f})\delta} \gamma \Gammapp
\end{align}
which can be rewritten explicitly as
\begin{align}
	&\Gamma_{\bk^{}\omega^{};\bk'\omega'}=\nonumber\\
 &\Gammapp^{*}_{\bk^{}\omega^{};\bk'\omega'}-\int_{\bk''\omega''} \gamma_{\bk^{}\omega^{};\bk''\omega''}\left(\tilde{g}_{\bk''\omega''}\ln {T}+ \tilde{f}_{\bk''\omega''}\right) \Gamma_{0;\bk'\omega'}  \nonumber\\
 &\Gammapp^{*}_{\bk^{}\omega^{};\bk'\omega'} = \Gammapp_{\bk^{}\omega^{};\bk'\omega'} - \int_{\bk''\omega''} \phi_{\bk^{}\omega^{};\bk''\omega''} \Gammapp^{*}_{\bk''\omega'';\bk'\omega'}
 \label{eq:Gammapp*}
\end{align}
Setting $\bk\omega = 0$, we find
\begin{align}
\label{eq:fullGamma_scaling}
	\Gamma_{0;\bk'\omega'}=\frac{\Gammapp^{*}_{0;\bk'\omega'}}{1 - g\ln {T} - f}\nonumber\\
\end{align}
Notice that $\Gammapp^*$ does not have any $\ln T$ scaling, as $\phi$ in Eq.\eqref{eq:Gammapp*} is a regular function.

\subsection{Emergent BCS of Dynamical Screened Coulomb interaction}
\label{sec:app_eliashberg}
As we mentioned, the plasmon contribution $W^\txs$ in the screened Coulomb interaction diverges at $\omega-\omega' \neq 0$,  $|\bk^{}-\bk'|=0$. Therefore,
We have to verify that the $\ln T$ divergence from $\Pi_{\bk'\omega'}$ is not changed by this additional singularity.

In Eq.\eqref{eq:Gammapp_sep}, the first regular term $\Gammapp^\txr_{\bk^{}\omega^{};\bk'\omega'}$ is smooth at $\bk'\omega' = 0$. Applying Eq.\eqref{eq:G2}, we can write
\begin{align}
	\int_{\bk'\omega'}&\Gammapp^{\txr}_{\bk^{}\omega^{};\bk'\omega'}\Pi_{\bk'\omega'}=\int_{\bk'\omega'}\left(\Gammapp^{\txr}_{\bk^{}\omega^{};\bk'\omega'} - \Gammapp^{\txr}_{\bk\omega;0} +\Gammapp^{\txr}_{\bk\omega;0} \right)\Pi_{\bk'\omega'} \nonumber\\
	&=A\Gammapp^{\txr}_{\bk\omega;0}\ln T + B\Gammapp^{\txr}_{\bk\omega;0} + \int_{\bk'\omega'} \Delta\Gammapp^{\txr}_{\bk^{}\omega^{};\bk'\omega'}\Pi_{\bk'\omega'}
 \label{eq:reg}
\end{align}
where $\Delta\Gammapp^{\txr}_{\bk^{}\omega^{};\bk'\omega'} = \Gammapp^{\txr}_{\bk^{}\omega^{};\bk'\omega'}-\Gammapp^{\txr}_{\bk\omega;0}$ vanishes as polynomials of $|\bk'|-k_\txF$ and $\omega'$ when $\bk'\omega'\rightarrow 0$. This means the singularity of $\Pi_{\bk'\omega'}$ in the last term is cancelled by $\Delta\Gammapp^{\txr}_{\bk^{}\omega^{};\bk'\omega'}$, yielding a regular function of $\bk\omega$ and $T$. Clearly all terms in Eq.\eqref{eq:reg} are finite when $\bk\omega$ goes to zero, which satisfies the emergent BCS condition. 

The analysis of the second term $W^\txs$ is more tricky. For simplicity, we assume the Fermi surface is spherical, so that dispersion in Eq.\eqref{G} reduces to scalar product $v_\txF^*(k'-k_\txF)$ with no angle dependence. Integrating the angle dependence explicitly, we have
\begin{align}
  &\int_{\bk'\omega'}W^{\txs}_{\bk^{}-\bk', \omega-\omega'}\Pi_{\bk'\omega'} \nonumber =\nonumber\\ 
    &\int _{k'\omega,\chi} k'^2\frac{e^2}{\pi(k^2 +k'^2 - 2k^{}k' \chi)} \frac{(\omega - \omega')^2}{(\omega - \omega')^2+\omega_p^2} \frac{(z^\txe)^2}{\omega'^2 + \epsilon_{k'}^2}\nonumber \\
    = &\int_{k'\omega'} (k_\txF+\Delta k')^2 \frac{2 e^2}{\pi} \ln\left| \frac{\Delta k-\Delta k'}{2k_\txF + \Delta k+ \Delta p}\right| \nonumber\\
    &\times \frac{(\omega - \omega)^2}{(\omega - \omega')^2+\omega_p^2}\frac{(z^{\txe})^2}{\omega'^2 + v_\txF^{*2}\Delta k'^2}
\end{align}
where $\chi \in [-1, 1]$ and $\Delta k' = k'-k_\txF$.  The key observation is that 
singular region of $W^\txs$ and $\Pi$ overlaps in the limit $\Delta k\rightarrow 0$, $\Delta p \rightarrow 0$, $\omega_n \rightarrow 0$. Therefore, the only contribution, omitting a constant factor $k_\txF^2 z^{2} e^2/ \pi$, that could potentially break the emergent BCS condition is
\begin{align}
S_{k\omega}&= \frac{ \omega^2}{\omega^2+\omega_p^2} \int_{\Delta k'\omega'} \ln\left|\frac{\Delta k-\Delta k'}{2k_\txF}\right| \frac{1}{\omega'^2 + v_\txF^{*2} \Delta k'^2} \nonumber\\
&= \frac{ \omega^2}{\omega^2+\omega_p^2}\int_{\Delta k'} \ln\left|\frac{\Delta k-\Delta k'}{2k_\txF}\right| \frac{\tanh(\frac{\epsilon}{2T})}{\epsilon}
\label{eq:regular}	
\end{align}

Here $\epsilon = v_\txF^{*} \Delta k'$, and we explicitly sum over $\omega'$. The $\tanh(\frac{\epsilon}{2T})$ is equivalent to introducing an infrared cutoff $T/v_\txF$ to momentum integral. In the low-energy limit $k\omega\rightarrow0$, the momentum integral yields
\begin{align}
\lim\limits_{k\omega\rightarrow 0} S_{k\omega} &=\frac{\pi ^2T^2}{\pi^2T^2+\alpha^2\omega_{p}^2}\ln \frac{p_{\txc} T}{4 k_\txF v_\txF^*}\ln \frac{p_{\txc} v_\txF^*}{T} \nonumber
\end{align}
Here $p_{\txc}$ is the momentum UV cutoff. This term vanishes as $(T \ln T)^2$ when $T$ approaches zero. Therefore it does not break the emergent BCS  condition. The dynamic structure of $\Gammapp^\txs$ is crucial here, since it regularizes the $(\ln T)^2$ dependence that originates from the singular behavior of Coulomb interaction at small transfer momentum $q$. 

\section{Finite Temperature Scaling: Electron-Phonon BCS}
\label{app:eliashberg}

Since the electron-phonon interaction is a regular function in both momentum and frequency, the PCF temperature scaling we derived for generic cases applies to systems where screened Coulomb and electron-phonon interactions are equally important.
In this section, we discuss how to simplify the PCF, taking advantage of the separation of energy scales in electron-phonon systems.

The key observation is that, in most simple metals with moderate  $r_\txs$ ranges between $1\sim 5$, the characteristic energy of phonon and plasmon are well separated. The plasmon frequency $\omega_\txp$ is
of the same order of magnitude as the Fermi energy $E_\txF$, while the Debye frequency $\omega_\txD$ of
phonons is generally two orders of magnitude smaller than $E_\txF$.
Below we aim to show, without further assuming the details of the model, that
the high energy part of the PCF equation above a cut-off frequency $\omega_\txc$,
which is chosen between the two characteristic frequencies $\omega_\txD \ll
\omega_\txc \ll \omega_\txp$,
could be integrated out up to the accuracy of two small parameters,
$\eta_\txp=\omega_\txc/\omega_\txp$ and $\eta_\txe=\omega_\txD/\omega_\txc$. 

\subsection{Phonon effects on the vertices of Quasi-particles}

In this section, we discuss the effects of phonon-mediated interaction on the Green's function and vertices under the assumption that $\omega_\txD/E_\txF\ll 1$ and that the electron part is already solved. 
The electron properties relevant to our discussion are the quasi-particle Green's function
\begin{equation}
\label{eq:Ge}
  G^{\txe}_{\bk\omega} = \frac{z^\txe}{-i\omega + \varepsilon_{\bk}} + \text{reg}
\end{equation}
and the improper 3-vertex function
\begin{equation}
    \Gamma^{3\txe}_{\bk\omega;\bq \nu} = \frac{1}{z^\txe}\frac{1}{1-(v_{\bq} + f_{\bq}^{\txxc})\Pi^0_{\bq \nu}} + \text{reg},
    \label{eq:ver3_ansatz}
\end{equation}
Here $\varepsilon_{\bk}=\mathbf{v}_\txF^*\cdot (\bk-\bk_\txF)$ is the quasiparticle dispersion of electrons,  $\Pi_0$ is the frequency-momentum Lindhard function with the mass $m^*$, $f_\bq^{\txxc}$ is the exchange-correlation kernel,
$q$ is the transfer frequency-momentum, and $k$ is the frequency-momentum of the incoming electron.

These functions can be prepared from first-principle calculation of UEG without phonons. The phonon effects, then, could be included in a simplified way according to the Migdal's theorem, as long as we only concerns about them up to $O(\omega_\txD/E_\txF)$ accuracy.

We start by considering the phonon-induced interaction between quasi-particles that enters the particle-particle irreducible vertex $\Gammapp$
\begin{align}
\label{eq:gammapp_ph}
 \Gammapp&=\Gammapp^\txe+W_{\txph}+O(m/M)\\
W^{\txph}_{\bk\omega; \bk'\omega'; \bq\nu} &= \Gamma^{3\txe}_{\bk\omega;\bq\nu} {g_{\bq}{D}_{\kappa}(\bq,\nu) g_{\bq}}\Gamma^{3\txe}_{\bk'\omega';\bq\nu}.
\label{eq:Wph}
\end{align}
Notice that as discussed in previous sections, the phonon propagator already includes the full electron density-density correlation $\chi_{\txnn}$, thus no additional resummation over bubble diagrams is required in this form.
Also, the 3-vertex includes no phonon correction, as dictated by Migdal's theorem.

Next, we parametrize the phonon modification to the quasi-particle Green's function Eq.~\eqref{eq:Ge}. As demonstrated in Ref(PHONON-SELFENERGY), the changes to dispersion $\epsilon_{\bk}$ and Fermi velocity $v_{\txF}^*$  are of order $\mathcal{O}(m/M)$, which gives
\begin{equation}
  G_{\bk\omega} = \frac{z^\txe}{-i\omega/z^{\txph}_{\omega} + \varepsilon_{\bk}} +O(\eta_\txe)+\text{reg},
\end{equation}
Here $z^{\txph}_{\omega}$ is the frequency dependent factor obtained from analysing the first order corrections of self-energy by $W_{\txph}$.

\subsection{4-vertex and pseudo-potential}

With $\Gammapp$ parametrized as Eq.~\eqref{eq:gammapp_ph}, the PCF equation of the electron-phonon problem can be expressed as:
\begin{eqnarray}
  \pcf_{\bk^{}\omega^{}} = I_{\bk^{}\omega^{}} + \int_{\bk'\omega'}(\tilde{\Gamma}^\txe_{\bk^{}\omega^{},\bk'\omega'}+W^{\txph}_{\bk^{}\omega^{},\bk'\omega'})\Pi_{\bk'\omega'}\pcf_{\bk'\omega'}+O(\eta_\txe),
  \label{eq:pcf_eph}
\end{eqnarray}
where $I_{\bk\omega}\equiv 1$ is the source term. Starting from this point, variables and internal integrals are again omitted for brevity, wherever context permits. The above equation can be rewritten as,
\begin{eqnarray}
  \pcf = I + (\tilde{\Gamma}^\txe+W^{\txph})\Pi \pcf+O(\eta_\txe).
\end{eqnarray}

We replace the electron irreducible vertex function with the electron full vertex function,
\begin{eqnarray}
  \Gamma^{\txe} = \tilde{\Gamma}^{\txe} + \tilde{\Gamma}^{\txe}\Pi^{\txe}\Gamma^{\txe},
\end{eqnarray}
where $\Pi^{\txe}$ is the pair-field propagator without phonon. We obtain
\begin{eqnarray}
  \label{eq:pcfephfull}
  \pcf = I + \Gamma^{\txe}\Pi^{\txe}I + \Gamma^{\txe}\delta\Pi \pcf  +
  (1+\Gamma^{\txe}\Pi^{\txe})W^{\txph}\Pi \pcf +O(\eta_\txe),\nonumber\\
\end{eqnarray}
where $\delta\Pi=\Pi -\Pi^{\txe}$.

In the equation above, the first two terms collectively represent the PCF for
the pure electron liquid, denoted as $\pcf^{\txe}=I + \Gamma^{\txe}\Pi^{\txe}I$. The above
equation can be formally expressed as: 
\begin{eqnarray}
  \label{eq:pcfephfull2}
  \pcf = \pcf^{\txe} + \Gamma^{\txe}\delta\Pi \pcf  +
  (1+\Gamma^{\txe}\Pi^{\txe})W^{\txph}\Pi \pcf +O(\eta_\txe),\nonumber\\
\end{eqnarray}
Here, the electron liquid PCF $\pcf^{\txe}$ serves as the source term that is
subsequently modified by the electron-phonon interactions to yield the full
electron-phonon PCF $\pcf$.

We first define the low-energy part of the pair-field propagator. Turning off the electron-phonon coupling, the low-energy pair-field propagator is defined by,
\begin{equation}
\label{eq:pi_es}
    \Pi^{\txe,\txs}_{\bk,\omega} \equiv \frac{(z^{\txe})^2}{\omega^2 + \varepsilon_{\bk}^2}\Theta(\omega_\txc-|\varepsilon_{\bk}|), 
\end{equation}
The electron-phonon coupling renormalizes the low-energy pair-field propagator into,
\begin{equation}
\label{eq:pi_s}
    \Pi^{\txs}_{\bk,\omega} \equiv \frac{(z^{\txe})^{2}}{\left(\frac{\omega}{z^{\txph}_\omega}\right)^2 + \varepsilon_{\bk}^2}\Theta(\omega_\txc-|\varepsilon_{\bk}|), 
\end{equation}
where the weight $z^{\txph}_\omega$ approaches to one above the Debye frequency $\omega_\txD$, reducing Eq. \eqref{eq:pi_s} to Eq. \eqref{eq:pi_es}.

The low-energy difference of the pair-field propagator due to the phonon-induced
correction, is then
\begin{eqnarray}
\begin{split}
  \delta\Pi^{\txs}_{\bk,\omega} &= \Pi^{\txs}_{\bk,\omega} -\Pi^{\txe,\txs}_{\bk,\omega}\\
  &=\frac{(z^{\txe})^2\left(\frac{1}{{z^{\txph}_\omega}}-1\right)\omega^2}{(\omega^2+\varepsilon_{\bk}^2)\left(\left(\frac{\omega}{z^{\txph}_\omega}\right)^2+\varepsilon_{\bk}^2\right)}\Theta(\omega_\txc-|\varepsilon_{\bk}|).
\end{split}
\end{eqnarray}
The weight $z^{\txph}_\omega$ approaches to one above the Debye frequency $\omega_\txD$, introducing a frequency cutoff to $\delta \Pi$.

Consider two functions $A_{\bk\omega}$ and $B_{\bk\omega}$ that are regular functions of the momentum near the Fermi surface below the scale of $\omega_\txD/v_{\txF}^*$. Then we can prove two equations,
\begin{equation}
\label{eq:approx1}
    \int_{\bk\omega} A_{\bk\omega}\Pi_{\bk\omega}W^{\txph}_{\bk\omega; \bk'\omega'} = \int_{\bk\omega} A_{\bk_{\txF}\omega}\Pi^{\txs}_{\bk\omega}W^{\txph}_{\bk_{\txF}\omega; \bk'\omega'} + O(\eta_\txe),
\end{equation}
and
\begin{equation}
\label{eq:approx2}
    \int_{\bk\omega} A_{\bk\omega}\delta \Pi_{\bk\omega}B_{\bk\omega} = \int_{\bk\omega} A_{\bk_{\txF}\omega} \delta \Pi^{\txs}_{\bk\omega} B_{\bk_{\txF}\omega}  + O(\eta_\txe),
\end{equation}
The proof of Eq.~\eqref{eq:approx1} follows two steps:

First, we prove the following equation,
\begin{equation}
\label{eq:pi_minus_pis}
    \int_{\bk\omega} A_{\bk\omega}\Pi_{\bk\omega}W^{\txph}_{\bk\omega; \bk'\omega'} = \int_{\bk\omega} A_{\bk\omega}\Pi^{\txs}_{\bk\omega}W^{\txph}_{\bk\omega; \bk'\omega'} + O(\eta_\txe)
\end{equation}
The phonon propagator $W^{\txph}$ imposes a UV frequency cutoff $\omega_\txD$ to the internal frequency $\omega$. As a result, the total contribution is suppressed by the small parameter $\eta_\txe$.

Consider the renormalized phonon propagator,
\begin{equation}
    W^{\txph}_{\bk\omega;\bk'\omega'} \equiv \frac{\omega_{\bk-\bk'}^2}{(\omega-\omega')^2+\omega_{\bk-\bk'}^2}\le \frac{\omega_\txD^2}{(\omega-\omega')^2+\omega_\txD^2},
\end{equation}
where we assume that the phonon dispersion is upper bounded by the Debye frequency $\omega_{\bk-\bk'}\le \omega_\txD$.

We show that any bounded and absolutely integrable function $\psi_{\bk\omega}$ convoluted with the phonon propagator is of the order $O(\eta_\txe)$,
\begin{equation}
\label{eq:theorem1}
\begin{split}
    \left|\int_{\omega\bk} \psi_{\bk\omega}W^{\txph}_{\bk\omega; \bk'\omega'}\right| 
   & \le  \int_{\omega\bk} \frac{\left|\psi_{\bk\omega} \right|\omega_\txD^2}{(\omega-\omega')^2+\omega_\txD^2} \\ 
    &\le \pi \omega_\txD \sup_{\omega''}\int_{\bk} {|\psi_{\bk\omega''}|}   
    =  O(\eta_\txe)
\end{split}
\end{equation}

We then observe that 
\begin{equation}
\label{eq:delta_pi}
    \Pi_{\bk\omega} - \Pi^{\txs}_{\bk\omega} = \frac{\left(z^\text{e}\right)^2}{\left(\frac{\omega}{z^{\txph}_\omega}\right)^2+\varepsilon_{\bk}^2}\Theta(\varepsilon_\bk - \omega_\txc) + \text{reg}
\end{equation}
is a bounded function that satisfies Eq.~\eqref{eq:theorem1}. Therefore, replacing $\Pi_{\bk\omega}$ with $\Pi_{\bk\omega}^s$ only leads to corrections of the order $O(\eta_\txe)$, as written in Eq.~\eqref{eq:pi_minus_pis}.

We further prove
\begin{equation}
\label{eq:pis_to_kF}
    \int_{\bk\omega} A_{\bk\omega}\Pi^{\txs}_{\bk\omega}W^{\txph}_{\bk\omega;\bk'\omega'} = \int_{\bk\omega} A_{\bk_{\txF} \omega}\Pi^{\txs}_{\bk\omega}W^{\txph}_{\bk_{\txF} \omega;\bk'\omega'} + O(\eta_\txe).
\end{equation}
Consider the first order corrections 
from the momentum dependence of $A$ and $W^{\txph}$. We assume that 
\begin{eqnarray}
    A_{\bk\omega} = A^{(0)}_{\omega}+{\mathbf{A}}^{(1)}_{\omega}\cdot{\delta\bk}/k_{\txF} + O({|\delta\bk|^2/k_{\txF}^2}),
\end{eqnarray}
and similar relation for $W^{\txph}$. Then the leading correction is 
\begin{equation}
\begin{split}
 &\left|\int_{\bk\omega} (\mathbf{A}^{(1)}_{\omega}\cdot\delta{\bk}/k_\txF)\Pi^{\txs}_{\bk\omega}W^{\txph}_{\bk_{\txF} \omega;\bk'\omega'}\right|\\
 \le &|\mathbf{A}^{(1)}_{0}| \int_{\omega} \frac{\omega_\txD^2}{(\omega-\omega')^2+\omega_\txD^2}
 \int_{\varepsilon_\bk<\omega_\txc}\frac{k_\txF^2}{v_\txF^*E_\txF}\frac{|\varepsilon_\bk|}{\omega^2+{\varepsilon_{\bk}}^2}+O(\eta_\txe)\\
  \le &|\mathbf{A}^{(1)}_{0}|\frac{k_\txF^2}{v_\txF^*E_\txF}
  \int_{\omega} \frac{\omega_\txD^2}{(\omega-\omega')^2+\omega_\txD^2} \ln\left(1+\frac{\omega_\txc^2}{\omega^2}\right)+O(\eta_\txe)\\
 =&O(\eta_\txe)
\end{split}
\end{equation}
Notice that the integrand vanishes for $\omega \gg \omega_\txD$ in the last equation, such that the first term is proportional to $\omega_\txD/E_\txF<\eta_\txe$ .

To prove Eq.~\eqref{eq:approx2}, we taylor expand both $\mathbf{A}$ and $\mathbf{B}$. Since the equation is symmetric with respect to them, we only consider the following leading correction
\begin{equation}
\begin{split}
 &\left|\int_{\bk\omega} (\mathbf{A}^{(1)}_{\omega}\cdot\delta{\bk}/k_F)\delta\Pi^{\txs}_{\bk\omega}B_{k_{\txF}\omega}\right|\\
 \le &|\mathbf{A}^{(1)}_{0}| \int_{\omega} B_{k_{\txF}\omega}
 \int_{\varepsilon_\bk<\omega_\txc}\frac{k_\txF^2}{v_\txF^*E_\txF}\frac{(z^{\txe})^2\left|\varepsilon_\bk\right|\left(\frac{1}{{z^{\txph}_\omega}}-1\right)\omega^2}{(\omega^2+\varepsilon_{\bk}^2)\left(\left(\frac{\omega}{z^{\txph}_\omega}\right)^2+\varepsilon_{\bk}^2\right)}\\
 &+O(\eta_\txe)\\
 \le &|\mathbf{A}^{(1)}_{0}|\frac{k_\txF^2}{v_\txF^*E_\txF}
  \int_{\omega}   \frac{(z^{\txe})^2 B_{k_{\txF}\omega}}{(1+\frac{1}{z^{\txph}_\omega})}\ln{ \left( \frac{\omega^2+\omega_\txc^2}{\omega^2+(z^{\txph}_{\omega}\omega_\txc)^2}\right)}+O(\eta_\txe)\\
 =&O(\eta_\txe)
\end{split}
\end{equation}
Again, notice that the integrand vanishes as $z^{\txph}_{\omega}$ approaches $1$ for $\omega \gg \omega_\txD$. Therefore, the first term is proportional to $\omega_\txD/E_\txF<\eta_\txe$.

In the next step, we extract the low-energy contribution from the electron four-vertex function $\Gamma^{\txe}$. The first observation is that $\Gammapp^{*}$ in Eq.~\eqref{eq:Gammapp*}  inherits the Coulomb singularity in $\Gammapp$. This can be proved by contradiction if we rewrite
\begin{equation}
    \Gammapp^{*}_{\bk^{}\omega^{};\bk'\omega'}  + \int_{\bk''\omega''} \phi_{\bk^{}\omega^{};\bk''\omega''} \Gammapp^{*}_{\bk''\omega'';\bk'\omega'}= \Gammapp^{\txe}_{\bk^{}\omega^{};\bk'\omega'}
\end{equation}
If $\Gammapp^{*}$ is a regular function, the l.h.s is also regular, which contradicts the Coulomb singularity in the r.h.s. Therefore, following Eq.~\eqref{eq:fullGamma_scaling},  we separate $\Gamma^{\txe}$ into three distinct contributions as follows:
\begin{eqnarray}
\label{eq:components}
  \Gamma^{\txe}_{\bk^{}\omega^{};\bk'\omega'}=\Gamma^{\txe}_0 + W^s_{\bk^{}-\bk', \omega^{}-\omega'} + \delta{\Gamma}_{\bk^{}\omega^{};\bk'\omega'}.
\end{eqnarray}

Here, $\Gamma^{\txe}_0$ represents the Fermi surface-averaged electron vertex function, formally given by 
\begin{equation}
\Gamma^{\txe}_0 \equiv \left<\Gamma^{\txe}_{\bk^{}\omega^{};\bk'\omega'}\right>_{|\bk^{}\omega^{}|=|\bk'\omega'|=(k_{\txF}, \pi T)}.
\label{eq:gamma4average}
\end{equation}
This term scales as $\propto \frac{1}{1+g_0\ln(\omega_\txp/T)}$ at temperatures
much lower than $\omega_\txp$ as given by Eq.~\eqref{eq:fullGamma_scaling}. The Coulomb singularity $W^{s}$ follows the same parametrization in Eq.~\eqref{eq:plasmonsing}. The remaining part of the full electron vertex function is denoted as $\delta{\Gamma}$. It should be regular up to the scale the momentum scale $k_{\txF}$ and the energy scale $v_{\txF}^* k_{\txF}$.

The problem could be simplified significantly if the plasmon mode is gapped as in 3D electron system with moderate $r_\txs$ or the interaction between electrons are short-range.
In what follows, we demonstrate that the singular component $W^s_{\bk^{}-\bk', \omega^{}-\omega'}$ contributes terms that scale as $O(\eta_\txp^2,\eta_\txe)$, where $\eta_\txp=\omega_\txc/\omega_\txp$.  This scaling behavior implies that both the $W^s\delta\Pi \pcf$ and $W^s\Pi^{\txe}W^{\txph}\Pi \pcf$ terms make negligible contributions to the overall result. It is important to note, however, that this does not diminish the relevance of the plasmon mode itself—rather, its contribution is already effectively captured within $\Gamma_0^{\txe}$.

To estimate the contributions from $W^s\delta\Pi \pcf$ and $W^s\Pi^{\txe}W^{\txph}\Pi \pcf$, we first consider their low frequency components. 
Considering a spherical Fermi surface, the momentum integration in $W^s\delta\Pi \pcf$ takes the form:
\begin{equation}
\int k'^2dk' \ln(|k'-k_\txF|)\delta\Pi_{k' \omega'}\pcf_{k'\omega'}
=\frac{C}{|\omega'|}\Theta(\omega_\txD-|\omega'|)+\text{reg},
\end{equation}
where $C$ represents a constant factor. 
This result arises from two key observations. 
First, $\pcf_{\bk',\omega'}$, as derived in Eq.~\eqref{eq:RK_final}, remains smooth near $\omega=0$ and $|\bk'|=k'_{\txF}$. 
Second, $\delta\Pi_{k' \omega'}$ decays at high frequencies ($\omega' \gg \omega'_\txD$) as ${z_{\txph}^2(\omega')}$ approaches unity, leading to the $\Theta(\omega'_\txD-|\omega'|)$ constraint. 
For frequencies below $\omega'_\txD$, $\delta\Pi_{k' \omega'}$ consists of two terms scaling as $1/(\omega'^2 + (v_{\txF}^*(k'-k_\txF)^2)$, leading to the singular $1/|\omega'|$ behavior.

We denote the low frequency components ($\omega \lesssim  \omega_\txD$) by $\txL$ and the high frequency components ($\omega_\txD \ll \omega \lesssim \omega_\txc$) by $\txH$.
Then we have
\begin{equation}
  \begin{split}
  \left[W_\txs\delta\Pi \pcf\right]_\text{L}&\sim\sum_{\omega'}\frac{{\omega'}^2}{{\omega'}^2+\omega_\txp^2}
  \frac{1}{|\omega'|}\Theta(\omega_\txD-|\omega'|) \\
  &\propto \frac{\omega_\txD^2}{\omega_\txp^2}=O(\eta_\txp^2\eta_\txe^2).
    \end{split}
\end{equation}
Similarly, the smallness of $W_\txs\Pi^{\txe} W^{\txph}\Pi \pcf$ at low frequencies can be demonstrated through
\begin{equation}
    \begin{split}
\label{eq:WPiWPiR}
  \left[W_\txs\Pi^{\txe} W^{\txph}\Pi \pcf\right]_{L} &\sim\sum_{|\omega'|, |\omega''|<\omega_\txc}^{|\omega'-\omega''|<\omega_\txD}\frac{{\omega'}^2}{{\omega'}^2+\omega_\txp^2}
  \frac{1}{|\omega'|}\frac{1}{|\omega''|}\\
  &\sim\sum_{|\omega'|<\omega_\txc}\frac{\omega_\txD}{{\omega'}^2+\omega_\txp^2}\\
  &\propto \frac{{\omega_\txD}}{\omega_\txp}=O(\eta_\txe\eta_\txp),
    \end{split}
\end{equation}
Here, the $1/|\omega'|$ and $1/|\omega''|$ terms arise from the momentum integration of $W_\txs\Pi^{\txe}$ and $\Pi \pcf $, whereas the frequency summation constraint is due to the vanishing of $W^{\txph}$ at $|\omega'-\omega''|\gg\omega_\txD$. 

In addition, we must estimate the contribution at frequencies $\omega\sim\omega_\txc$, since $\Lambda$ appears in the self-consistent equations for frequencies below $\omega_\txc$.
Similar to the low frequency components,we have
\begin{equation}
  \begin{split}
  \left[W_\txs\delta\Pi \pcf\right]_\text{H}&\sim\sum_{\omega'}\frac{{\left(\omega'-\omega_\txc\right)}^2}{{\left(\omega'-\omega_\txc\right)}^2+\omega_\txp^2}
  \frac{1}{|\omega'|}\Theta(\omega_\txD-|\omega'|) \\
  &\propto \frac{\omega_\txc^2}{\omega_\txp^2}=O(\eta_\txp^2),
    \end{split}
\end{equation}
and
\begin{equation}
  \begin{split}
\label{eq:WPiWPiR_H}
\left[W_\txs\Pi^{\txe} W^{\txph}\Pi \pcf\right]_{H} &\sim\sum_{|\omega'|, |\omega''|<\omega_\txc}^{|\omega'-\omega''|<\omega_\txD}
\frac{{\left(\omega'-\omega_\txc\right)}^2}{{\left(\omega'-\omega_\txc\right)}^2+\omega_\txp^2}
\frac{1}{|\omega'|}\frac{1}{|\omega''|}\\
&\propto \frac{{\omega_\txc^2}}{\omega_\txp^2}=O(\eta_\txp^2).
  \end{split}
\end{equation}
These contributions are also small; however, they are controlled by $\eta_\txp$ alone, contrary to the low frequency contributions that are controlled by $\eta_\txp\eta_\txe$.
Although these high-frequency contributions do not affect the phonon physics, they do affect the renormalization of the electron-electron interaction. 
If the condition $\eta_\txp \ll 1$ is not met, these terms will contribute a correction to the pseudopotential proportional to $\eta_\txp^2$.

After applying this approximation
we can simplify Eq.~\eqref{eq:pcfephfull2} to
\begin{equation}
    \begin{split}
  \pcf  = \pcf^{\txe}&+ (\Gamma_0^{\txe}+\delta{\Gamma})\delta\Pi \pcf  \\
  &+
  [1+(\Gamma_0^{\txe}+\delta{\Gamma})\Pi^{\txe}]W^{\txph}\Pi \pcf+O(\eta_\txp^2,\eta_\txe).
    \end{split}
\end{equation}

Now using Eqs.~\eqref{eq:approx1} and \eqref{eq:approx2}, we have
\begin{eqnarray}
  (\Gamma_0^{\txe}+\delta{\Gamma})\delta\Pi \pcf =\Gamma_0^{\txe}\delta\Pi^{\txs} \bar{\pcf}  + O(\eta_\txe),
\end{eqnarray}
\begin{eqnarray}
  W^{\txph}\Pi \pcf =\bar{W}^{\txph}\Pi^{\txs}\bar{\pcf}  + O(\eta_\txe),
\end{eqnarray}
and
\begin{eqnarray}
  (\Gamma_0^{\txe}+\delta{\Gamma})\Pi^{\txe} W^{\txph}\Pi \pcf=\Gamma_0^{\txe}\Pi^{\txe,\txs}\bar{W}^{\txph}\Pi^{\txs}\bar{\pcf}+ O(\eta_\txe),
\end{eqnarray}
where $\bar{W}^{\txph}_{\omega^{ },\omega'}=W^{\txph}_{\bk_{\txF}\omega^{ };\bk_{\txF}\omega'}$, and
$\bar{\pcf}^{\txe} \equiv \pcf^{\txe}_{\omega} \equiv \pcf^{\txe}_{\bk_{\txF}\omega}$.
And the equation becomes
\begin{eqnarray}
  \bar{\pcf}  = \bar{\pcf}^{\txe}+ \Gamma_0^{\txe}\delta\Pi^{\txs} \bar{\pcf} +(1+\Gamma_0^{\txe}\Pi^{\txe,\txs})\bar{W}^{\txph}\Pi^{\txs}\bar{\pcf} + O(\eta_\txe,\eta_\txp^2),
\end{eqnarray}
where $\bar{\pcf}_{\omega}=\pcf_{k_{\txF},\omega}$. Notice that in this equation the momentum dependences only appear in $\Pi^{\txs}$, $\Pi^{\txe,\txs}$ and $\delta\Pi^{\txs}$, and can be integrated out.

Further substitute $\Gamma_0^{\txe}$ with
\begin{eqnarray}
  \label{eq:UfromGamma}
  U = \Gamma_0^{\txe} - \Gamma_0^{\txe}\Pi^{\txe,\txs}U,
\end{eqnarray}
where by subtracting the Cooper logarithm from $\Gamma_0^{\txe}$, $U$ is now temperature independent and serves as the TMA pseudopotential in our equation below.
And the Eq.~\eqref{eq:pcfephfull2} is further simplified as
\begin{align}
  \label{eq:rpcf}
  \bar{\pcf}_{\omega} = \eta_{\omega}
                      + \sum_{\omega'}(U+\bar{W}^{\txph}_{\omega,
                          \omega'})\bar{\Pi}^s_{\omega'} \bar{\pcf}_{\omega'}+O(\eta_\txe,\eta_\txp^2),
\end{align}
where 
\begin{equation}
    \eta_{\omega}=\bar{\pcf}^{\txe}_{\omega}-U\int_{\omega'}\Pi^{\txe,\txs}_{\omega'}\pcf^{\txe}_{\omega'}
\end{equation}
and
\begin{equation}
 \bar{\Pi}^s_{\omega}=\frac{2\left(z^\text{e}\right)^2z^{\txph}_{\omega}\tan^{-1}(\frac{z^{\txph}_{\omega}\omega_\txc}{|\omega|})}{v_{\txF}^*|\omega|}
 \label{eq:pis_softcutoff}
\end{equation}
The source term $\eta_{\omega}$ is temperature independent in the low frequency
part, and the form of this term has no effects on the resulting $T_c$, thus we
can simply set this quantity to unity in calculations. This results in correct
$T_c$, but the obtained $\pcf_0$ will differ by a factor from the exact one. 

While the $\bar{\Pi}^s_{\omega}$ defined in Eq.~\eqref{eq:pis_softcutoff} does not contain an explicit cut-off in frequency domain, the additional inverse tangent factor guarantees the decay of this function above $\omega_\txc$. 
In numerical calculations, an alternative form of $\bar{\Pi}^s_{\omega}$ with hard cut-off would be more convenient:
\begin{equation}
 \bar{\Pi}^s_{\omega}=\frac{2\left(z^\text{e}\right)^2z^{\txph}_{\omega}}{v_{\txF}^*|\omega|}\Theta(\omega_\txc-\left|\omega\right|).
 \label{eq:pis_hardcutoff}
\end{equation}
There should be no additional difficulties to show all the derivation above still holds for this choice of form.

\section{First-Principle Calculations of the Vertex Functions}

Now we explain how to compute the relevant vertex functions in the UEG model with the help
of VDiagMC approach.
These include the Coulomb pseudopotential derived in the previous section by sampling the
two-quasiparticle scattering amplitude averaged on Fermi surface [see Eq.~\eqref{eq:gamma4average}],
and the 3-vertex appearing in the phonon-induced interaction $W^{\txph}$.
Here we adopt the same convention for three dimension with angular momentum quantum number $\ell=0$, which corresponds to the s-wave case.

\subsection{Spin and Angular Momentum Conventions}
The UEG model pertains full translational and rotational symmetries, 
and the superconducting properties depends on the symmetry channel.
In this work, we focused on the s-wave spin singlet superconductivity, 
although the conventions presented below were generic.
The generic discussion of angular momentum decomposition of the PCF equation could be found in Appendix A of Ref.~\cite{pengcheng}.

To illustrate the spin convention, we use Greek letter $\alpha,\beta,\gamma,\delta$ to represent the spin degree of freedom and abbreviate the space coordinates of particles with numbers. Then, the response function $\pcf_{\alpha\beta}(12)$, like the gap-function of
superconductivity, has the following property:
\begin{equation}
  \pcf_{\alpha\beta}(12) = -\pcf_{\beta\alpha}(21).
\end{equation}
It is more convenient to decompose the spin configuration of the response
function into the singlet and triplet components, 
\begin{equation}
  \begin{split}
    \pcf_{\txs}(12) &= \frac{1}{2}(\pcf_{\uparrow\downarrow}(12)-\pcf_{\downarrow\uparrow}(12))\\
    \pcf_{\txt}(12) &= \frac{1}{2}(\pcf_{\uparrow\downarrow}(12)+\pcf_{\downarrow\uparrow}(12))\\
  \end{split}
\end{equation}
such that
$\pcf_{\txs}(12) = \pcf_{\txs}(21),\pcf_{\txt}(12) = -\pcf_{\txt}(21)$.
And similarly we have $F_{\txs/\txt}(12)=\int_{34}G(13)\pcf_{\txs/\txt}(34)G(42)$.

For 4-vertices, the direct and exchange components contribute equally in Cooper
channel, thus we needed to consider only the direct part. For direct part we had
\begin{equation}
  \Gamma^\txd_{\alpha\beta\gamma\delta}(12;34) \equiv \Gamma^{+}_{12;34}\delta_{\alpha\beta}\delta_{\gamma\delta} + \Gamma^{-}_{12;34} \vec{\sigma}_{\alpha\beta}\cdot \vec{\sigma}_{\gamma\delta}.
\end{equation}
Then the equation
\begin{equation}
  \pcf_{\alpha\gamma} = \eta_{\alpha\gamma} + \Gamma^\txd_{\alpha\beta\gamma\delta} F_{\alpha\gamma}
\end{equation}
could be decomposed as
\begin{equation}
  \begin{split}
    \pcf_\txs &= \eta_\txs + (\Gamma^\txs-3\Gamma^\txa)F_\txs, \\
    \pcf_\txt &= \eta_\txt + (\Gamma^\txs+\Gamma^\txa)F_\txt. \\
  \end{split}
\end{equation}

In diagMC implementation, it is more convenient to work with the following
convention for 4-vertices:
\begin{equation}
  \begin{split}
    \Gamma_{\txuu}(12;34) &= \Gamma^+_{12;34}+\Gamma^-_{12;34}, \\
    \Gamma_{\txud}(12;34) &= \Gamma^+_{12;34}-\Gamma^-_{12;34}.
  \end{split}
\end{equation}
Thus
\begin{equation}
  \begin{split}
    \Gamma^{\txs}&=\Gamma^+-3\Gamma^-=-\Gamma_{\txuu}+2\Gamma_{\txud}, \\
    \Gamma^{\txt}&=\Gamma^++\Gamma^-=\Gamma_{\txuu}.
  \end{split}
\end{equation}

Since the Coulomb pseudopotential is defined for s-wave superconductivity, only the singlet case was considered in this work. However, there is no restriction to extend the definition to higher angular momentum cases, where even $\ell$ corresponds to spin singlet state and odd $\ell$ corresponds to triplet.

In addition, we defined dimensionless versions of $\Gamma^{\txe}_0$ and
$U$ as $\gamma_0 = z_\txe^2N_{\txF}^*\Gamma^{\txe}_0 $ and $u = UN_{\txF}^*$, with $N_{\txF}=\frac{m^*k_{\txF}}{2\pi^2}$.
Note that the spin factor is omitted as the spin degree of freedom is accounted as described above.
We evaluated the direct and exchange components on equal footing, 
thus the $\Gamma_0$ is obtained by taking half the value.

\subsection{First-principle approach to the pseudopotential}

As mentioned in  Sec.[main], we computed the self-energy and four-point vertex function for the renormalized UEG theory with VDiagMC. 
In this section we explained the details of these calculations.

\subsubsection{Self-energy expansion}
The self-energy calculation followed the approach explained in Ref.~\cite{hou_feynman_2024}. Here we only briefly illustrated the procedure.

After performing the $\xi$ expansion mentioned in Sec.[main] in the main text, 
the self-energy was expanded as a series
\begin{equation}
  \Sigma(\xi) = \sum_{n} {\xi^{n}}\Sigma^{(n)}, \label{eq:sigma_xi_dmu_dlambda}
\end{equation}
where each term $\Sigma^{(n)}$ represents a group of diagrams that could be evaluated stochastically as explained in Ref.\cite{hou_feynman_2024}.
The desired quantities, quasiparticle weight and the effective mass,
was then given by the equation
\begin{flalign}
  z^\txe = \left(1 - \left.\frac{\partial\text{Im}\Sigma(k_{\rm F}, i\omega)}{\partial (i\omega)}\right|_{\omega = 0}\right)^{-1},
\end{flalign}
and
\begin{flalign}
  \label{eq:eff_mass_ratio}
  \frac{m_\txe^*}{m} = \frac{1}{z^\txe} \cdot 
  \left(1 + 
  \frac{m}{k_{\rm F}}\left.\frac{\partial \text{Re}\Sigma(k, 0)}{\partial k}\right|_{k = k_{\rm F}}\right)^{-1}.
\end{flalign}
Plugging in Eq.\eqref{eq:sigma_xi_dmu_dlambda} resulted in a series expansion of these quantities with respect to $\xi$:
\begin{equation}
  z^\txe = 1 + \sum_{n=1}^\infty\delta z^{(n)}\xi^n,
\end{equation}
and
\begin{equation}
  \bar{m}=\frac{m_\txe^*}{m} = 1 + \sum_{n=1}^\infty \delta m^{(n)}\xi^n.
\end{equation}

\subsubsection{Four-point vertex expansion}

The same diagrammatic expansion as the self-energy was performed for the four-point vertex, and this resulted in a series expansion of $\Gamma^\txe$.
To obtain the dimensionless quasiparticle scattering amplitude $\gamma_T$, we needed to multiply $z^\txe$ and $\frac{m_\txe^*}{m}$, and also $N_\txF$ onto it.
The effective mass part was straightforward.
We simply multiply the series of $\Gamma^\txe$ and $\frac{m_\txe^*}{m}$ term-by-term and collect the results as a series of $\xi$. 
The quasiparticle residue part was less trivial. 
To optimize the convergence, we multiplied the $\left[z^\txe(\xi)\right]^2$ not on the final series of $\Gamma^\txe$, but on the interaction $v_\txR$ within the expansion. 
The diagrams thus generated was again re-grouped as a series of $\xi$.
By doing so, the higher order counterterms generated by the series expansion of $z^\txe$ canceled the contributions from the vertex correction and electron's self-energy, yielding a better convergence. The relation between $z^\txe$ and 3-vertex correction could be found in Sec.~\ref{sec:app:ver3}. 

\subsubsection{Frequency cut-off shift}
\label{sssec:omegashift}

While the converged result of $\psp$ was determined solely by the chosen frequency cut-off $\omega_c$, the finite-order expansion was less-trivial. 
In a finite-order expansion as given in previous sections, the Green's functions in the Cooper ladder diagrams are only resummed up to a finite-order, resulting in a finite-order effective mass appearing in it. 
This finite-order effective mass differs from the converged one, thus affected the logarithmic divergent terms generated from the ladder diagram.
Specifically, when the cut-off was introduced from the momentum space, the momentum cut-off was determined by $\Delta k=\omega_c / v_\txF^* \propto \bar{m}\omega_c$.
The cut-off $\omega_c$ should be fixed for all orders, but for finite order diagrams, the effective mass appears only as a finite order correction, resulting in a different $\Delta k$ for different order.
Thus, in order to cancel out this effect, an additional $\bar{m}^{(n)}$ needed to appear in the logarithmic divergence factor, resulting in $\ln(\frac{\bar{m}^{(n)}\omega_c}{T})$ instead of $\ln(\frac{\omega_c}{T})$.
This could in turn be regarded as an effective cut-off shift to 
\begin{equation}
  \omega_{\txc}^{(n)} = \omega_{\txc}(1+\sum_{i=1}^{n}\delta m^{(i)}).
\end{equation}

The n-th order result $\psp^{(n)}(\omega_{\txc}^{(n)})$ could be obtained by subtracting the Cooper instability logarithm from the $\gamma$ series as Eq.\eqref{eq:psp_resum} showed, with proper frequency cut-off inserted:
\begin{equation}
  \begin{split}
    \psp^{(1)} &= \gamma_T^{(1)}, \\
    \psp^{(2)} &= \gamma_T^{(2)} - \gamma_T^{(1)}\ln(\omega_{\txc}^{(n)}/T)\gamma_T^{(1)},\\
    &\ldots.
  \end{split}
\end{equation}
The series of $\psp$ could then be shifted back to the desired cutoff by
\begin{equation}
  \psp^{(n)}(\omega_{\txc})=\frac{\psp^{(n)}(\omega_{\txc}^{(n)})}{1+\psp^{(n)}(\omega_{\txc}^{(n)})\ln(\omega_{\txc}^{(n)}/\omega_{\txc})},
\end{equation}
and sum up to the highest computed order. This leads to an estimation of $\psp$ to this order with cutoff $\omega_{\txc}$.

By doing so, the contribution of effective mass to the logarithmic
divergent terms could be canceled out order by order.
To see this, consider third order (contribution does not appear until the third
order) expansion of $\psp$. 

For the third order, we have
\begin{eqnarray}
  \psp^{(3)}
     &=& \gamma^{(3)} + \delta m^{(1)}\gamma^{(2)} + \delta m^{(2)}\gamma^{(1)}\nonumber\\
            && -2\gamma^{(1)}L\gamma^{(2)}-2\gamma^{(1)}L\gamma^{(1)}\delta m^{(1)}\nonumber \\
            && +\gamma^{(1)}L\gamma^{(1)}L\gamma^{(1)},
\end{eqnarray}
where $L=\ln(\omega_\txc/T)$.
The logarithmic divergent contribution of third order particle-particle
reducible diagram with one particle propagator dressed by first order
self-energy, and of $\delta m^{(1)}\gamma^{(2)}$,  are canceled by 
$-2\gamma^{(1)}L\gamma^{(1)}\delta m^{(1)}$ term.

\subsubsection{Mathematical Status of the Resummation Protocol}
\label{sec:app:math_status}

The mathematical status of the implemented resummation protocol relies on the analyticity of $\gamma_T(\xi)$. Our only assumption---not fully controlled but reasonable at the level of self-consistency---is the analyticity of $\gamma_T(\xi)$ as a function of complex $\xi$ within a certain domain containing points $\xi=0$ and $\xi=1$. This implies that the r.h.s. of Eq.~(\ref{eq:gamma_T_series}) is the Taylor expansion (convergent at small enough $|\xi|$) of $\gamma_T(\xi)$ at the point $\xi=0$ that uniquely defines---by analytic continuation---the desired value of $\gamma_T(\xi =1)$. The rest is a mathematically rigorous method of resummation of a Taylor series of an analytic function by the conformal map technique outlined in the following. 

Let $f(\xi)$ be a certain analytic function represented by the (Taylor) expansion in powers of the complex argument $\xi$. In the most general case, the conformal map analytically transforms both the function $f$ and its argument $\xi$ so that we deal with a new analytic function, $g$, of a new argument, $w$:
\begin{equation}
g(w) = Q(f(\xi(w)),w) \, .
\label{g_of_w}
\end{equation}
Here, the old argument $\xi$ is replaced with a new argument $w$ by introducing an analytic function $\xi \equiv \xi(w)$, such that $\xi(w=0) = 0$, $\xi(w=1) = 1$; and the old function $f(\xi)$ is replaced with the new function $g(w)$ with the help of the function $Q(f,w)$, the form of which can be quite arbitrary provided the resulting function $g(w)$ is analytic at any $|w|\leq 1$. The resummation procedure amounts to finding the value of $g(w=1)$ by summing up its Taylor expansion in powers of $w$; the coefficients of the series being related to the coefficients of the original series for $f(\xi)$ by expanding the r.h.s. of Eq.~(\ref{g_of_w}) in powers of $w$. The estimated value of $g(w=1)$ then allows one to find the desired value of $f(\xi=1)$ from Eq.~(\ref{g_of_w}).

The most commonly used particular form of conformal map resummation does not involve introducing a new function. Less common but most relevant for our work is yet another particular case in which one introduces a new function while working with the old variable $\xi$:
\begin{equation}
g(\xi) = Q(f(\xi),\xi) \, .
\label{g_of_xi}
\end{equation}
This type of conformal map is natural when the function $g(\xi)$, rather than the series-generating function $f(\xi)$, is of our main interest, so that there is no need to restore $f$ from $g$. Our case is precisely like that: The series-generating function $\gamma_T(\xi)$ is conformally mapped onto the function of our interest, $\mu_{\omega_\txc}(\xi)$, and the estimated value of $\mu_{\omega_\txc}(\xi=1)$ is then used to extract $\mu_{_{E_\txF}}$.

In the absence of a proof of the analyticity of $\gamma_T(\xi)$ as a function of $\xi$, we formulate an \textit{a posteriori} argument in favor of the consistency of our resummation protocol. The argument is two-fold: (i) the fact of the convergence of the series Eq.~(\ref{eq:psp_series}) at $\xi=1$ within a certain range of the values of parameter $\omega_\txc$, defining the particular form of the coefficients of the series; and (ii) the independence of the extracted result for $\mu_{_{E_\txF}}$ from the particular choice of $\omega_\txc$.

\subsubsection{Results}
\begin{figure}
    \centering
    \includegraphics[width=0.98\linewidth]{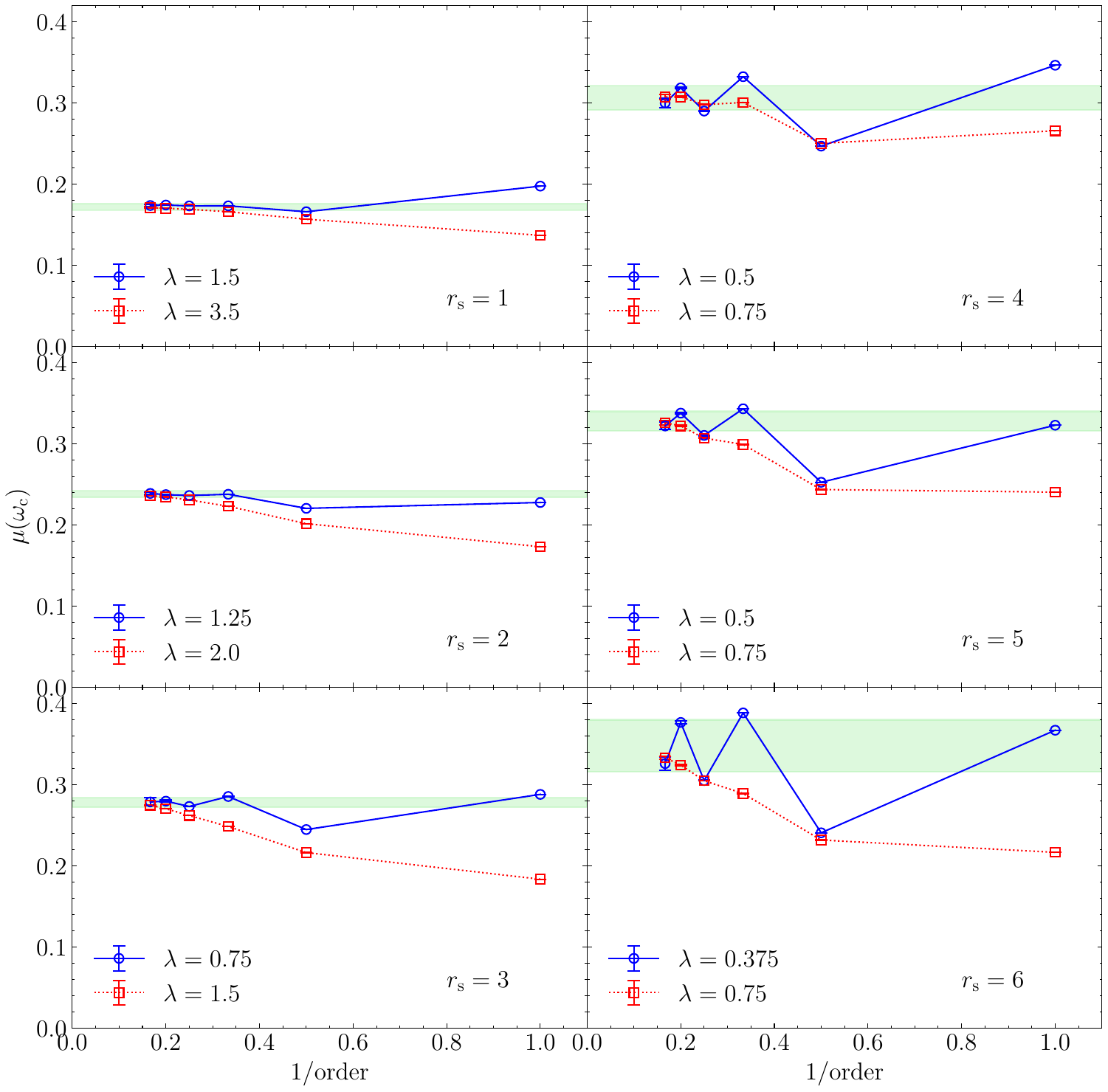}
    \caption{Sixth order series of the pseudopotential $\psp(\omega_{\txc}=0.1E_{\txF})$ for $r_\txs=1$ to $6$.  For each $r_\txs$, results from two $\lambda$ were presented. The light green labels the range of estimated convergent result from the data.}
    \label{fig:psp_converg}
\end{figure}

An example of temperature dependence of series expansion of $\gamma_T$ and $\psp$, computed at $r_\txs=1.0$ with $\lambda_\txR=3.5$, were presented in Fig.\ref{fig:lnT_m_renorm}.
The converged results for $r_\txs=1$ to $6$ were shown in Fig.~\ref{fig:psp_converg}.

\subsection{First-principle approach to the 3-Vertex}
\label{sec:app:ver3}

In this section, we show the validity of the ansatz Eq.~\eqref{eq:ver3_ansatz} in the context of UEG by comparing it with the VDMC results. 
Two additional series expansions need to be obtained in order to give the desired quantity $W\Gamma^\txe_3$: the proper 3-vertex $\Gamma^\txe_3$ and the polarization $P$ (we use $P$ to represent polarization in this section to distinguish it from the quasiparticle pair-field propagator). 
In addition, a series of exchange-correlation kernel $f_{\txxc}$ with its convergent value are also needed for better convergence of the series.
All these functions could be obtained with VDMC sampling in a similar way as described in the last subsection, with the diagrammatic expansion altered to the corresponding ones.

The post-processing of these results need to be done cautiously such that terms with bad convergence cancels each other order by order. 
Specifically, the polarization and proper 3-vertex is characterized by the following ansatz for moderate $r_\txs$:
\begin{equation}
  P=\frac{P}{1-f_{\txxc}P^{(0)}}+\text{correction},
\end{equation}
and
\begin{equation}
  z^\txe\Gamma^{\txe}_3=\frac{1}{1-f_{\txxc}P^{(0)}}+\text{correction},
\end{equation}
where $P^{(0)}$ is the bare polarization.
At around $r_\txs=5$, the denominator began to vanish at small $\bq$, resulting in a divergent series for both quantities.
However, if handled correctly, the divergence cancels each other.

First, we compute the series of $z^\txe\Gamma^{\txe}_3$ and $P$, which are not converging around $r_\txs\approx 5$.
Then, we compute two quantities, $z^\txe\Gamma^{\txe}_3/P$ and $WP$. 
By carefully cancel the divergent terms order by order, the series of these two quantities could be made convergent.
In the end, we multiply them to obtain the final result $z_\txe\Gamma^{3e}v_{\bq}$.
Below we discuss step-by-step how this could be done.

Suppose we have series expansions
\begin{equation}
    z^\txe(\xi)\Gamma^{\txe}_3(\xi) = 1 + \xi\delta\verthree^{(1)} + \xi^2\delta\verthree^{(2)}+\ldots,
\end{equation}
and
\begin{equation}
    P(\xi) = P^{(0)} + \xi\delta P^{(1)} + \xi^2\delta P^{(2)} + \ldots,
\end{equation}
and we write $\verthree^{{(n)}}=\sum_{i=0}^{n}\xi^i\delta\verthree^{(i)}$ with $\verthree^{(0)}=\delta\verthree^{(0)}=1$, 
and $P^{(n)}=\sum_{i=0}^{n}\xi^i\delta P^{(i)}$ with $P^{(0)}=\delta P^{(0)}$.
Then we have the expansion series of $\verthree/P$ given by
\begin{equation}
    \delta\left(\verthree/P\right)^{(i)}=\frac{\delta\verthree^{(i)}P^{(i-1)}-\verthree^{(i-1)}\delta P^{(i)}}{{P}^{(i)}{P}^{(i-1)}},
\end{equation}
with $\left(\verthree/{P}\right)^{(0)}=1/{P}^{(0)}$.
The diverging part  in the numerator cancels, resulting in a convergent series of $\verthree/{P}$.

The situation is different for the $W{P}$ term. For $W{P}$ we have
\begin{equation}
    W{P} = -\frac{{P}}{{P}-v_{\bq}^{-1}},
\end{equation}
which expands with respect to the series of ${P}$ as
\begin{equation}
    \delta\left(W{P}\right)^{(i)}=\frac{v_{\bq}^{-1}\delta{P}^{(i)}}{({P}^{(i)}-v_{\bq}^{-1})({P}^{(i-1)}-v_{\bq}^{-1})},
\end{equation}
with $\left(W{P}\right)^{(0)}=-\frac{{P}^{(0)}}{{P}^{(0)}-v_{\bq}^{-1}}$.
It could be seen that a bad convergence of ${P}$ leads to a bad convergence of $W{P}$ unless $\bq=0$.
Thus we need to first obtain a convergent series of ${P}$.
To do this, we first compute ${P}(1-f_{\txxc}{P}^{(0)})$ order by order. 
Suppose we have the series expansion of $f_{\txxc}$ with \textit{the same expansion parameters} as ${P}$,
\begin{equation}
    f_{\txxc}=\xi f^{(1)}+\xi^2\delta f^{(2)} + \xi^3\delta f^{(3)} + \ldots,
\end{equation}
and $f^{(n)}=\sum_{i=1}^{n}\delta f^{(i)}$ with $f^{(1)}=\delta f^{(1)}, f^{(0)}=0$.
Then the series expansion of ${P}(1-f_{\txxc}{P}^{(0)})$ is given by
\begin{align}
  \delta\left[{P}(1-f_{\txxc}{P}^{(0)})\right]^{(i)}&=\delta{P}^{(i)}(1-f^{(i-1)}{P}^{(0)})\nonumber\\ 
  &-{P}^{(i-1)}\delta f^{(i)}{P}^{(0)}-\delta{P}^{(i)}\delta f^{(i)}{P}^{(0)}.
\end{align}
The convergent series of ${P}$ is then obtained by divide $(1-f_{\txxc}{P}^{(0)})$ from the convergent series ${P}(1-f_{\txxc}{P}^{(0)})$ with \textit{converged result of} $f_{\txxc}$.
This converged result of $f_{\txxc}$ could be obtained from other expansion series with different parameters.
The convergent series of $W{P}$ could then be computed and the final result could be obtained by multiplying the above results together. 
Results of this convergent series were shown in Fig.\ref{fig:ver3angle}.

\begin{figure}[H] 
    \centering
    \includegraphics[width=0.95\linewidth]{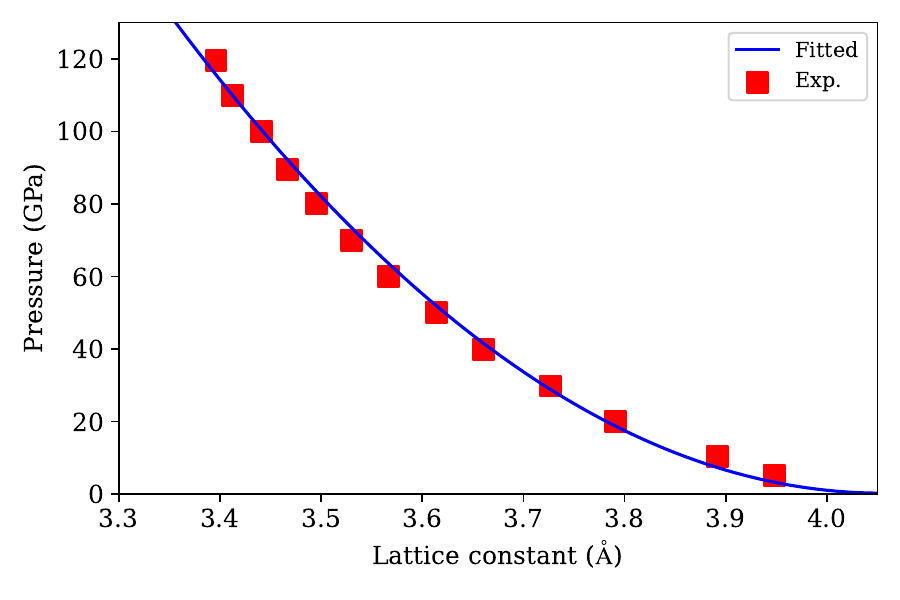}
    \caption{
    Pressure dependence of the Aluminum lattice constant. The red squares represent experimental data from Ref.~\cite{Al_EOS}, and the solid blue line denotes the quadratic fit used in this work to determine the structural parameters for finite-pressure calculations.
    }
    \label{fig:eos_al}
\end{figure}

\section{Equation of State for Aluminum}
\label{sec:app:al_pressure}
To determine the lattice parameter of FCC-Al as a function of pressure, we performed a fit to the experimental data from Ref.~\cite{Al_EOS} using a second-order polynomial. The resulting equation of state (EOS) allows us to interpolate the lattice constants used in our DFT calculations at arbitrary pressures. FIG.~\ref{fig:eos_al} displays the experimental data alongside our fitted curve, showing excellent agreement over the pressure range of interest.


\end{appendices}

\end{document}